\documentclass[longauth,desactivate]{aa} 

\usepackage{graphicx}
\usepackage{txfonts}
\usepackage{xcolor}
\usepackage[utf8]{inputenc}
\usepackage{float}
\usepackage{natbib}
\usepackage[colorlinks=true,linkcolor=blue, citecolor=blue, urlcolor=blue]{hyperref}

\newcommand{\kms}{{km\,s}$^{-1}$}

\newcommand{\teff}{$T_\mathrm{eff}$\,}
\newcommand{\logg}{$\log g$\,}

\newcommand{\vsini}{$\varv\sin i$}
\newcommand{\vsinift}{$\varv\sin i_{\rm FT}$}
\newcommand{\vsinifwhm}{$\varv\sin i_{\rm FWHM}$}

\newcommand{\vcrit}{$\varv_{\rm{crit}}$}

\begin{document}

   \title{Binarity at LOw Metallicity (BLOeM)\thanks{Based on observations collected at the European Southern Observatory under ESO program ID 112.25R7}  }

   \subtitle{Projected rotational velocities}

   \author{D.~J.~Lennon\inst{\ref{inst:iac}, \ref{inst:ull}}
        \and S.~R.~Berlanas\inst{\ref{inst:iac}, \ref{inst:ull}, \ref{inst:esac}}
        \and A.~Herrero\inst{\ref{inst:iac}, \ref{inst:ull}}
        \and N.~Britavskiy \inst{\ref{inst:rob}}
        \and P.~L.~Dufton \inst{\ref{inst:qub}}
        \and N.~Langer \inst{\ref{inst:bonn}}
        \and H.~Jin \inst{\ref{inst:mpa}}
        \and A. Schootemeijer \inst{\ref{inst:bonn}}
        \and A.~Menon \inst{\ref{inst:columbia}}
        \and J.~Bestenlehner \inst{\ref{inst:sheffield1}, \ref{inst:sheffield2}}
        \and P.~Crowther\inst{\ref{inst:sheffield1}}
        \and J.~S.~Vink \inst{\ref{inst:armagh}}
        \and J.~ Bodensteiner \inst{\ref{inst:antonpannekoek}}
        \and T.~Shenar \inst{\ref{inst:TelAv}}
        \and K.~Deshmukh \inst{\ref{inst:kul}}
        \and J.~I.~Villase\~nor \inst{\ref{inst:mpia}}
        \and L.~Patrick \inst{\ref{inst:cab}}
        \and F.~Najarro \inst{\ref{inst:cab}}
        \and A.~de Koter \inst{\ref{inst:antonpannekoek}, \ref{inst:kul}}  
        \and L.\ Mahy \inst{\ref{inst:rob}}
        \and D.\ M.\ Bowman\inst{\ref{inst:newcastle}, \ref{inst:kul}}
        \and A.\ Bobrick \inst{\ref{inst:bobrick1}. \ref{inst:bobrick2}}
        \and C.\ J.\ Evans \inst{\ref{inst:esa_stsci}}
        \and M.~Gull\inst{\ref{inst:carn}, \ref{inst:cit}}
        \and G.\ Holgado\inst{\ref{inst:iac}, \ref{inst:ull}}
        \and Z.~Katabi \inst{\ref{inst:TelAv}}
        \and J.\ Kub\'at\inst{\ref{inst:ondrejov}}
        \and P.\ Marchant\inst{\ref{inst:gent}}
          \and D.\ Pauli\inst{\ref{inst:kul}}
          \and M.\ Pawlak\inst{\ref{inst:lund}}
          \and M.\ Renzo\inst{\ref{inst:AZ}}
          \and D.\ F. Rocha\inst{\ref{inst:LN_Br}} 
          \and A.\ A.\ C.\ Sander\inst{\ref{inst:ari}, \ref{inst:iwr}}
          \and T.\ Sayada\inst{\ref{inst:TelAv}} 
          \and S.\ Sim\'on-D\'iaz\inst{\ref{inst:iac}, \ref{inst:ull}} 
          \and M.\ Stoop\inst{\ref{inst:antonpannekoek}}
          \and R.\ Valli\inst{\ref{inst:mpa}} 
           \and C.\ Wang\inst{\ref{inst:chen1}, \ref{inst:chen2}}
           \and X.-T.\ Xu\inst{\ref{inst:xu}}
          }

\institute{
{Instituto de Astrof\'isica de Canarias, C. V\'ia L\'actea, s/n, 38205 La Laguna, Santa Cruz de Tenerife, Spain\label{inst:iac}}
\and {Universidad de La Laguna, Dpto. Astrof\'isica, Av.\ Astrof\'sico Francisco S\'anchez, 38206 La Laguna, Santa Cruz de Tenerife, Spain\label{inst:ull}}
\and {Centro de Astrobiolog\`ia (CSIC–INTA), Camino Bajo del Castillo, s/n, E-28692 Villanueva de la Ca\~nada, Madrid, Spain \label{inst:esac}}
\and {Royal Observatory of Belgium, Avenue Circulaire/Ringlaan 3, B-1180 Brussels, Belgium} \label{inst:rob}
\and {Astrophysics Research Centre, School of Mathematics \& Physics,  Queen’s University, Belfast, BT7 1NN, UK\label{inst:qub} }    
\and {Argelander-Institut f\"{u}r Astronomie, Universit\"{a}t Bonn, Auf dem H\"{u}gel 71, 53121 Bonn, Germany\label{inst:bonn}}
\and {Max-Planck-Institute for Astrophysics, Karl-Schwarzschild-Strasse 1, 85748 Garching, Germany\label{inst:mpa}}
\and{Department of Astronomy, Pupin Hall, 538 West 120th Street,  Columbia University, New York City, NY 10027, U.S.A\label{inst:columbia}}
\and {Astrophysics Research Cluster, School of Mathematical \& Physical Sciences, University of Sheffield, Hicks Building, Hounsfield Road, Sheffield, S3 7RH, UK \label{inst:sheffield1}} 
\and {School of Chemical, Materials and Biological Engineering, University of Sheffield, Sir Robert Hadfield Building, Mappin Street, Sheffield, S1 3JD, UK \label{inst:sheffield2}}
\and {Armagh Observatory, College Hill, Armagh, BT61 9DG, Northern Ireland, UK\label{inst:armagh}}
\and {Anton Pannekoek Institute for Astronomy, University of Amsterdam, Science Park 904, 1098 XH Amsterdam, the Netherlands\label{inst:antonpannekoek}}
\and {The School of Physics and Astronomy, Tel Aviv University, Tel Aviv 6997801, Israel\label{inst:TelAv}}
\and {Institute of Astronomy, KU Leuven, Celestijnenlaan 200D, 3001 Leuven, Belgium\label{inst:kul}}
\and {Max-Planck-Institut f\"{u}r Astronomie, K\"{o}nigstuhl 17, D-69117 Heidelberg, Germany\label{inst:mpia}}
\and {Centro de Astrobiolog\'ia (CSIC-INTA), Ctra.\ Torrej\'on a Ajalvir km 4, 28850 Torrej\'on de Ardoz, Spain\label{inst:cab}}
\and {School of Mathematics, Statistics and Physics, Newcastle University, Newcastle upon Tyne, NE1 7RU, UK\label{inst:newcastle}}
\and {School of Physics and Astronomy, Monash University, Clayton, VIC 3800, Australia \label{inst:bobrick1}}
\and {Australian Research Council Centre of Excellence for Gravitational Wave Discovery, Clayton, VIC 3800, Australia \label{inst:bobrick2}} 
\and {{European Space Agency (ESA), ESA Office, Space Telescope Science Institute, 3700 San Martin Drive, Baltimore, MD 21218, USA}\label{inst:esa_stsci}}
\and {The Observatories of the Carnegie Institution for Science, 813 Santa Barbara Street, Pasadena, CA 91101, USA \label{inst:carn}}
\and {Department of Astronomy, California Institute of Technology, Pasadena, CA 91125, USA \label{inst:cit}}
\and {Astronomical Institute, Academy of Sciences of the Czech Republic, Fri\v{c}ova 298, CZ-251 65 Ond\v{r}ejov, Czech Republic \label{inst:ondrejov}}
\and {Sterrenkundig Observatorium, Universiteit Gent, Krijgslaan 281 S9, 9000 Gent, Belgium \label{inst:gent}}
\and {Lund Observatory, Division of Astrophysics, Department of Physics, Lund University, Box 43, SE-221 00, Lund, Sweden}\label{inst:lund}
\and {{Department of Astronomy \& Steward Observatory, 933 N. Cherry Ave., Tucson, AZ 85721, USA}\label{inst:AZ}}
\and {Laborat\'orio Nacional de Astrof\'isica, Rua Estados Unidos 154, 37504-364, Itajub\'a, MG, Brazil\label{inst:LN_Br}}
\and {Zentrum f{\"u}r Astronomie der Universit{\"a}t Heidelberg, Astronomisches Rechen-Institut, M{\"o}nchhofstr. 12-14, 69120 Heidelberg\label{inst:ari}}
\and {Interdisziplin{\"a}res Zentrum f{\"u}r Wissenschaftliches Rechnen, Universit{\"a}t Heidelberg, Im Neuenheimer Feld 225, 69120 Heidelberg, Germany\label{inst:iwr}}
\and {School of Astronomy and Space Science, Nanjing University, Nanjing, 210023, People's Republic of China \label{inst:chen1}}
\and {Key Laboratory of Modern Astronomy and Astrophysics, Nanjing University, Ministry of Education, Nanjing, 210023, People's Republic of China \label{inst:chen2}}
\and {Tsung-Dao Lee Institute, Shanghai Jiao-Tong University, 1 Lisuo Road, Shanghai 201210, People’s Republic of China \label{inst:xu}}
}

   \date{}

 
 \abstract{
The Binarity at LOw Metallicity (BLOeM) survey is an ESO large programme designed to obtain multi-epoch spectroscopy for 929 massive stars in the Small Magellanic Cloud (SMC). 
It will provide binary fractions and orbital configurations of binary systems and search for dormant black hole binary candidates (OB+BH). 
We present projected rotational velocities (\vsini) of all sources, and, using the multiplicity properties presented in previous papers, we derive the \vsini\ distributions of apparent single stars, single-lined spectroscopic (SB1) binaries, and SB2 systems.
We identify a locus in the Hertzsprung-Russell diagram where rotational velocities decrease significantly; we interpret this feature as broadly corresponding to the terminal-age main sequence.
The main-sequence cohort is distinguished by a broad range of \vsini\ values, but with a strong peak in the distribution in the range 30--60\,\kms, which is close to the resolution limit of $\sim$30\,\kms, indicating the presence of many upper limits.
Sources in this low \vsini\ peak are distributed throughout the main sequence and are also present in the SB1 sample, though less prominent than in the single-star distribution. 
A preliminary analysis of the lowest \vsini\ cohort, which includes SB1 systems, implies that roughly one-third may be nitrogen rich, and we speculate that this cohort is a mix of pristine single stars, long-period binaries, and merger products.
The SB2 systems appear to be mostly short-period binaries in synchronous rotation, and their \vsini\ estimates are distributed around a mean value of $\sim$140\,\kms.
Higher \vsini\ sources are also present in the single and SB1 systems, all of which have a tail to higher \vsini\ values. This is consistent with tidal and mass-transfer effects.
The supergiants, with a few exceptions, have low \vsini, and the bulk of these systems is essentially unresolved at the current spectral resolution ($\sim$30\,\kms).
 }

   \keywords{stars: massive -- stars: rotation -- stars: early-type -- stars: evolution -- binaries: spectroscopic --  Magellanic Clouds}

   \maketitle
%

\section{Introduction}
\label{introduction}

In the massive star domain ($M_*\gtrsim8M_\odot$), rotational mixing can have important consequences for a star's evolution \citep{maeder2000}, its surface composition \citep{hunter2008a,hunter2009,bouret2013,jin2024}, its final endpoint as a supernova (SN) explosion or gamma-ray burst (GRB; \citealt{yoon2005,groh2013}), and their remnants \citep{hirschi2025}.
The distribution of the rotation speeds of young massive stars can provide clues on the star formation process \citep{oscar2013,bastian2020,britavskiy2023} or constrain the mode of star formation in a given environment \citep{wolff2006}.
The rotation speed can also be used as a diagnostic for a variety of processes, for example, mass transfer in binary evolution \citep{demink2013},  as a possible indicator of a previous stellar merger \citep{schneider2019,frost2024}, to determine the presence of magnetic fields \citep{ud-Doula2009,fossati2015,petit2019}, or as a constraint on the history of runaway stars ejected as a result of a SN explosion \citep{renzo2019b,sana2022}.
Binary population synthesis also predicts how the distribution of rotational speeds evolves, which provides interesting predictions for the rotation velocities to be expected for products such as X-ray quiet OB+BH binaries \citep{xu2025,schuermann2025}.
Therefore, the distribution of rotation speeds of massive stars is of fundamental importance for understanding their formation and evolution.

\begin{figure*}
    \centering
    \includegraphics[width=0.9\linewidth]{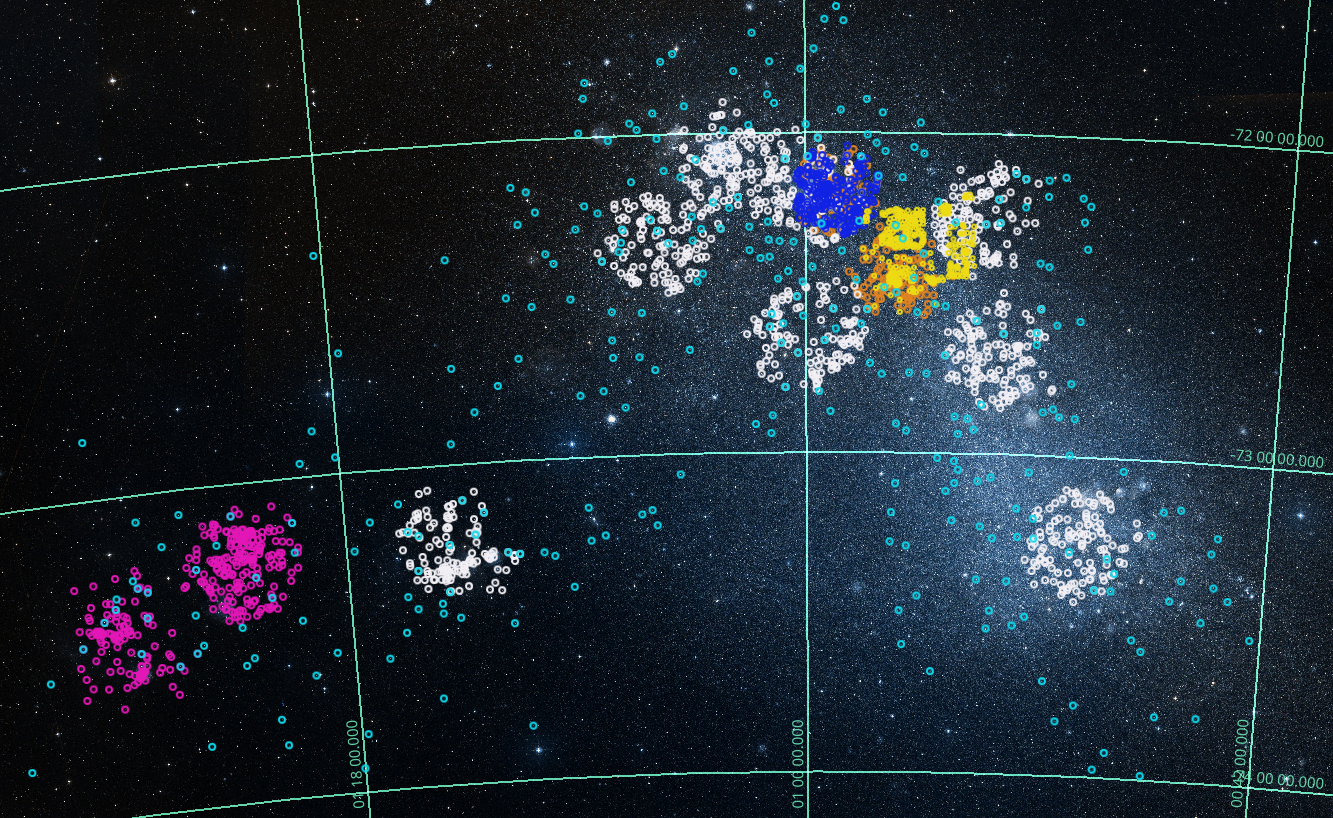}
    \caption{Distribution of BLOeM targets (white circles) compared with those of some other major \vsini\ surveys: FSMS \citep[][orange]{hunter2008b}, Martayan \citep[][yellow]{martayan2007}, Ramachandran \citep[][magenta]{ramachandran2019}, Dufton \citep[][blue]{dufton2019}, and RIOTS4 \citep[][cyan]{dorigojones2020}. }
    \label{fig:smc}
\end{figure*}

\citet{howarth2004} reviewed early work on the projected rotational velocities (\vsini) of massive stars in the Milky Way, while more recent studies focused on the \vsini\ distributions of OB stars \citep[e.g.][]{berlanas2025,holgado2022,huang2006,braganca2012} and B-type supergiants \citep[e.g.][]{fraser2010,deburgos2023}.
The lower metallicity ($Z$) of the Magellanic Clouds offers an opportunity to examine the effect of $Z$ on \vsini\ distributions, and several such studies already exist \citep[examples include][]{mokiem2006,martayan2007,hunter2008b,penny2009,dufton2013,oscar2013,dufton2019,ramachandran2019,dorigojones2020,bodensteiner2023}. 

We derive \vsini\ estimates of the 929 massive stars in the Small Magellanic Cloud (SMC) that have been observed by the survey called Binarity at LOw Metallicity (BLOeM) \citep{shenar2024}.
In addition to adding to the body of work that already exists for the SMC, two unique aspects of this survey are that its multi-epoch nature is tailored to detect and characterize binaries \citep[see][for details]{shenar2024}, while the eight surveyed fields are distributed throughout the SMC bar and wing, as shown in Fig. \ref{fig:smc}.
\citet{bestenlehner2025} have presented preliminary stellar parameters for 778 BLOeM sources that also included \vsini\ estimates. 
Their \vsini\ estimates are based on profile fitting (a goodness-of-fit (GOF) method) using a discrete grid of \vsini\ values. 
However, their test results for a limited number of sources highlighted differences between the GOF and Fourier transform (FT) methods that were attributed to the FT approach being relatively insensitive to broadening effects that may include macroturbulence, intrinsic line breadth, and weak-line blending in binary systems.
In this paper, we therefore extend the FT approach to the full BLOeM sample.

\section{Observational data}
\label{observations}

\begin{figure}
    \centering
    \includegraphics[width=0.9\linewidth]{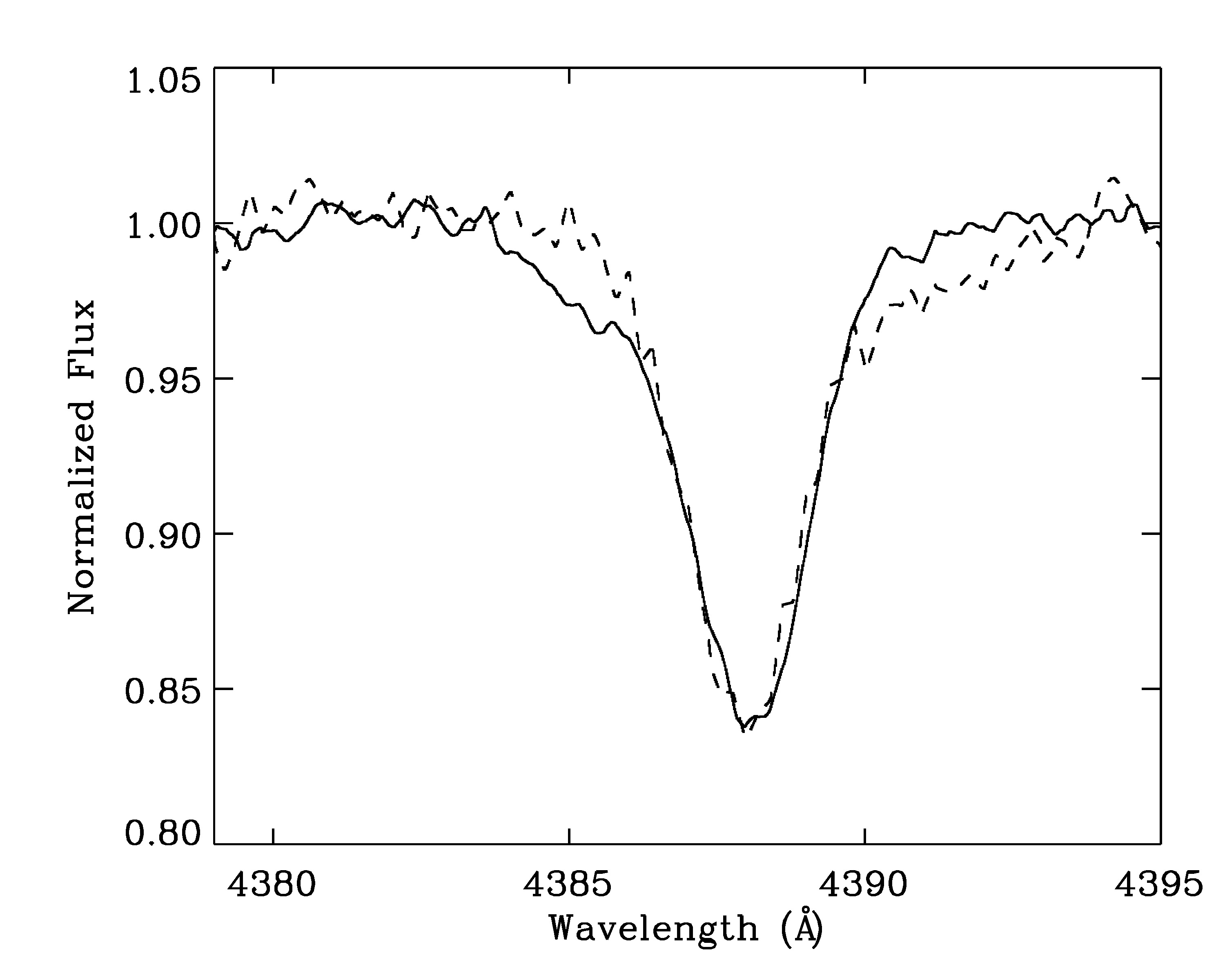}
    \caption{Two \ion{He}{i} 4387\,\AA\ line profiles for the source BLOeM\,1-020, which is classified as an SB1 system and has a peak-to-peak velocity of $\sim$100\,\kms. The full and dashed lines show the profiles obtained at each quadrature. A faint secondary is visible on opposing sides of each profile.}
    \label{fig:1-020}
\end{figure}

We used the BLOeM data release 4 (DR4) normalized spectra comprising 21 epochs for the eight fields, except for fields 4 and 8, which have 22 and 23 epochs, respectively. All epochs were merged to produce combined spectra for the analysis, and individual spectra were corrected for radial velocity shifts relative to combined nine-epoch DR3 spectra that were derived using the cross-correlation of the 4360--4560\,\AA\ region \citep[see][]{patrick2025}, and weighted by their observed counts.
The multi-epoch nature of the combined spectra has some implications for the subsequent analysis.
Spectroscopic binaries with a weak secondary spectrum (SB1 systems) for example are velocity shifted to the rest frame of their primary star, and hence, weak absorption lines of the secondary star may be smeared over the wings of the lines of the primary (see Fig.\,\ref{fig:1-020}). 
Similarly, any line centroid shifts due to pulsation, such as $\beta$-Cepheid pulsation \citep[see the review by][]{bowman2020}, can result in some smearing of the line profile.
Double-lined spectroscopic binaries (SB2) present a particular problem, and our approach to these systems is described further in Sect.\,\ref{methodology}. 


\section{Method}
\label{methodology}

The FT approach adopted here for the BLOeM sample has been widely discussed in the literature (see Sect. \ref{otherwork} for several applications to massive stars in nearby galaxies, while practical considerations for its application to hot stars have been discussed in detail by \citet{simondiazherrero}). 
Briefly, the FT of the Uns\"old-Struve rotational broadening function \citep[see][for assumptions and limitations]{collins1995} is a Bessell function of the first kind that has zeros at positions that are characteristic of the rotational velocity and limb-darkening coefficient of the star \citep{carroll1928,carroll1933}.
Formally, \vsini $=x_0/f_0$, where $f_0$ is the position of the first minimum, and $x_0$ is a constant depending on the limb-darkening coefficient $\alpha$.
In common with most comparable FT work in the literature, we adopted a limb-darkening coefficient of $\alpha=0.6$, the grey atmosphere solution, in the standard linear limb-darkening law  \citep{gray1976}, 
implying that \vsini$=0.66/f_0$.
Direct determination of $\alpha$ using model atmospheres \citep{howarth2011,reeve2016} demonstrates that $\alpha\sim$0.2--0.4 for main-sequence OB stars, and appropriate $x_0$ values \citep[see][]{collins1995,jankov1990} then imply that \vsini\ estimates for these stars are overestimated by $\sim$5--3\%, respectively.
The median error of our results from other sources is $\sim$20\%, and we therefore did not apply any corrections to our data.
Furthermore, given the moderate resolving power of the BLOeM data of $\sim$7\,000 and the effect of binaries on GOF results (discussed below), we primarily focus on rotational velocities here and ignore the potential effect of macroturbulence \citep[see][]{ryans2002,simondiaz2014,simondiaz2017}.

In the current approach, we derived the FT for a number of absorption lines for each BLOeM source. Their first minima yield separate \vsini\ measurements, the mean value of which yielded a measurement we refer to as \vsinift. 
The choice of lines for a given star was dictated by a combination of signal-to-noise ratio (s/n), spectral type, and line width, resulting in three to ten lines that were identified per star.
The strong \ion{He}{i} lines ($\lambda\lambda$\,4143, 4387, 4471) were measured in all but a few of the hottest O-type stars, the coolest stars in the BAF group, and a number of the OBe stars. 
For stars with low \vsini\ , however, these values were supplanted by measurements from the weak metal lines (e.g. \ion{N}{ii} $\lambda$3995 if present, \ion{Si}{iii} $\lambda$4553), and \ion{He}{i} ($\lambda\lambda$\,4168, 4437) as permitted by s/n and spectral type (primarily B-type supergiants and bright giants). 
Given the small number of measurements per star, formal 68\% confidence limits for the FT zeros were estimated using the Student $t$-distribution.
For well-defined FTs, the results for all lines can be identical, and in these cases, formal error estimates were taken to be the resolution of the FT itself.
These formal error estimates should be regarded with some caution. Other work suggests that a general uncertainty of about 20\% might be appropriate \citep{simondiazherrero}.
In any case, some indication of problematic measurements can be derived by comparing FT and FWHM results. 
This is discussed further in Appendix \ref{app:errors}, where a sample of results are shown (Fig.\,\ref{fig:plotfts}).

The spectral groups within the BLOeM sample (O, BSG, B-dwarf, OBe, and BAF) required some variation in strategy.
For a few of the hottest O-type stars that lacked strong \ion{He}{i} $\lambda\lambda$\,4143, 4387 lines, we also used the \ion{He}{ii} $\lambda\lambda$\,4026, 4200, and 4542 lines. Since these lines are intrinsically broad, the approach was only used for stars with high \vsini.
The cool supergiants (the BAF group) possess numerous sharp lines, and for this group, we used up to ten isolated lines for \vsini\ measurements,
although identification of isolated lines for F-type supergiants is problematic.
The OBe group and other emission line stars such as the SgB[e] stars required special attention due to the presence of emission lines that contaminated many of the diagnostic lines.
For these stars, we manually identified lines that appeared devoid of emission. These typically were the weak \ion{He}{i} lines (noted above) and metal lines, if present.
One star, the SgB[e] star BLOeM 2-116, was devoid of absorption lines and is the only source for which we derived a \vsini\ estimate from its emission lines, as discussed in Appendix \ref{sgbe}.

The FWHM of each line was also derived using a Gaussian fit to the profile and \vsini\ derived using the conversion factor provided by \cite{collins1995}, with mean values that we refer to as \vsinifwhm, which is a useful check on \vsinift\ concerning the effect of other broadening mechanisms, such as pressure broadening of the \ion{He}{i} lines, or the presence of a weak secondary component.
We illustrate the latter effect on the B0\,III star BLOeM\,1-020, discussed by \citet[their Fig. 14]{bestenlehner2025} as having discrepant FT and GOF \vsini\ estimates, which is attributed to the possible impact of binarity.
This is an SB1 system with a measured peak-to-peak (p2p) velocity of $\sim$100\,\kms\, and by merging spectra at each quadrature, we enhanced the presence of the secondary component on opposing wings of each \ion{He}{i} line. This is shown in Fig.\,\ref{fig:1-020} for the \ion{He}{i} 4387\,\AA\ line.
These wings broaden the merged profile and increase the line width, which leads to an overestimate of the GOF/FWHM results with respect to the FT approach, which is relatively insensitive to their presence provided the profile is dominated by the primary.
For example in the case of BLOeM\,1-020, the \vsinift\ of the \ion{He}{i} 4143, 4387, and 4471\,\AA\ lines is 81\,\kms, compared with a \vsinifwhm\ of 127\,\kms.
The weaker lines, on the other hand, \ion{He}{i} 4168, 4437\,\AA\ and \ion{Si}{iii} 4553\,\AA, agree better between \vsinift\ and \vsinifwhm\ , with mean values of 88 and 92\,\kms, respectively.

\begin{figure}
    \centering
    \includegraphics[width=1.0\linewidth]{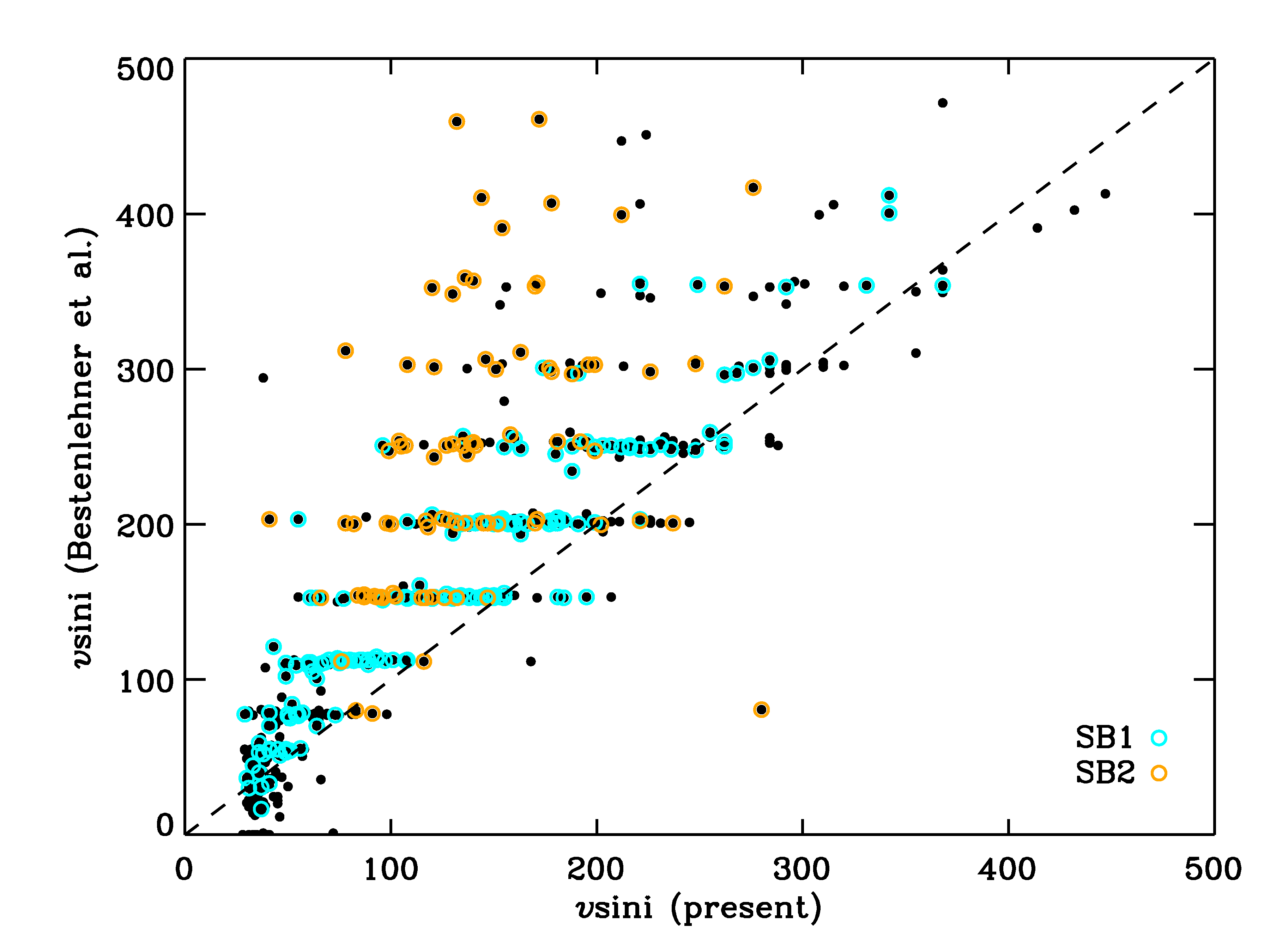}
    \caption{Comparison of pipeline \citet{bestenlehner2025} results with current measurements (black dots). The banding of the former results is a consequence of their adoption of a discrete grid of \vsini\ values. SB1 and SB2 systems are flagged as indicated in the legend, and component A \vsini\ values were adopted for the SB2 sources. The outlier at position (280,81) is the SB2 system BLOeM 5-057, which has very different line widths of 280 and 56\,\kms\ for components A and B. Clearly, the pipeline value is appropriate for component B. }
    \label{fig:compare_vsinis}
\end{figure}


As noted in \citet{shenar2024}, a significant number of double-lined spectroscopic binary (SB2) systems were detected during the spectral classification.
While we included these systems in our standard FT approach, we also made a preliminary estimate of \vsini\ for each individual component.
As orbital solutions are not yet available, we defined the stronger component as A and the weaker line system as B.
Inspection of the spectra for each epoch of a given source yielded a number of spectra observed with the two components partially separated to at least two-thirds of the line depth, the number of useful spectra depending on s/n, relative strengths of the components, and velocity separation.
The FWHM of these components was determined by fitting a double Gaussian and was converted into \vsini\ as explained above.
The lower s/n in single-epoch spectra compared to merged data led to a reliance on the stronger \ion{He}{I} and \ion{He}{II} lines for these measurements.
Furthermore, comparing results for the various lines, it is clear that the lines that are most affected by intrinsic line broadening, such as \ion{He}{i} 4471\,\AA\ and \ion{He}{ii} 4542, 4200\,\AA, tend to give higher \vsini\ values than \ion{He}{i} 4387\,\AA;
for example \ion{He}{i} 4387\,\AA\ is the least impacted by intrinsic broadening, with \ion{He}{i} 4471\,\AA\ giving results $\sim$20\% higher. We therefore restricted our results to the means of the estimates from the \ion{He}{i} 4387\,\AA\ line.
The approach outlined here is clearly approximate and should be refined when orbital solutions are available.
An important caveat concerns our designation of the A component with stronger lines, typically hydrogen.
Without orbital solutions, it is clear that the inferred primary and secondary natures are uncertain, and this should be kept in mind during our discussion of the overall results. 
Nevertheless, these results provide a first look at the rotation properties of the SB2 systems in BLOeM.

\section{Results}
\label{results}

\begin{table}
\fontsize{9pt}{9pt}\selectfont
    \centering
\caption{Mean projected rotational velocities in \kms\  (FT and FWHM values), with their 68\% confidence limits, and the number of measured lines.  }\label{tab:vsinis} 
\begin{tabular}{crrrrrrr} \\ \hline
BLOeM & \vsini\ & $\sigma+$ & $\sigma-$ & $n_{\rm FT}$ & \vsini\ &  $\sigma$ & $n_{\rm FWHM}$ \\
ID & FT & & & & FWHM & & \\ \hline
1-001  &    31  &    3  &    2 & 10  &   32  &    2  & 10 \\
1-002  &    55  &   23  &   14 &  3  &   57  &    8  &  3 \\ \hline
    \end{tabular}
    \tablefoot{The complete table is published online at the CDS. \vsini\ estimates for SB2 systems should be taken from Table\,\ref{tab:vsinisb2}.}
\end{table}

\begin{table}
\fontsize{7pt}{7pt}\selectfont
\centering
        \caption{\vsini\ values for A and B components of SB2 systems, with 1$\sigma$ uncertainties, number of used spectra (n), wavelengths of the \ion{He}{i} 4387|,\AA\ lines for the A and B components (waveA and waveB), and the epoch number of the spectrum from which they were measured. }
    \label{tab:vsinisb2}
    \begin{tabular}{lrrrrcccc}
    \hline
    BLOeM & \vsini\ & $\sigma$ & \vsini\ & $\sigma$ & n & waveA & waveB & nepoch\\
    Id.   & A & & B & & & \AA\ & \AA\ &   \\ \hline
 1-006 & 196 &  22 & 164 &  27 & 4 & 4388.90 & 4394.41 & 1 \\
 1-011 & 115 &   2 &  87 &  15 & 4 & 4388.67 & 4393.00 & 4 \\ \hline
    \end{tabular}
    \tablefoot{The complete table is published online at the CDS.}
\end{table}

We list the \vsinift\ and \vsinifwhm\ results in Table \ref{tab:vsinis}. 
The SB2 procedure outlined in Sect. \ref{methodology} was applied to 70 of the 91 SB2 systems, the other systems having a too small velocity separation, too little contrast between components, or an insufficient signal-to-noise ratio.
This latter group was treated as SB1 systems.
Table\,\ref{tab:vsinisb2} lists  results for the A and B components, but we recall our previously discussed caveat for inferred primary or secondary natures.
Figure\,\ref{fig:compare_vsinis} compares the present results with those by \citet{bestenlehner2025}, analogous to their Fig.\,13, reflecting their conclusion that a GOF approach overestimates \vsini. We note that SB2 systems are frequent outliers in this plot
because we used the current A component \vsini\ values in this comparison.
This figure also indicates two important features of the different approaches. The FT results converge to a value of $\sim$30\,\kms, as expected given the spectral resolution of the instrument, while the GOF results converge to zero, although with significant uncertainty. This latter result is a consequence of detecting small perturbations of the instrumental broadening using a GOF approach. Therefore, the comparison of the \citet{bestenlehner2025} results with our values might provide an indication of which sources in the low \vsini\ domain have values below our \vsinift\ limit, noting also that the former were not corrected for the possible effect of macroturbulence. The low \vsini\ domain can be explored further using higher resolution ($R\sim20\,000$) VLT/Flames data (P.I. L.\,Mahy). 

\begin{figure}
    \centering
    \includegraphics[width=0.9\linewidth]{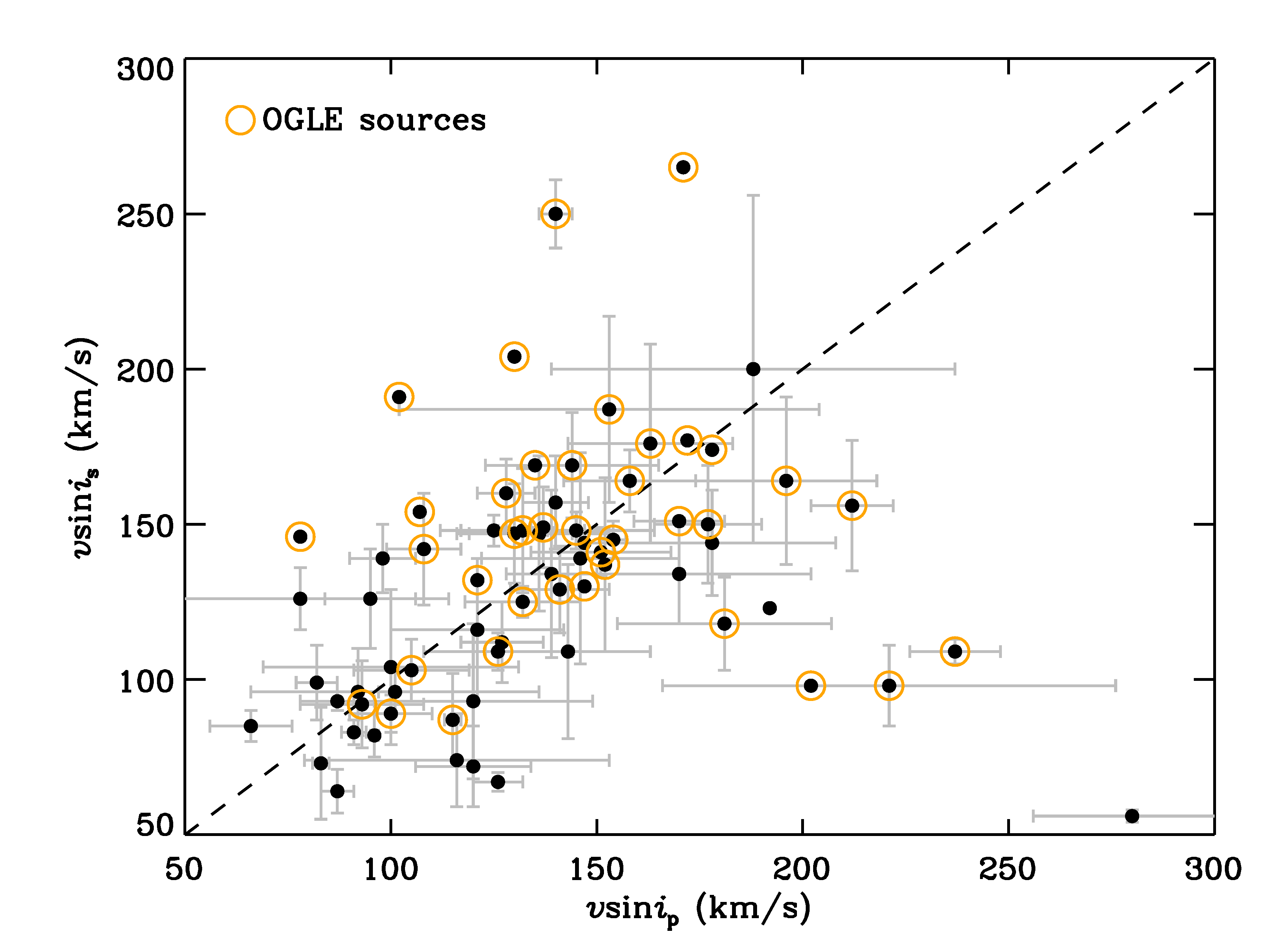}
    \caption{Comparison of \vsini\ measurements for designated A and B components of SB2 systems. The orange circles denote eclipsing or ellipsoidal binaries with already well-determined periods from the OGLE project \citep{pawlak2016}.}
    \label{fig:sb2}
\end{figure}

Figure\,\ref{fig:sb2} compares \vsini\ measurements of components A and B, showing a significant correlation, as expected for synchronized systems with similar radii. 
The periods from the OGLE survey \citep{pawlak2016} for 39 of these systems are well determined. The periods for 33 of them are 1--5\,d, the periods of 4 are between 5--10\,d, and the periods for the last 2 are between 15-20\,d.
The short periods together with the range and correlation of rotational velocities is most likely a consequence of tidal synchronization \citep[see also][]{lennon2024} for most of these systems.
There are a number of outliers in the correlation, which might indicate some systems with super-synchronous rotation as a result of recent or ongoing mass transfer. 
Two examples of such systems with well-defined separated components are BLOeM 1-055 and 5-057, which have \vsini$_{p}$=221$\pm$55 and \vsini$_{s}$=98$\pm$13, and  \vsini$_{p}$=280$\pm$24 and \vsini$_{s}$=56$\pm$2 \kms, respectively.
The former is known to have a period of 3.04\,d \citep{pawlak2016}. 

\begin{figure*}
        \centering
        \includegraphics[width=0.9\linewidth]{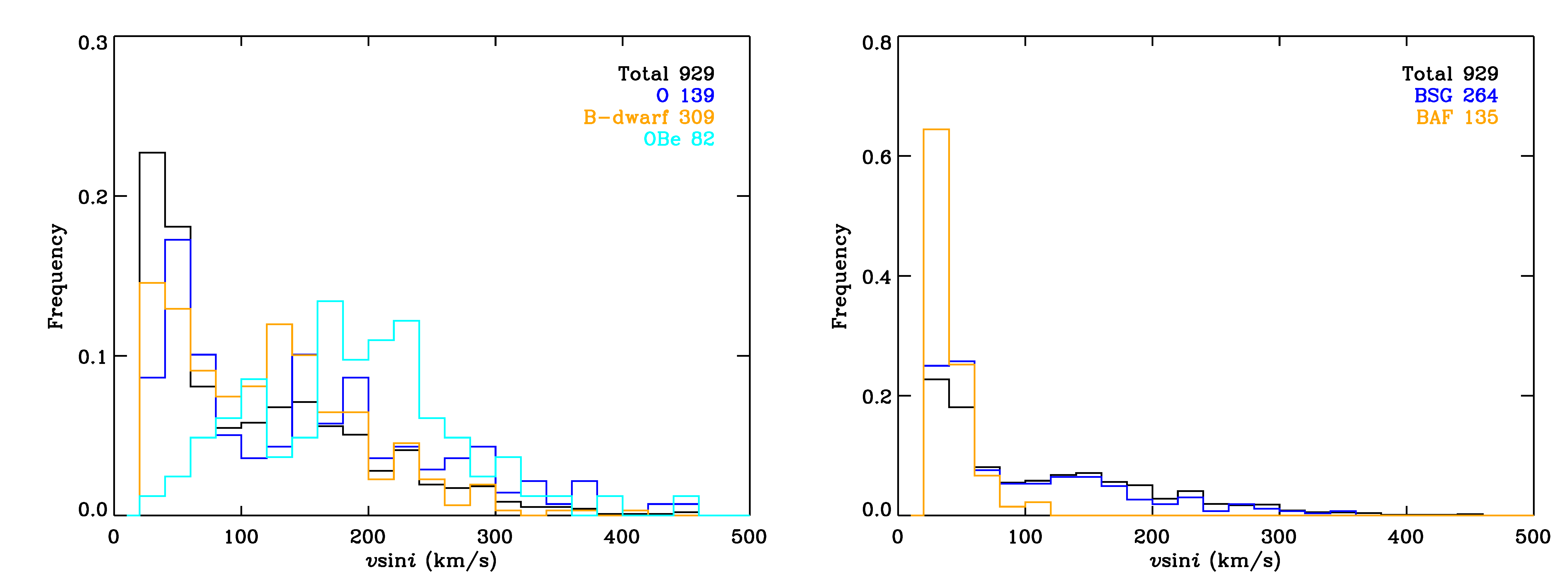}
        \caption{Normalized histograms for the total sample and each of the BLOeM groups, split into two panels for clarity. The groups are identified as indicated by the insets, which also indicate the number of sources in each group.}
        \label{fig:hist_groups}
\end{figure*}

In the following discussion of the demographics of the rotation properties of the BLOeM sample, we adopt the \vsinift\ results for the bulk of the sources, but we replace the 70 SB2 values with their relevant A \vsini\ results.
Henceforth, we refer to their projected rotational velocities as simply \vsini.
We also make use of single SB1 and SB2 designations as assigned in previous BLOeM papers \citep{shenar2024,villasenor2025,britavskiy2025,bestenlehner2025,sana2025,bodensteiner2025}, reproduced in Table\,C.1 for convenience. 
Broadly speaking, SB2 systems were identified by visual inspection of the spectra, SB1 systems are those (non-SB2 systems) that passed the two radial velocity (RV) variability criteria\footnote[1]{Briefly, the criteria for OB stars are that the peak-to-peak RV amplitude must exceed 20\,\kms, and at least one of the RV pairs differ by more than 4$\sigma$.}, and the single designation refers to stars that failed both criteria. 
Some stars (87) were assigned ambiguous binary status flags: SB?, SB1/lpv (lpv = line profile variable), SB2?, etc. 
These systems are excluded from the single/SB1/SB2 groups in the following discussion, but this exclusion affects the overall statistical picture little.
One important exception arises because \citet{villasenor2025} subdivided their single cohort (as described above) into two sub-types, RV variable (RVvar) if any pair of RV measurements differed by more than 2$\sigma$, otherwise defined as RV constant.
However, for consistency with the O and BSG samples, the RVvar sample was absorbed into the single category discussed above.
Nevertheless, we recall that many systems with the single label might be long period binaries, the components of which are evolving as single stars, or are (almost) pole-on systems.
An important caveat to the above, which should also be kept in mind in the following discussion, is that only a handful of sources have been formally identified as triple systems. Based on previous work \citep[for example][]{moe2017}, however, we expect the average OB primary to have between 1.5 and 2.0 companions (for $q>0.1$). Therefore, the full effect of triples and higher-order systems on the analysis must await the full analysis of the radial velocity data.

\begin{figure*}
    \centering
    \includegraphics[width=1.0\linewidth]{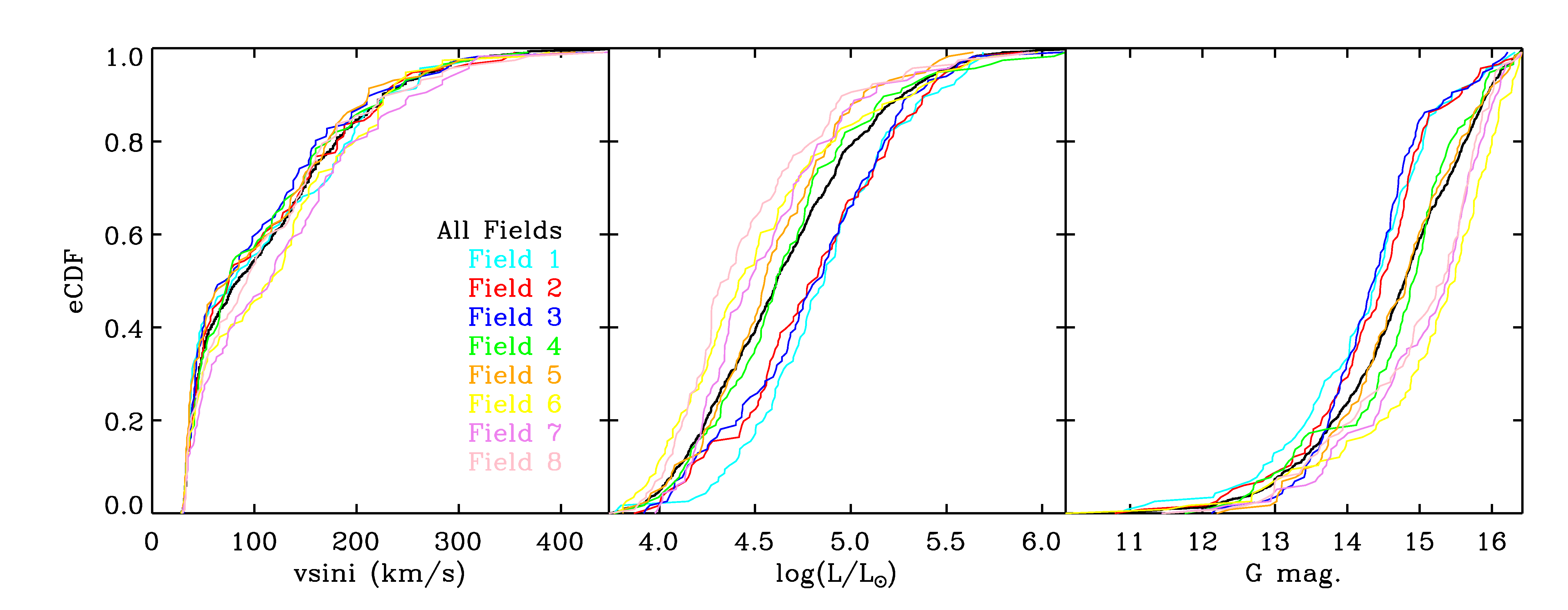}
    \caption{Left: Empirical \vsini\ cumulative distribution function (eCDF) for the full BLOeM sample (black line) compared with those of each of the eight BLOeM fields (coloured lines). Middle and right: Equivalent eCDFs of luminosity and the $Gaia$ $G$-band magnitudes.}
    \label{fig:plot_edfs}
\end{figure*}

\begin{figure*}
        \centering
        \includegraphics[width=0.9\linewidth]{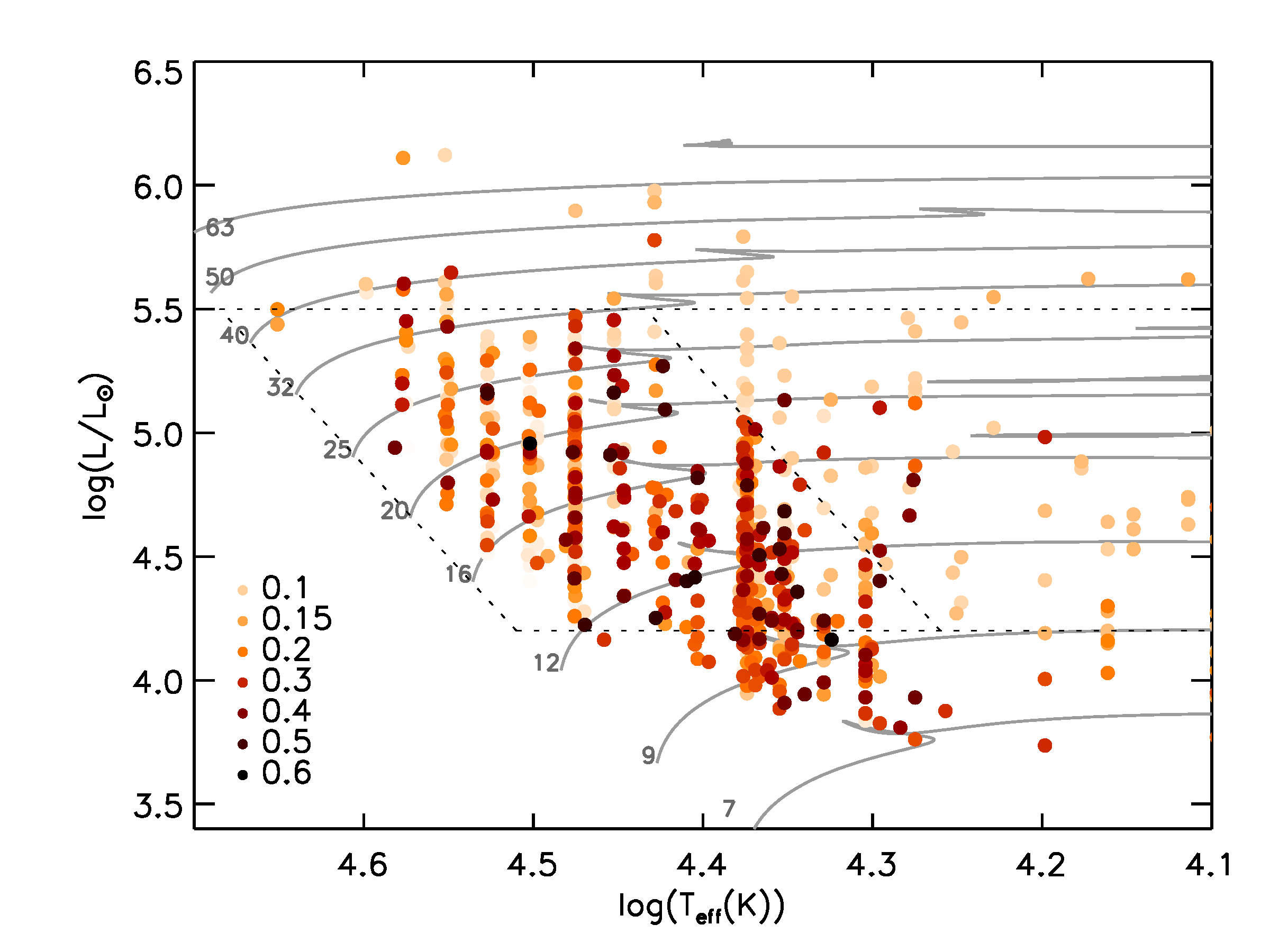}
        \caption{Hertzsprung-Russel diagram of the BLOeM OB sources colour-coded according to the ratio \vsini/\vcrit\, as indicated in the inset. The stellar parameters are taken from \citet{bestenlehner2025}, and the evolutionary tracks, for illustration, are those of \citet{schootemeijer2019} assuming $\alpha_{\rm ov}=0.33$ and $\alpha_{\rm sc}=10$. The tracks are labelled with their initial mass in solar units. The large trapezium bounded by diagonal dashed lines between luminosities 4.2 and 5.5 indicates the approximate extent of the fast rotators and main sequence, as discussed in Sect. \ref{demographics}.}
        \label{fig:hrd}
\end{figure*}

\begin{figure}
        \centering
        \includegraphics[width=0.95\linewidth]{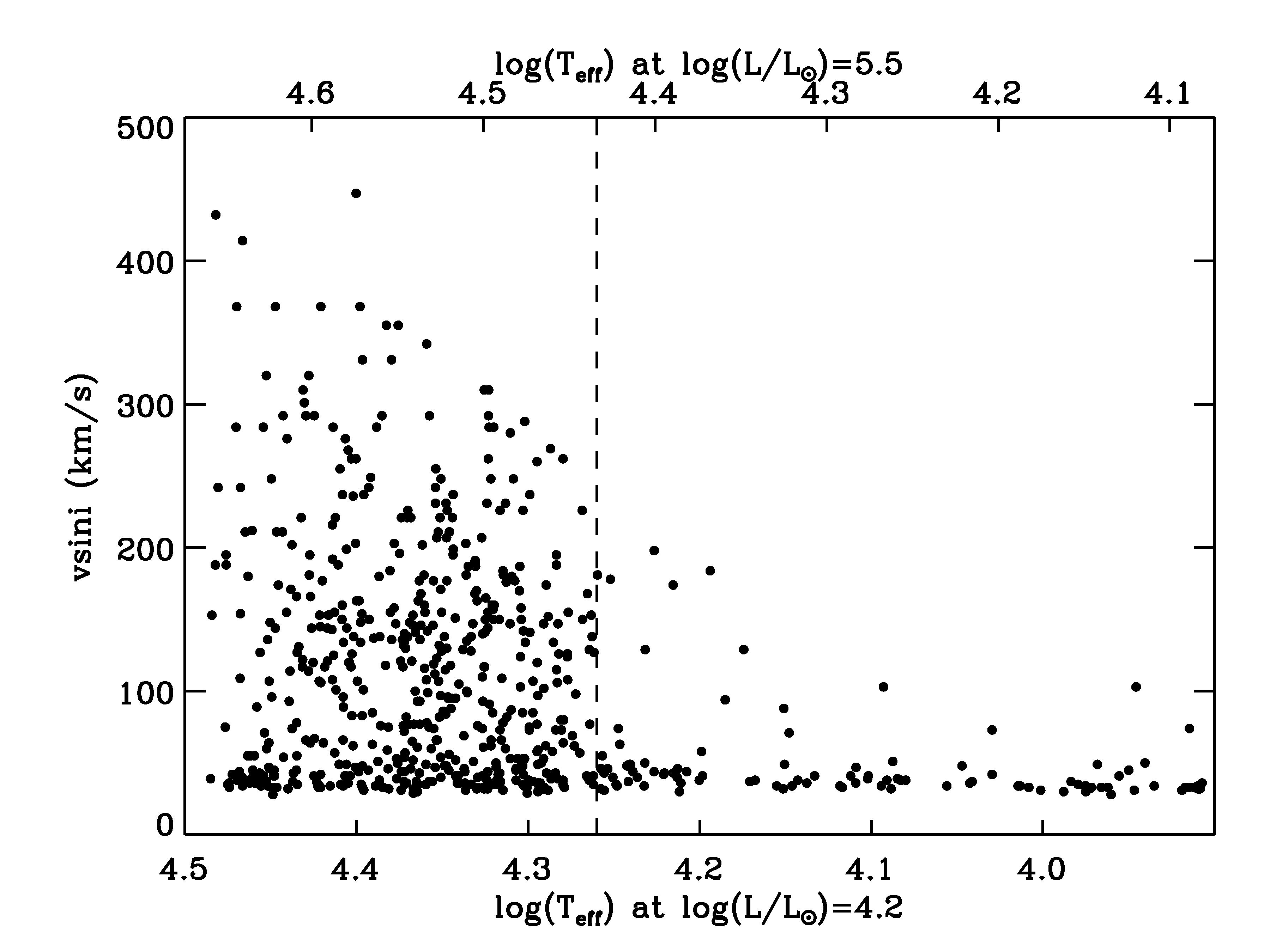}
    \caption{Distribution of \vsini\ as a function of distance from the ZAMS as indicated by \teff\ as defined by the lower and upper luminosity boundaries illustrated in Fig.\,\ref{fig:hrd}. 
    The vertical dashed line indicates the TAMS at these luminosities as defined by the approximate position of the decrease in the \vsini\ distribution.}
        \label{fig:vsinidrop}
\end{figure}

\subsection{Demographics of rotation speeds}
\label{demographics}

Figure\,\ref{fig:hist_groups} shows histograms for the full sample, plus the five main BLOeM groups \citep{shenar2024}; O-stars, the BSG group (luminosity class I and II B-type stars), B-dwarf (luminosity class III-V), OBe (mostly classical Oe and Be stars), and the BAF group (supergiant stars of spectral type B5 and cooler). 
The strong peak at low \vsini\ in the full sample is also strongly present in the O and B-dwarf groups, indicating that this feature is not solely due to the supergiants in the full sample. 
An indication of bi-modality is visible in the O, B-dwarf, and BSG groups, and, as expected, the OBe sample has a moderate to high \vsini.

It is worth noting that there are significant differences between individual fields.
The left panel of Fig.\,\ref{fig:plot_edfs} compares the empirical cumulative distribution functions (eCDFs) of \vsini\ for each field with the full sample.
They show varying fractions of low \vsini\ sources between the fields. A simple K-S test reveals that only fields 2, 4, and 8 have probabilities greater than 0.5 of being drawn from the mean eCDF. 
The middle and right panels of Fig.\,\ref{fig:plot_edfs} show that the stellar luminosities and $Gaia$ \citep{gaia} $G$-band magnitude distributions are also significantly different. 
For example, the ratios of massive to intermediate-mass stars and/or stellar age distributions. Only fields 4 and 5 have K-S probabilities greater than 0.1 of being drawn from the mean distributions for these quantities.
This field-to-field variation is to be expected, given the patchy location of massive stars in the SMC, for example as indicated by the distribution of its UV sources \citep{cornett1997}.
This is reflected in the differences between the luminosity and $G$-magnitude eCDFs of each field. For example fields 2 and 3 are incomplete in the $G$-band magnitude at approximately $G=15$ because the bright blue sources in these fields are denser than others, as also indicated by their luminosity functions (central panel). One important bias to be aware of is that the central regions of NGC\,346 and NGC\,330 were avoided due to crowding issues that affect fields 4 and 7, respectively.
In the following discussion, we discuss merged samples across all fields as representing an SMC average picture of the \vsini\ demographics, but we recall that there are likely local variations compared with other work.

\subsection{The main sequence}
\label{mainsequence}

\begin{figure}
        \centering
        \includegraphics[width=0.95\linewidth]{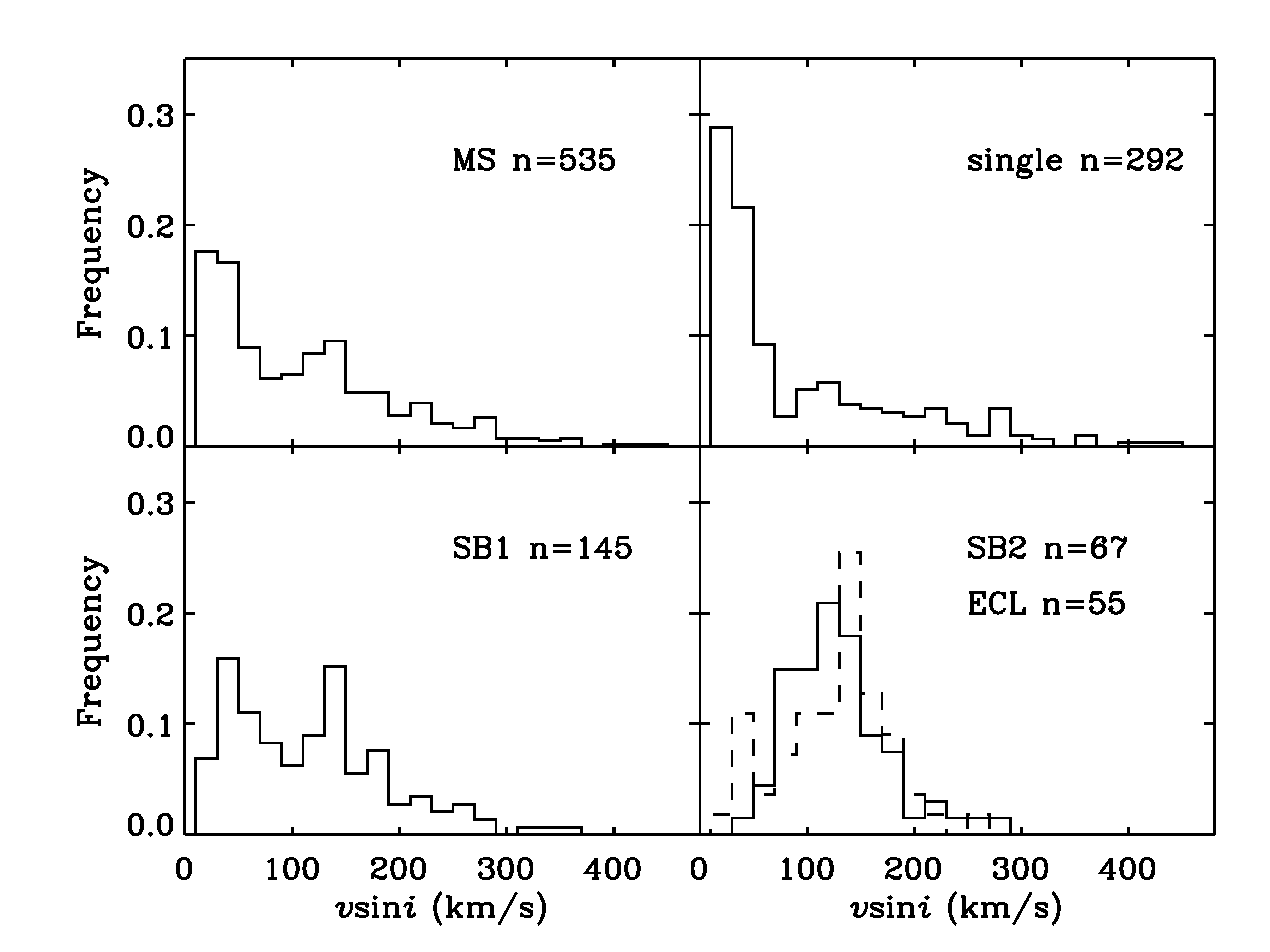}
        \caption{\vsini\ distributions of non-OBe main-sequence systems together with single SB1 and SB2 systems. The sample sizes are as indicated in the insets. In the SB2 panel (lower right), the \vsini\ distribution of the eclipsing systems (ECL) in BLOeM is also shown (dashed line). The histogram bin size is 20\,\kms. The total of main-sequence sources is greater than the sum of single SB1 and SB2 as some main-sequence sources have unclear multiplicity designations, as discussed at the beginning of Sect. \ref{results}. }
        \label{fig:hist_binaries}
\end{figure}

The definition of this region of the HRD that is populated by main-sequence core hydrogen-burning stars, and in particular, the location of the terminal age main sequence (TAMS), is a long-standing problem in stellar evolution. 
Previous attempts to address this issue used properties such as stellar density, rotational velocity, and binary frequency \citep{dufton2006,hunter2008a,vink2010,brott2011,mcevoy2015,vink2025,deburgos2025}.
Following \citet{dufton2006} and \citet{hunter2008a}, we used stellar rotation as an indicator of the main-sequence extent. However, a given \vsini\ has different implications for stars of differing radii, for example main-sequence versus supergiant stars.
Therefore, we used the ratio of \vsini/\vcrit as the metric for defining `fast' or `slow` (projected) rotation, where \vcrit\ is the critical velocity. 
The critical velocities were estimated using the stellar parameters from \citet{bestenlehner2025} and \citet{patrick2025} where available, otherwise, we adopted the data from \citet{shenar2024}.
To give an example, Fig.\,\ref{fig:hist_groups} indicates that \vsini$\sim$100\,\kms\ is a reasonable estimate for differentiating fast/slow cohorts in the BLOeM sample, which corresponds to \vsini/\vcrit$\sim$0.15 on the main sequence.

The distribution of BLOeM systems in the Hertzsprung-Russell diagram (HRD) is shown in  Fig.\,\ref{fig:hrd}, colour-coded according to the above ratio. 
Known OBe stars were removed from this HRD as their parameters are highly uncertain due to disc contamination of their spectra and spectral energy distribution.
Their luminosities of most OBe stars were derived by \citet{shenar2024} using K-band magnitudes, and it is well established that OBe stars have a significant K-band and visible excess due to the presence of their disks \citep{bonanos2010,dunstall2011}.

Interpreting the over-density of `fast' rotators as indicating the extent of the main sequence, we see a sharp drop in number density towards the cool side of the HRD.
This is better visualized in Fig.\,\ref{fig:vsinidrop}, which plots \vsini\ as function of distance in \teff from the approximate position of the ZAMS (left dashed diagonal line in Fig.\,\ref{fig:hrd}).
We only counted stars between the horizontal dashed lines indicating the luminosity range log($L/L_\odot$)=4.2--5.5 because outside this range, completeness becomes a serious issue.
The x-axes of Fig.\,\ref{fig:vsinidrop} indicate the \teff\ a star would have at these luminosity boundaries.
This figure shows a significant decrease in the \vsini\ range at a locus of approximately log(\teff)=4.26 at log($L/L_\odot$)=4.2 to log(\teff)=4.434 at log($L_\odot$)=5.5, as indicated by the vertical dashed line.
This locus is also illustrated as the right dashed diagonal line in Fig.\,\ref{fig:hrd}, which we interpret as indicating the possible position of the TAMS.

It is worth noting here that an alternative explanation for the decrease in \vsini\ in Fig.\,\ref{fig:vsinidrop} is bi-stability braking during the core hydrogen-burning main sequence \citep{vink1999}. 
The appeal of this explanation is that it predicts numerous stars that are cooler than the transition temperature, with the break occurring at approximately the correct location in the stellar HRD.
However, a key limitation is that the mass-loss bi-stability jump has so far only been confirmed in luminous blue variables (LBVs), and not more broadly among OB supergiants, in the SMC \citep{trundle2004,trundle2005,bernini2024}, LMC \citep{verhamme2024}, or the Milky Way \citep{crowther2006,deburgos2024}.

\citet{menon2024} argued that the surface compositions and locations of luminous B-type supergiants in the HRD can be explained as a result of post-main-sequence mergers, as proposed by \citet{podsiadlowski1990} as the origin of the B-supergiant progenitor of SN1987A. This avoids the need to extend the main sequence to cooler effective temperatures.
Clearly, the location of the TAMS is still under debate, and inspection of Fig.\,\ref{fig:hrd} indeed shows that the TAMS implied by the stellar evolution tracks used here by way of illustration \citep{schootemeijer2019} does not match our notional TAMS location. 
Further characterization of the BLOeM binaries will shed further light on this subject and will be the subject of future BLOeM papers (in preparation), but in the following, we continue to refer to main-sequence stars as described above.

The \vsini\ distributions of the main-sequence stars, effectively single stars, SB1, and SB2 systems are shown in Fig.\,\ref{fig:hist_binaries}. 
The single systems have a  strong peak at low \vsini,  with a hint of a secondary peak at intermediate values. 
This secondary peak is much more pronounced in the SB1 systems, likely indicating differing binary characteristics within these two groups.
The SB2 systems lack the low \vsini\ peak, which is expected given our earlier conclusion that many of these systems are likely synchronized by tides and have short periods \citep[see also][]{blex2024}.
We also show the \vsini\ distribution of known eclipsing binary systems \citep[from][]{pawlak2016}, and, assuming common axes of orbital and spin angular momentum, we can infer that this distribution closely reflects that of the equatorial rotation velocities of the A component stars.

A potential implication of Fig.\,\ref{fig:hist_binaries} is that stars with moderate rotation are binaries that have been spun up by tides, while slowly rotating stars are most likely effectively single stars (which may include some fraction of mergers), while stars with the highest \vsini\ have been spun up by previous mass accretion.
Fig.\,\ref{fig:hrd} also shows that the low \vsini\ cohort pervades the main sequence. 

\subsection{The supergiants}
\label{supergiants}

\begin{figure}
    \centering
    \includegraphics[width=0.95\linewidth]{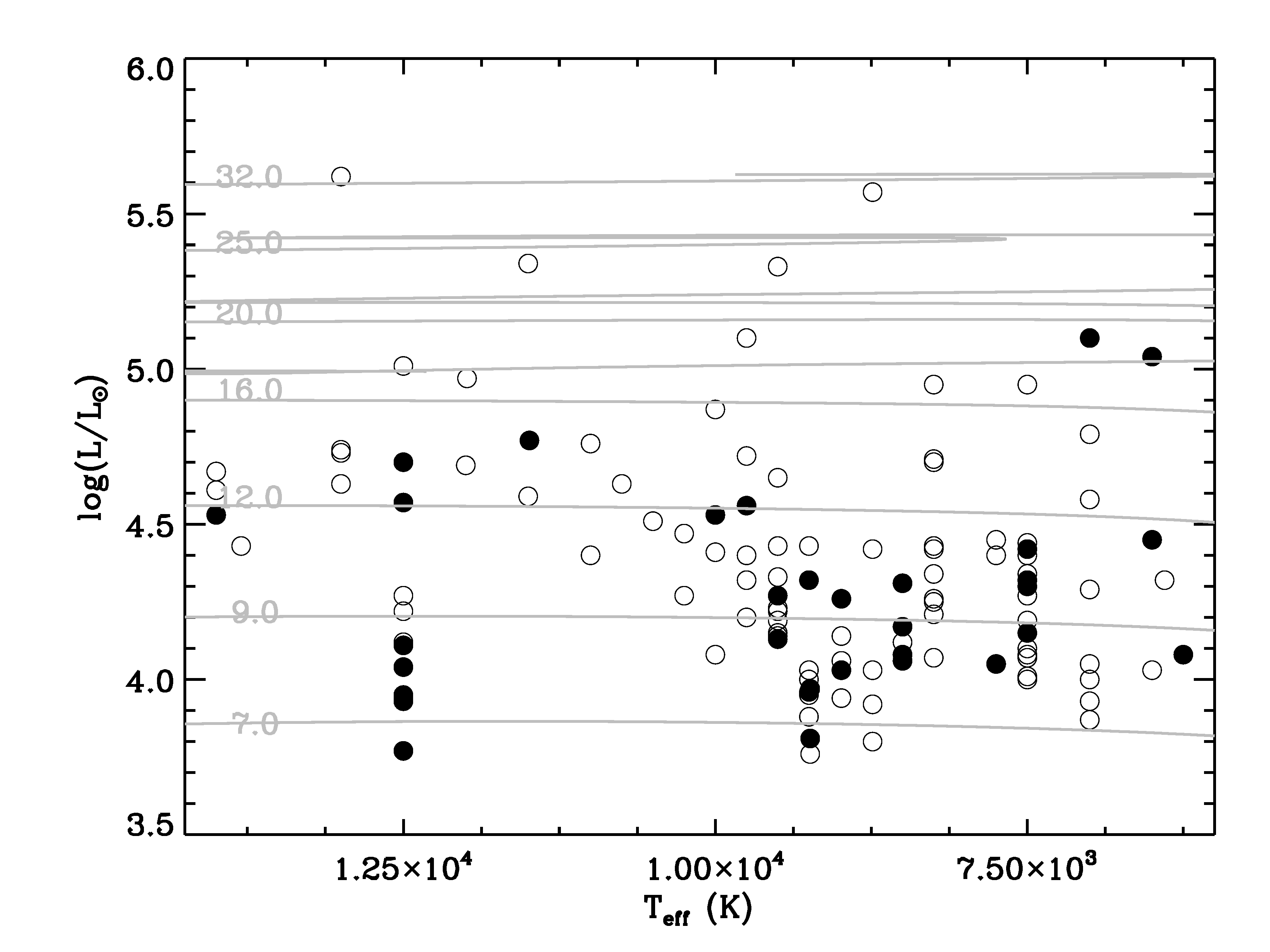}
    \caption{Hertzprung-Russel diagram for the B5 to F-type stars using parameters from \citet{patrick2025}, with stellar evolution tracks from \citet{schootemeijer2019} labelled according to their initial mass. The filled symbols represent sources whose \vsini\ measurement is greater than the lower limit of 33\,\kms\ by at least $3\sigma$. There is some small overlap with the cool side of Fig.\,\ref{fig:hrd}.}
    \label{fig:baf_vsini}
\end{figure}

The post-TAMS B-type supergiants, those with luminosity class I, have mainly low \vsini, as expected, although there are some exceptions, as shown in Fig.\,\ref{fig:vsinidrop}.
Those near the TAMS may well just reflect uncertainties in stellar parameters or the TAMS position, or both.
A few systems with high \vsini\ are also found at cooler temperatures, however, for example BLOeM\,1-042 (B2\,Ib), 4-042 (B2.5\,Ib) and 4-105 (B2\,Ib:), which are not radial velocity variables. 
This is consistent with other work, however, because a small number of similar outliers were found in previous \vsini\ surveys of B-type supergiants in the SMC \citep{dufton2006}, LMC \citep{lennon2010,mcevoy2015}, and Milky Way \citep{fraser2010,deburgos2025}.
A notable high \vsini\ system in this group is BLOeM 3-073 (B3\,Ib, \vsini=103\,\kms), which is a known eclipsing binary (OGLE SMC-ECL-1476; \citealt{pawlak2016}), and one of the very few binaries cool-wards of the TAMS.
BLOeM 4-096 (B1.5\,II:) also has high \vsini, but this system is flagged as an H$\alpha$ emission line source from $Gaia$ data \citep{shenar2024}.

The BAF supergiants are even more strongly peaked at lower \vsini. The bulk of these stars have unresolved \vsini\ estimates at the current resolution \citep{patrick2025}.
Typically, we derive \vsini\ upper limits for most BAF supergiants of approximately 30\,\kms, which is similar to the results of \citet{verdugo1999} for 32 A-type supergiants in the Milky Way.
The measurements imply a modal value of $\sim$33\,\kms\ as the upper limit in this temperature range, although for F-type supergiants, the values are likely overestimated due to the effect of line crowding and blending on defining a true continuum \citep[see][for discussion of these issues for A and F type stars]{royer2002}.
For the early B-type supergiants, there is a scattering of objects at higher \vsini\ that was also noted by \citet{patrick2025}. 
We illustrate their locations in the HRD in Fig.\,\ref{fig:baf_vsini} using parameters from \citet{patrick2025}, and we flag sources whose \vsini\ values lie significantly above the \vsini\ limit.
Most of these outliers have initial masses lower than $\sim$10\,M$_{\odot}$, with the higher-mass supergiants being essentially unresolved. 
The two cool outliers below 8\,000\,K have \vsini\ values lower that 50\,\kms\ , and we discounted these for the reasons noted above.

\subsection{The OBe stars}
\label{obe}

\begin{figure}
    \centering
    \includegraphics[width=0.95\linewidth]{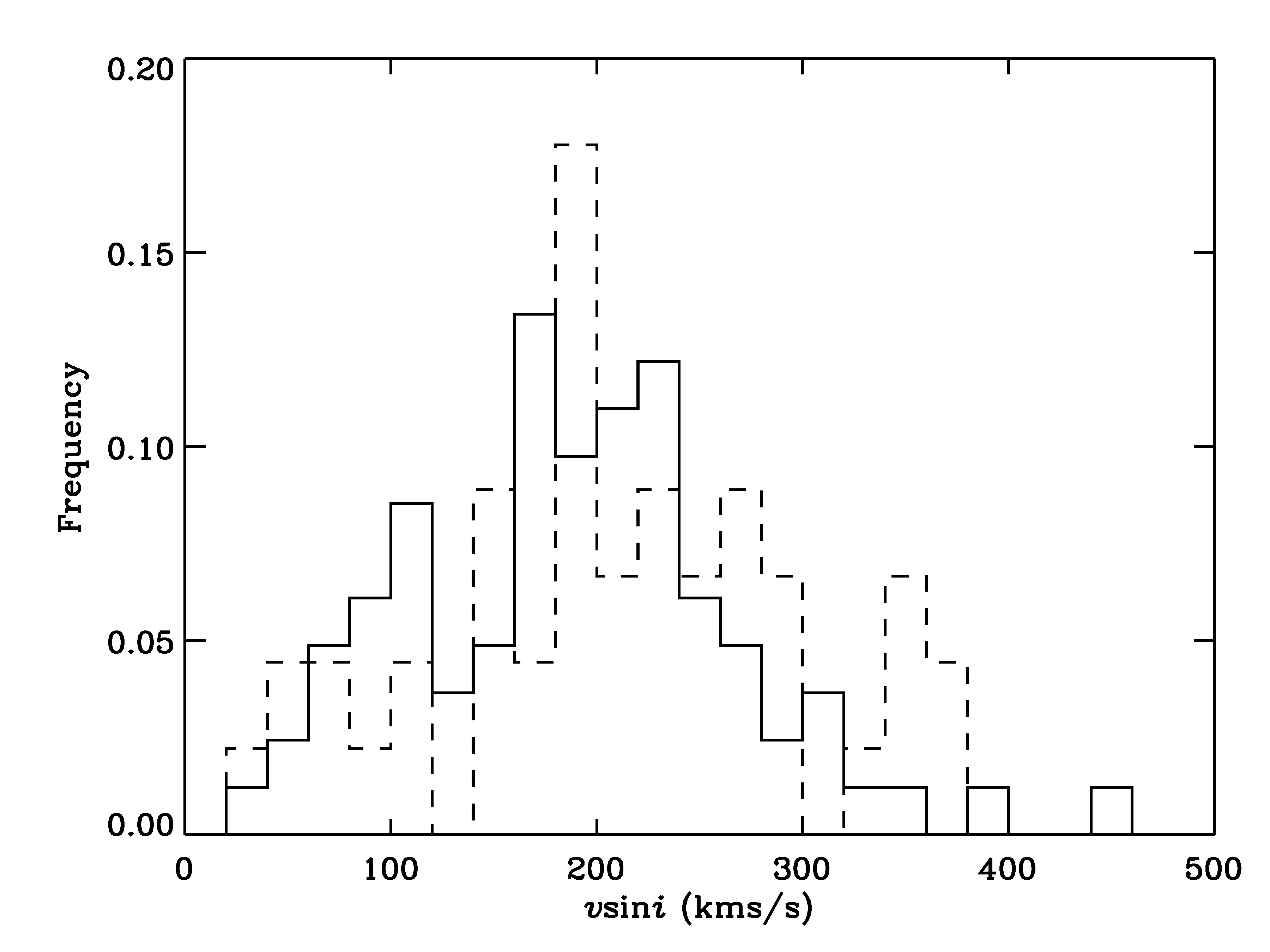}
   \caption{Histograms of \vsini\ measurements for BLOeM OBe stars (solid line) compared with those of \citet[dashed line]{dunstall2011}. The bin size is 20\,\kms.}
    \label{fig:plot_obe}
\end{figure}

The \vsini\ distribution of the BLOeM OBe stars is similar to that in other studies of the SMC and LMC. When the Of?p stars are excluded, the mean and standard deviation of their \vsini\ measurements are 199 and 77 \kms, respectively, to be compared with 208 and 88\,\kms\ of \citet{dunstall2011} for a combination of NGC\,346 and NGC\,330 sources (see Fig.\,\ref{fig:plot_obe}).  
These are only slightly below the values obtained for young OBe populations, NGC\,346 alone and the Tarantula Nebular region in the LMC \citep{dufton2022}.
Because the OBe designation is based primarily on the current spectral coverage that excludes H$\alpha$, however, we postpone a detailed discussion of their properties to a future paper that will include these data.

\section{Comparison with previous work}
\label{otherwork}

\begin{figure}
    \centering
    \includegraphics[width=1.0\linewidth]{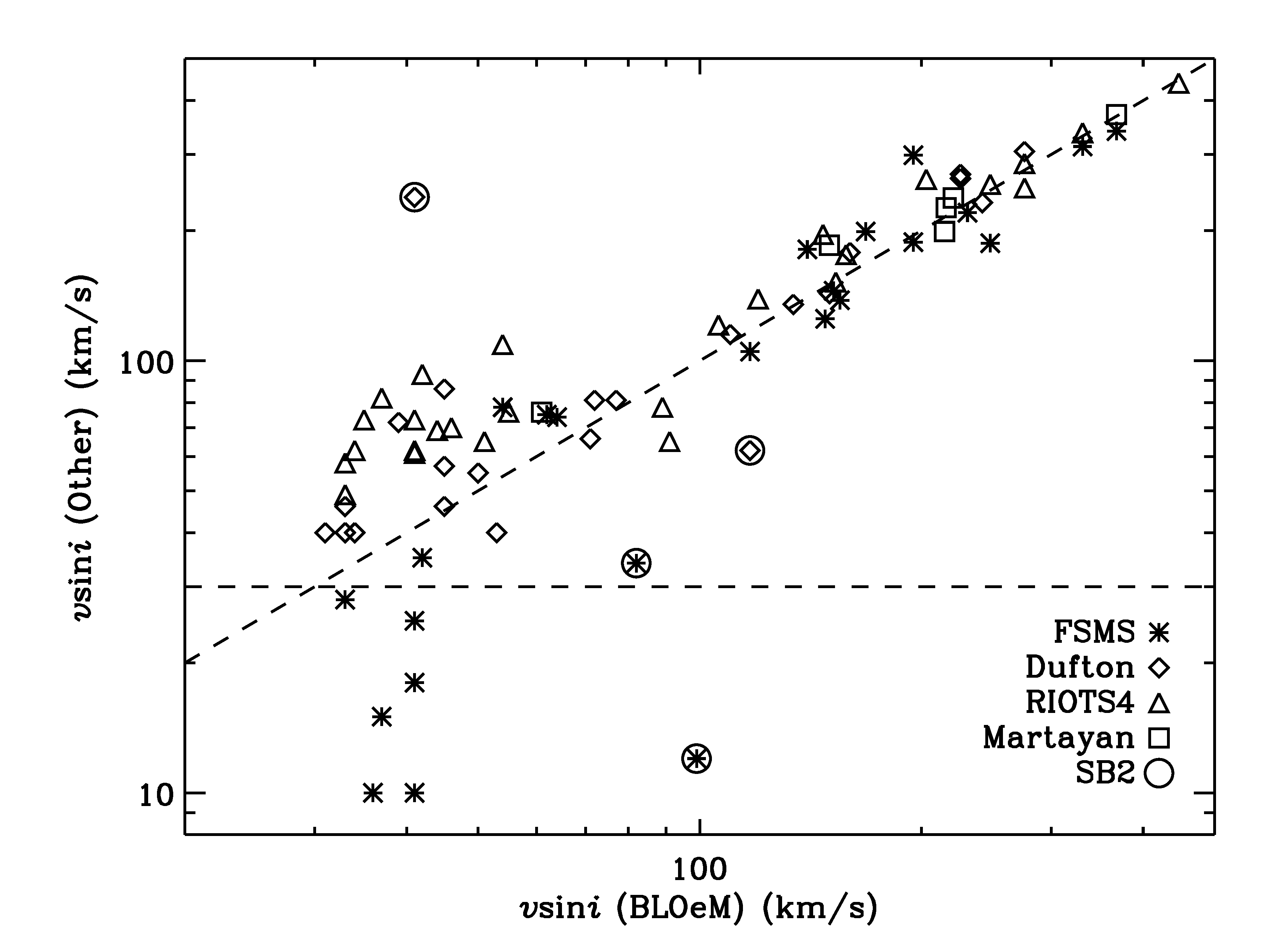}
    \caption{Comparison of this work with \vsini\ values in the literature, as noted in the inset and described in Sect. \ref{otherwork}, where references can also be found. The outliers in this plot are known SB2 systems (circled). The diagonal dashed line indicates the 1:1 correlation, and the horizontal line indicates the approximate lower limit of the BLOeM measurements.}
    \label{fig:comparison_vsini}
\end{figure}

In the introduction, we noted previous \vsini\ surveys in the SMC that we briefly summarize below. 
\begin{itemize}
    \item FSMS \citep{hunter2008a} obtained higher-resolution (R$\sim$20\,000) VLT/Flames data for two fields in the SMC centred on NGC\,346 and NGC\,330, which overlap fields 4 and 7 in BLOeM.
    \item Martayan \citep{martayan2007} obtained VLT/Flames data at similar resolution to BLOeM for two fields in the vicinity of NGC\,330.
    \item Ramachandran \citep{ramachandran2019} obtained similar resolution VLT/Flames data for two fields in the SMC wing.
    \item Dufton \citep {dufton2019} presented data with a similar resolution for three additional Flames fields centred on NGC\,346.
    \item RIOTS4 \citep{dorigojones2020} contains data for a range of resolutions for stars distributed throughout the SMC that were selected as being relatively isolated.
\end{itemize}
A comparison of \vsini\ measurements for stars in common with BLOeM in Fig.\,\ref{fig:comparison_vsini} illustrates the effects of differing resolutions in the low \vsini\ limit. For example the lower-resolution RIOTS4 data do not resolve values below $\sim$80\,\kms, while the higher-resolution FSMS results extend below the similar resolution BLOeM/Dufton limit of $\sim$30--40\,\kms. The BLOeM fields do not overlap the SMC wing fields \citep{ramachandran2019}. Overall, the agreement across all surveys is reasonable. The small number of outliers above a \vsini\ of about 60\,\kms\ are SB2 systems.

\begin{figure}
    \centering
    \includegraphics[width=1.0\linewidth ]{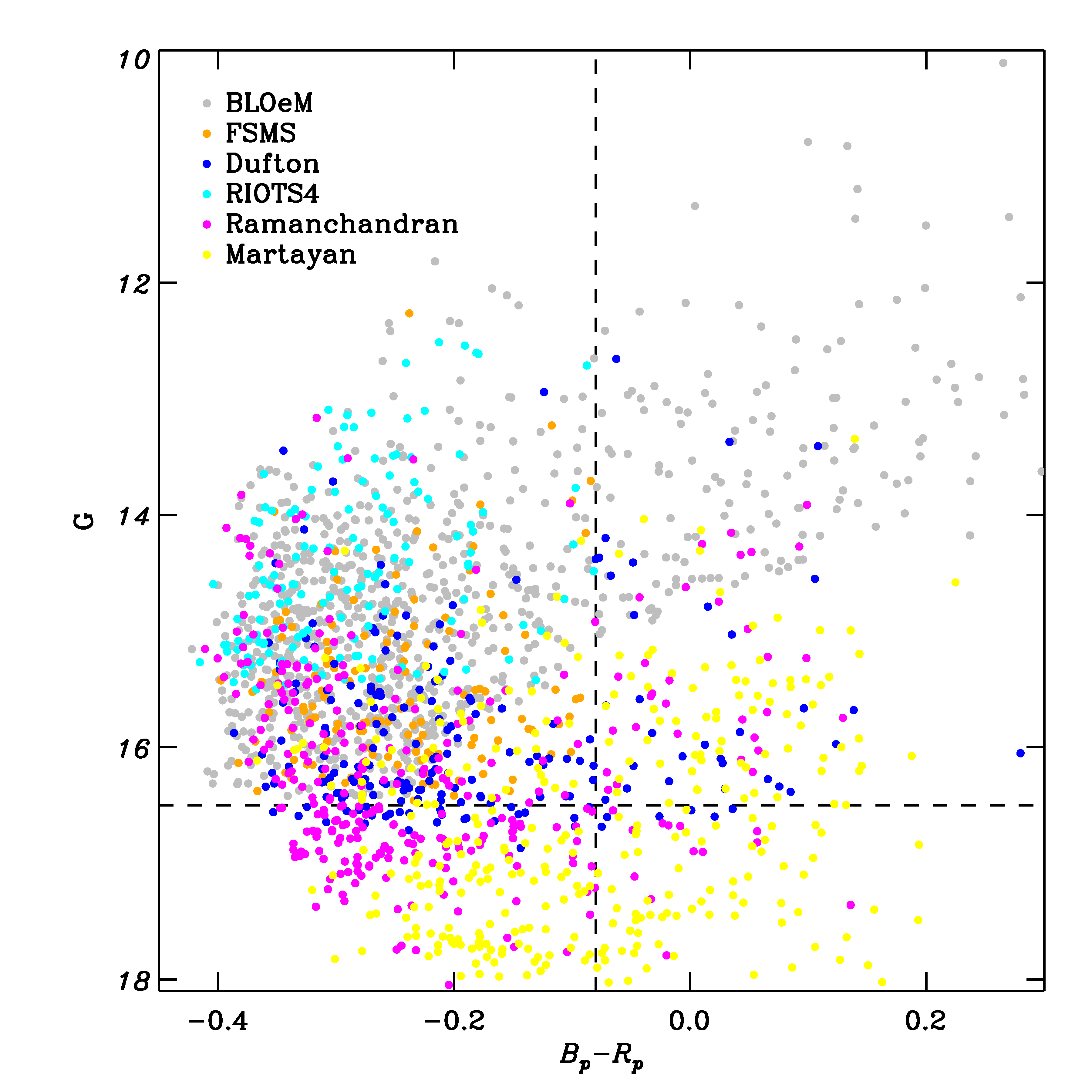}
    \caption{Colour-magnitude diagram for the BLOeM targets (grey squares) compared with those of other major \vsini\ surveys in the SMC: FSMS \citep[][]{hunter2008a}, Martayan \citep[][]{martayan2007}, Ramachandran \citep[][]{ramachandran2019}, Dufton \citep[][]{dufton2019}, and  RIOTS4 \citep[][]{dorigojones2020}, as indicated in the panel. The dashed lines indicate limits in colour and magnitude assumed for the \vsini\ comparison. }
    \label{fig:cmd}
\end{figure}

\begin{figure}
    \centering
    \includegraphics[width=1.0\linewidth]{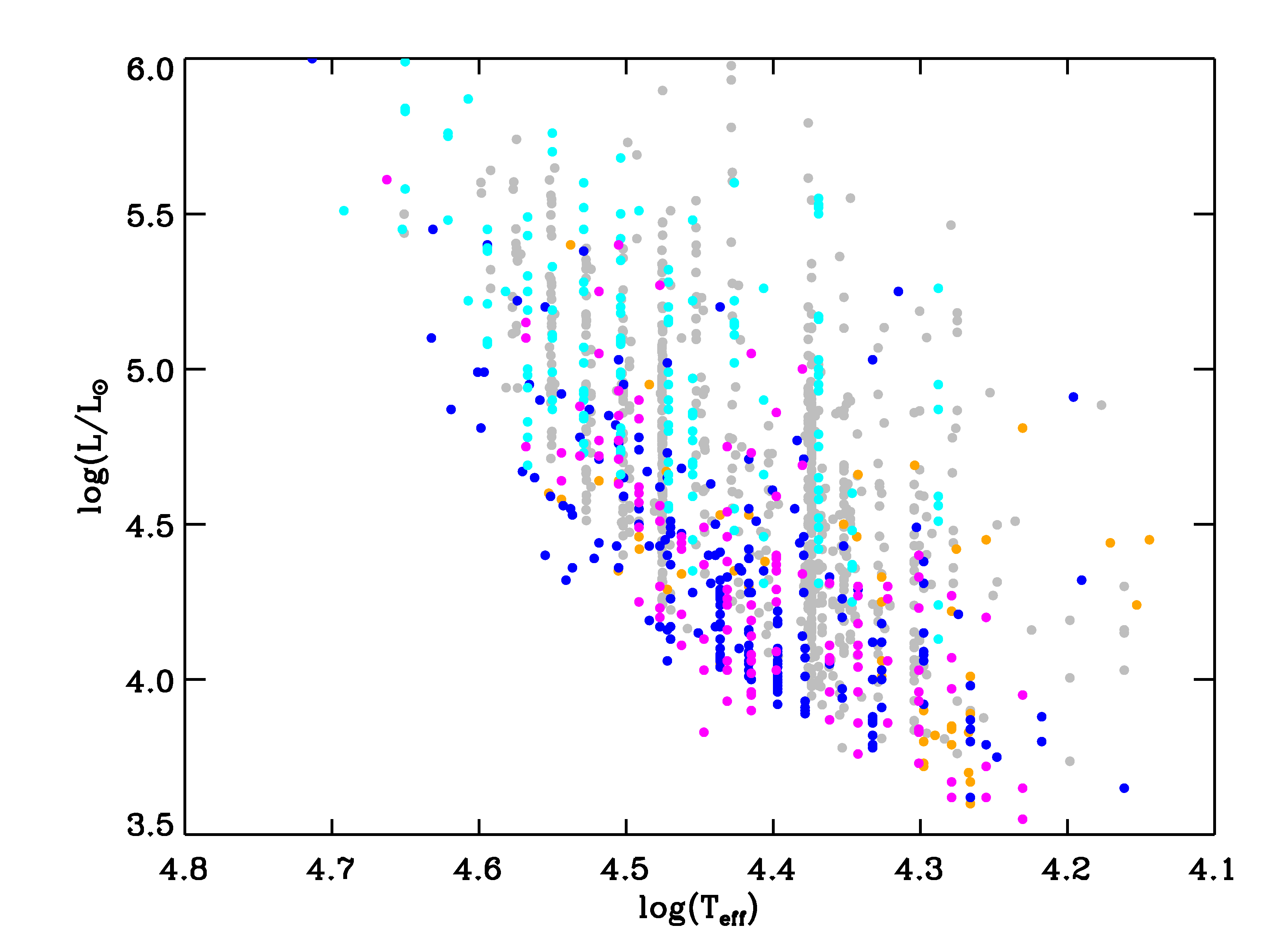}
    \caption{Hertzsprung-Russel diagram of bright blue sources from Fig.\,\ref{fig:cmd}, colour-coded as in that figure. The Martayan sample is not shown because no stellar luminosities were published.}
    \label{fig:plot_hrds}
\end{figure}

When the various samples are compared, it is important to consider selection effects and biases. For example Fig.\,\ref{fig:cmd} shows the $Gaia$ colour-magnitude diagram (CMD) for the BLOeM targets compared with these other surveys.
The various depths and colour cuts are apparent, and we therefore imposed magnitude ($G$ < 16.5 mag) and colour ($B_P-R_P$ < –0.08; approximately B3) cuts to enable a consistent comparison across all samples for massive main-sequence stars and B-type supergiants.
This colour cut excludes the BLOeM BAF, which is unique in this respect, although as discussed above, all but a few BAF stellar spectra are unresolved at our resolution.
The comparison of these populations in the HRD in Fig.\,\ref{fig:plot_hrds} illustrates the large number of early to mid B-type giants/bright-giants in BLOeM that is unmatched in other surveys.
On the other hand, Fig.\,\ref{fig:plot_hrds} also illustrates that the young populous region NGC\,346 (blue points) populates the region of the HRD closest to the ZAMS, which is only sparsely populated by BLOeM (see Fig.\,\ref{fig:plot_hrds}).

\begin{figure}
    \centering
    \includegraphics[width=1.0\linewidth]{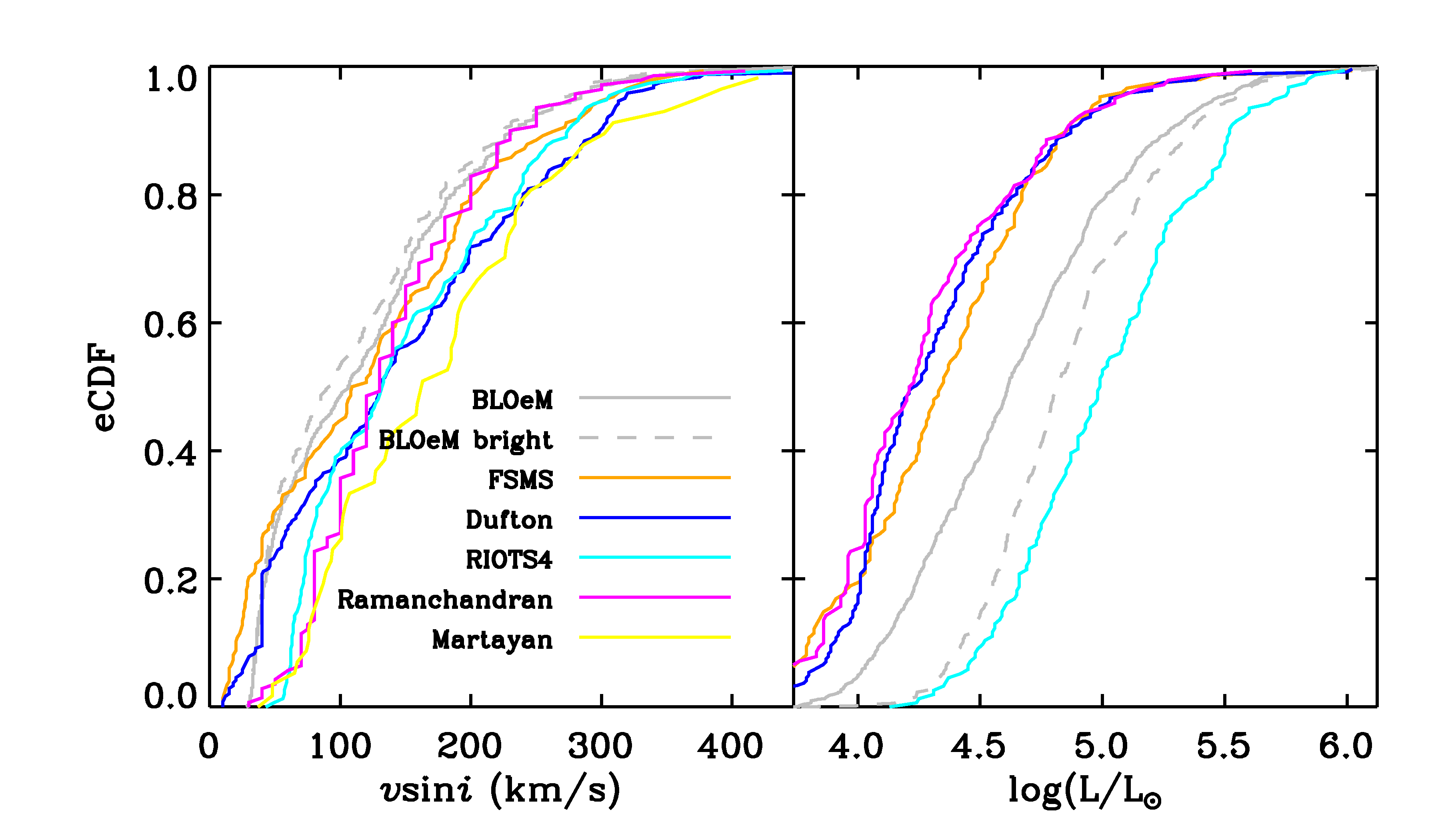}
    \caption{eCDFs for the samples illustrated in Fig.\,\ref{fig:plot_hrds}. The BLOeM sample (solid grey line) has had a colour cut applied as for the other surveys, as shown in Fig.\,\ref{fig:cmd}, while a magnitude cut (dashed grey line) was applied at 15.5 for comparison with RIOTS4. The Martayan dataset is missing from the right plot because no luminosity estimates were published for these sources.}
    \label{fig:plot_edfs_other}
\end{figure}

The eCDFs of these samples, Fig.\,\ref{fig:plot_edfs_other}, for \vsini\ and luminosity reflect their selection biases, as expected, even with the same colour and magnitude cuts. This is consistent with the significant differences found between the eight BLOeM fields themselves (Fig.\,\ref{fig:plot_edfs}).
The somewhat higher number of stars in the lowest \vsini\ bins of the FSMS and Dufton samples can be attributed to the higher resolution of the FSMS survey ($\sim$20\,000). These stars were also included in the NGC\,346 catalogue of \citet{dufton2019}.  
At $\sim$50\,\kms\ , however, the number of slow rotators in BLOeM exceeds those in FSMS and Dufton.  
By contrast, the RIOTS4 survey observed the bulk of their sample at lower resolution (2\,600--3\,700), although some sources were observed at high resolution (28\,000). 
This might explain the lack of stars in the lowest \vsini\ bins, but at moderate \vsini\ values, there are still fewer sources than in BLOeM.
As RIOTS4 preferentially targeted isolated massive stars, this difference might indicate that these objects are lacking the low \vsini\ cohort found in BLOeM,
although examination of their luminosity eCDFs reveals significant differences.

The eCDFs of Martayan and Ramachandran, who used a resolution identical to that of BLOeM, also have a significantly smaller fraction of stars at low \vsini.
The HRD/CMD coverages of these two surveys are very different from that of BLOeM, indicating that these surveys are biased towards near main-sequence stars. The lower-mass stars were removed from the eCDFs using the magnitude cut.
The Ramachandran and Martayan samples also lack slow rotators, although the former converges to BLOeM at higher \vsini\ values.
The similarity between the Dufton and Ramachandran luminosity functions may seem surprising in view of their differing \vsini\ eCDFs. The former focused on the NGC\,346 region of the SMC, however, which has a significant fraction of stars with ages $<$10\,Myr \citep{dufton2019}, while the Ramachandran sample has relatively few stars younger than $\sim$10\,Myr \citep{ramachandran2019}.

\begin{figure}
        \centering
        \includegraphics[width=1.0\linewidth]{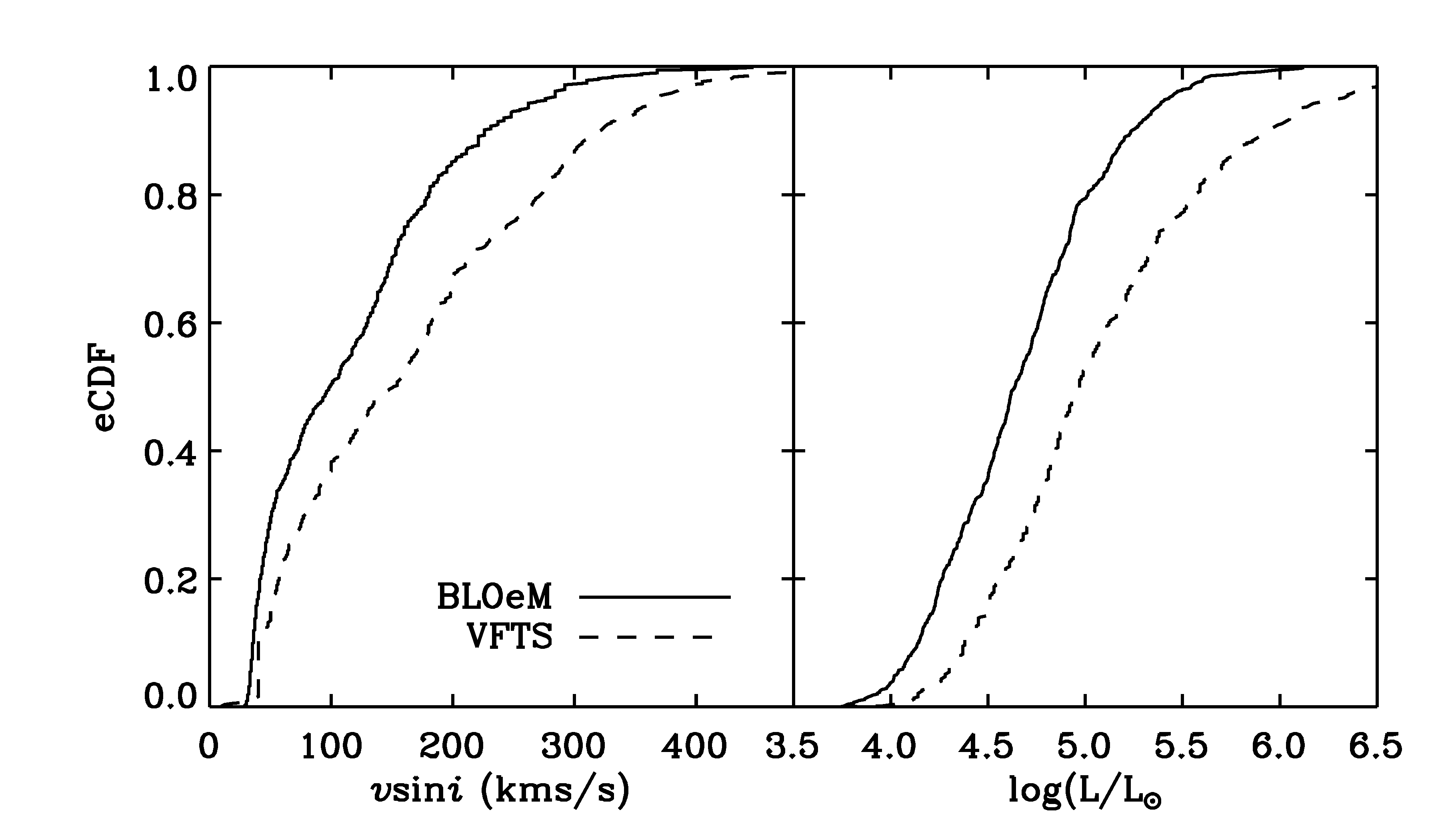}
        \caption{Comparison of \vsini\ and luminosity eCDFs for BLOeM and VFTS. Both datasets have been restricted to OB stars, but excluding OBe stars. The stellar parameters for the VFTS sources are from \citet{mcevoy2015,garland2017,schneider2018,dufton2018}. }
        \label{fig:plot_vfts}
\end{figure}

Similar issues of selection biases apply to comparisons with the LMC. For example we can compare with the VFTS survey \citep{evans2011} of the 30 Doradus region, which used the same instrument and resolution as BLOeM. 
Figure\,\ref{fig:plot_vfts} shows that while the overall shape of the \vsini\ distributions are similar, the BLOeM sample is shifted to lower values, while their luminosity distributions indicate that the VFTS sources are intrinsically more luminous.
The latter feature is easily understood when we consider that the 30 Doradus field is very young and many of those sources are closer to the ZAMS.
We refer again to Fig.\,\ref{fig:plot_edfs} and compare the solid and dashed lines. They clearly show that the incompleteness in the BLOeM sample is setting in around $G$=15\,mag.
Evidently, the BLOeM survey is focused more strongly on evolved massive stars.

To summarize this section, the metallicity (Z) appears to play a minor role in determining the \vsini\ distributions, at least compared to other environmental differences.
The change in \vsini\ distributions between regions within the SMC or with regions in other galaxies are strongly affected by the specific characteristics of these massive star populations: mass function, age distribution, etc.  
To some extent, this is contrary to previous expectations based on the fraction of Be stars in different environments \citep{maeder1999,bonanos2010,schootemeijer2022}, which suggests that the initial rotation rate is not an important driver in the production of Be stars.

\section{Discussion}
\label{discussion}

Perhaps the most striking result of this survey is the preponderance of stars with low \vsini\ throughout the main sequence, as discussed in Sect. \ref{demographics} and as illustrated by their distribution in the HRD (Figs.\,\ref{fig:hrd} and \ref{fig:hist_binaries}).
In typical binary population synthesis paradigms \citep{demink2013,xu2025,schuermann2025}, the slowly rotating cohort is primarily composed of stars that have retained their (modest) initial rotational velocity, unaffected by tides or mass transfer.
These are effectively single stars from an evolutionary perspective, even  though they may reside in long-period binary systems, as indicated by the low \vsini\ cohort in the SB1 panel of Fig.\,\ref{fig:hist_binaries}.
A fraction of the low \vsini\ cohort are also expected to be post-interaction systems, however, such as mergers  \citep{schneider2019,schneider2020,wang2022,xu2025}.
By contrast, the SB2 systems are consistent with being tidally spun up, are in synchronous rotation, and are a mix of pre- or post-interaction systems.
The presence of significant numbers of systems at higher \vsini\ in the single and SB1 systems would be consistent with a mix of undetected SB2s and post-interaction binaries.
Finally, the highest \vsini\ systems are thought to result from spin-up due to mass accretion in binary systems, perhaps together with the OBe stars, although there are relatively few of the latter in BLOeM compared to VFTS.

In the low \vsini\ cohort, the \ion{N}{ii} line at 3995\,\AA\ is present in many B-type stars and is suggestive of the N-rich slowly rotating B-type stars discussed by \citet[see also \citealt{dufton2018,dufton2020}]{hunter2008a}.
As a preliminary test for N-enhancement in the BLOeM slow rotators, we selected a sample of main-sequence B-type stars with \vsini$\leq$40\,\kms, 43 sources,  
and compared the \ion{N}{ii} 3995\,\AA\ equivalent widths with theoretical predictions for a range of nitrogen enhancements. The details are summarized in Appendix \ref{app:nii}.
This preliminary analysis implies that $\sim$30\% (17/43) of these slowly rotating stars are enhanced by at least 0.6 dex, while inspection of the radial velocities of the SB1 sources (20\%\ of the sample) suggests that they are long-period systems. 

The presence of nitrogen-rich slowly rotating main-sequence B-type stars in the Galaxy, LMC, and SMC is well established \citep[see for example][]{gies1992,lennon2003,hunter2008a,hunter2009,morel2008}. This is reflected in our data, although we cannot as yet comment on the sources with higher \vsini\ that would facilitate direct comparison with the Hunter diagram.
While a number of potential causes have been put forward to explain the nitrogen enrichment, it is not clear to which extent they might all contribute.
As previously noted, 
stellar mergers might explain some fraction of these systems, which appears to be higher when approaching the TAMS, according to population synthesis predictions \citep{xu2025}, although others may represent SMC analogues of the magnetic merger candidate $\tau$ Scorpii \citep{schneider2020}, which is closer to the ZAMS.
Similarly, \citet{jin2024} discussed evolutionary scenarios, including mergers, that might explain a large sample of boron-depleted and nitrogen-rich slowly rotating Galactic B-type stars, but they speculated in addition that mixing during the pre-main-sequence might explain some fraction of these nitrogen-enhanced objects.
Finally, some fraction of the slowly rotating SB1 sources might be partially stripped post-interaction systems \citep[][]{shenar2020,bodensteiner2020,villasenor2025,klencki2022}.

\section{Conclusions}

We have determined projected rotational velocities (\vsini) for the 929 systems in the BLOeM survey of massive stars in the SMC. We used the Fourier transform method for single stars, and for SB2 systems, a first attempt was made to determine \vsini\ for both components using a double-Gaussian fitting approach. We summarize the main results below.

\begin{itemize}
    \item Using the change in the \vsini\ distribution with \teff\, we can define a cool edge of the main sequence of likely core hydrogen-burning stars.
        \item The \vsini\ distributions of main-sequence single and SB1 systems exhibit a strong peak between 30--60\,\kms, close to the resolution limit, indicating the presence of many upper limits. These low \vsini\ systems pervade the main sequence, and some of them might be long-period systems.
        \item Preliminary inspection of the \ion{N}{ii} 3995\,\AA\ line in B-type stars revealed that many of these systems have an enhanced nitrogen abundance, although it is unclear whether this enhancement occurred pre- or post-main sequence. 
        \item The \vsini\ distribution of currently designated SB2 systems peaks at $\sim$140\,\kms\ , and together with known OGLE periods, this suggests that members of this cohort are mostly high mass-ratio systems in synchronous rotation.
        \item Only a very small number, 3\%, of the high \vsini\ stars ($>$300\,\kms) are found in the sample. Their potential runaway nature will be considered in a future paper.
        \item Internal comparisons between survey fields and with other surveys in the SMC, LMC, and Milky Way suggest that the variations in the derived \vsini\ distributions are mainly driven by intrinsic stellar population differences and not by metallicity.
\end{itemize}

Clearly, additional work that is aimed at determining orbital parameters and more precise surface compositions will provide much stronger constraints and further insight into the processes that shape the \vsini\ distributions presented in this paper.

\section*{Data availability}

Tables 1, 2, and C.1 are only available in electronic form at the CDS via anonymous ftp to cdsarc.u-strasbg.fr (130.79.128.5) or via http://cdsweb.u-strasbg.fr/cgi-bin/qcat?J/A+A/.

\begin{acknowledgements}
This work has made use of ESASky, TOPCAT, Simbad and Vizier.
SRB acknowledges support from the `Viera y Clavijo' postdoctoral program of the Consejer\`ia de Economía, Conocimiento y Empleo del Gobierno de Canarias through the Agencia Canaria de Investigaci\`on, Innovaci\`on y Sociedad de la Informaci\`on. 
SRWB, A.H., S.S-D. and G.H. acknowledge support from the State Research Agency (AEI) of the Spanish Ministry of Science and Innovation (MICIN) and the European Regional Development Fund, FEDER under grants PID2021-122397NB-C21 and PID2024-159329NB-C21.
G. H. received the support from the “La Caixa” Foundation (ID 100010434) under the fellowship code LCF/BQ/PI23/11970035.
DMB gratefully acknowledges UK Research and Innovation (UKRI) in the form of a Frontier Research grant under the UK government's ERC Horizon Europe funding guarantee (SYMPHONY; grant number: EP/Y031059/1), and a Royal Society University Research Fellowship (grant number: URF{\textbackslash}R1{\textbackslash}231631).
PM acknowledges support from the European Research Council (ERC) under the European Union’s Horizon 2020 research and innovation programme (grant agreement No. 101165213/Star-Grasp), and from the Fonds Wetenschappelijk Onderzoek (FWO) senior postdoctoral fellowship number 12ZY523N
T.~Shenar, ZK, and TS acknowledge support from the European Research Council (ERC) under the European Union's Horizon 2020 research and innovation program (grant agreement 101164755/METAL) and the Israel Science Foundation (ISF) under grant number 0603225041.
AACS acknowledges support by the Deutsche Forschungsgemeinschaft (DFG, German Research Foundation) in the form of an Emmy Noether Research Group -- Project-ID 445674056 (SA4064/1-1, PI Sander). AACS further acknowledges financial support by the Federal Ministry for Economic Affairs and Energy (BMWE) via the Deutsches Zentrum f\"ur Luft- und Raumfahrt (DLR) grant 50 OR 2306, co-funded by the European Union (Project 101183150 - OCEANS).
A.B.~acknowledges support from the Australian Research Council (ARC) Centre of Excellence for Gravitational Wave Discovery (OzGrav), through project number CE230100016.
JIV acknowledges support from the European Research Council for the ERC Advanced Grant 101054731.
LRP and FN acknowledge support by the Spanish Ministry of Science, Innovation and Universities/State Agency of Research MICIU/AEI/10.13039/501100011033 grants PID2022-137779OB-C41 and  PID2022-140483NB-C22, and by “ERDF A way of making Europe”. FN also acknowledges grant MAD4SPACE, TEC-2024/TEC-182 from Comunidad de Madrid (Spain).
DP acknowledges financial support from the FWO in the form of a junior postdoctoral fellowship No. 1256225N.
his research was supported in part by grant NSF PHY-2309135 to the Kavli Institute for Theoretical Physics (KITP)
\end{acknowledgements}

\bibliographystyle{aa}
\bibliography{literature.bib}

@ARTICLE{bonanos2010,
       author = {{Bonanos}, A.~Z. and {Lennon}, D.~J. and {K{\"o}hlinger}, F. and {van Loon}, J. Th. and {Massa}, D.~L. and {Sewilo}, M. and {Evans}, C.~J. and {Panagia}, N. and {Babler}, B.~L. and {Block}, M. and {Bracker}, S. and {Engelbracht}, C.~W. and {Gordon}, K.~D. and {Hora}, J.~L. and {Indebetouw}, R. and {Meade}, M.~R. and {Meixner}, M. and {Misselt}, K.~A. and {Robitaille}, T.~P. and {Shiao}, B. and {Whitney}, B.~A.},
        title = "{Spitzer SAGE-SMC Infrared Photometry of Massive Stars in the Small Magellanic Cloud}",
      journal = {\aj},
         year = 2010,
        month = aug,
       volume = {140},
       number = {2},
        pages = {416-429},
          doi = {10.1088/0004-6256/140/2/416},
archivePrefix = {arXiv},
       eprint = {1004.0949},
 primaryClass = {astro-ph.SR},
       adsurl = {https://ui.adsabs.harvard.edu/abs/2010AJ....140..416B}
}

@ARTICLE{menon2024,
       author = {{Menon}, Athira and {Ercolino}, Andrea and {Urbaneja}, Miguel A. and {Lennon}, Daniel J. and {Herrero}, Artemio and {Hirai}, Ryosuke and {Langer}, Norbert and {Schootemeijer}, Abel and {Chatzopoulos}, Emmanouil and {Frank}, Juhan and {Shiber}, Sagiv},
        title = "{Evidence for Evolved Stellar Binary Mergers in Observed B-type Blue Supergiants}",
      journal = {\apjl},
         year = 2024,
        month = mar,
       volume = {963},
       number = {2},
          eid = {L42},
        pages = {L42},
          doi = {10.3847/2041-8213/ad2074},
archivePrefix = {arXiv},
       eprint = {2311.05581},
 primaryClass = {astro-ph.SR},
       adsurl = {https://ui.adsabs.harvard.edu/abs/2024ApJ...963L..42M}
}

@ARTICLE{vink1999,
       author = {{Vink}, J.~S. and {de Koter}, A. and {Lamers}, H.~J.~G.~L.~M.},
        title = "{On the nature of the bi-stability jump in the winds of early-type supergiants}",
      journal = {\aap},
         year = 1999,
        month = oct,
       volume = {350},
        pages = {181-196},
          doi = {10.48550/arXiv.astro-ph/9908196},
archivePrefix = {arXiv},
       eprint = {astro-ph/9908196},
 primaryClass = {astro-ph},
       adsurl = {https://ui.adsabs.harvard.edu/abs/1999A&A...350..181V}
}

@ARTICLE{dufton2022,
       author = {{Dufton}, P.~L. and {Lennon}, D.~J. and {Villase{\~n}or}, J.~I. and {Howarth}, I.~D. and {Evans}, C.~J. and {de Mink}, S.~E. and {Sana}, H. and {Taylor}, W.~D.},
        title = "{Properties of the Be-type stars in 30 Doradus}",
      journal = {\mnras},
         year = 2022,
        month = may,
       volume = {512},
       number = {3},
        pages = {3331-3344},
          doi = {10.1093/mnras/stac630},
archivePrefix = {arXiv},
       eprint = {2203.02440},
 primaryClass = {astro-ph.SR},
       adsurl = {https://ui.adsabs.harvard.edu/abs/2022MNRAS.512.3331D}
}

@ARTICLE{dunstall2011,
       author = {{Dunstall}, P.~R. and {Brott}, I. and {Dufton}, P.~L. and {Lennon}, D.~J. and {Evans}, C.~J. and {Smartt}, S.~J. and {Hunter}, I.},
        title = "{The VLT-FLAMES survey of massive stars: Nitrogen abundances for Be-type stars in the Magellanic Clouds}",
      journal = {\aap},
         year = 2011,
        month = dec,
       volume = {536},
          eid = {A65},
        pages = {A65},
          doi = {10.1051/0004-6361/201117588},
archivePrefix = {arXiv},
       eprint = {1109.6661},
 primaryClass = {astro-ph.SR},
       adsurl = {https://ui.adsabs.harvard.edu/abs/2011A&A...536A..65D}
}

@ARTICLE{brott2011,
   author = {{Brott}, I. and {de Mink}, S.~E. and {Cantiello}, M. and {Langer}, N. and 
	{de Koter}, A. and {Evans}, C.~J. and {Hunter}, I. and {Trundle}, C. and 
	{Vink}, J.~S.},
    title = "{Rotating massive main-sequence stars. I. Grids of evolutionary models and isochrones}",
  journal = {\aap},
archivePrefix = "arXiv",
   eprint = {1102.0530},
 primaryClass = "astro-ph.SR",
     year = 2011,
    month = jun,
   volume = 530,
      eid = {A115},
    pages = {A115},
      doi = {10.1051/0004-6361/201016113},
   adsurl = {http://cdsads.u-strasbg.fr/abs/2011A%26A...530A.115B}
}

@ARTICLE{dufton2013,
   author = {{Dufton}, P.~L. and {Langer}, N. and {Dunstall}, P.~R. and {Evans}, C.~J. and 
	{Brott}, I. and {de Mink}, S.~E. and {Howarth}, I.~D. and {Kennedy}, M. and 
	{McEvoy}, C. and {Potter}, A.~T. and {Ram{\'{\i}}rez-Agudelo}, O.~H. and 
	{Sana}, H. and {Sim{\'o}n-D{\'{\i}}az}, S. and {Taylor}, W. and 
	{Vink}, J.~S.},
    title = "{The VLT-FLAMES Tarantula Survey. X. Evidence for a bimodal distribution of rotational velocities for the single early B-type stars}",
  journal = {\aap},
archivePrefix = "arXiv",
   eprint = {1212.2424},
 primaryClass = "astro-ph.SR",
     year = 2013,
    month = feb,
   volume = 550,
      eid = {A109},
    pages = {A109},
      doi = {10.1051/0004-6361/201220273},
   adsurl = {http://adsabs.harvard.edu/abs/2013A%26A...550A.109D}
}

@ARTICLE{collins1995,
       author = {{Collins}, II, George W. and {Truax}, Ryland J.},
        title = "{Classical Rotational Broadening of Spectral Lines}",
      journal = {\apj},
         year = 1995,
        month = feb,
       volume = {439},
        pages = {860},
          doi = {10.1086/175225},
       adsurl = {https://ui.adsabs.harvard.edu/abs/1995ApJ...439..860C}
}

@ARTICLE{dufton2006,
       author = {{Dufton}, P.~L. and {Ryans}, R.~S.~I. and {Sim{\'o}n-D{\'\i}az}, S. and {Trundle}, C. and {Lennon}, D.~J.},
        title = "{B-type supergiants in the Small Magellanic Cloud: rotational velocities and implications for evolutionary models}",
      journal = {\aap},
         year = 2006,
        month = may,
       volume = {451},
       number = {2},
        pages = {603-611},
          doi = {10.1051/0004-6361:20054600},
archivePrefix = {arXiv},
       eprint = {astro-ph/0511758},
 primaryClass = {astro-ph},
       adsurl = {https://ui.adsabs.harvard.edu/abs/2006A&A...451..603D}
}

@ARTICLE{simondiazherrero,
       author = {{Sim{\'o}n-D{\'\i}az}, S. and {Herrero}, A.},
        title = "{Fourier method of determining the rotational velocities in OB stars}",
      journal = {\aap},
         year = 2007,
        month = jun,
       volume = {468},
       number = {3},
        pages = {1063-1073},
          doi = {10.1051/0004-6361:20066060},
archivePrefix = {arXiv},
       eprint = {astro-ph/0703216},
 primaryClass = {astro-ph},
       adsurl = {https://ui.adsabs.harvard.edu/abs/2007A&A...468.1063S}
}

@ARTICLE{jankov1990,
       author = {{Jankov}, S.},
        title = "{Fourier analysis of rotationally broadened stellar spectra.}",
      journal = {Publications de l'Observatoire Astronomique de Beograd},
         year = 1995,
        month = aug,
       volume = {50},
        pages = {75-80},
       adsurl = {https://ui.adsabs.harvard.edu/abs/1995POBeo..50...75J}
}

@ARTICLE{carroll1928,
       author = {{Carroll}, J.~A.},
        title = "{The form of an absorption line in the spectrum of a rotating or expanding star}",
      journal = {\mnras},
         year = 1928,
        month = may,
       volume = {88},
        pages = {548-555},
          doi = {10.1093/mnras/88.7.548},
       adsurl = {https://ui.adsabs.harvard.edu/abs/1928MNRAS..88..548C}
}

@ARTICLE{carroll1933,
       author = {{Carroll}, J.~A.},
        title = "{The spectroscopic determination of stellar rotation and its effect on line profiles}",
      journal = {\mnras},
         year = 1933,
        month = may,
       volume = {93},
        pages = {478-507},
          doi = {10.1093/mnras/93.7.478},
       adsurl = {https://ui.adsabs.harvard.edu/abs/1933MNRAS..93..478C}
}

@ARTICLE{ryans2002,
       author = {{Ryans}, R.~S.~I. and {Dufton}, P.~L. and {Rolleston}, W.~R.~J. and {Lennon}, D.~J. and {Keenan}, F.~P. and {Smoker}, J.~V. and {Lambert}, D.~L.},
        title = "{Macroturbulent and rotational broadening in the spectra of B-type supergiants}",
      journal = {\mnras},
         year = 2002,
        month = oct,
       volume = {336},
       number = {2},
        pages = {577-586},
          doi = {10.1046/j.1365-8711.2002.05780.x},
       adsurl = {https://ui.adsabs.harvard.edu/abs/2002MNRAS.336..577R}
}

@ARTICLE{simondiaz2014,
       author = {{Sim{\'o}n-D{\'\i}az}, S. and {Herrero}, A.},
        title = "{The IACOB project. I. Rotational velocities in northern Galactic O- and early B-type stars revisited. The impact of other sources of line-broadening}",
      journal = {\aap},
         year = 2014,
        month = feb,
       volume = {562},
          eid = {A135},
        pages = {A135},
          doi = {10.1051/0004-6361/201322758},
archivePrefix = {arXiv},
       eprint = {1311.3360},
 primaryClass = {astro-ph.SR},
       adsurl = {https://ui.adsabs.harvard.edu/abs/2014A&A...562A.135S}
}

@BOOK{gray1976,
       author = {{Gray}, David F.},
        title = "{The observation and analysis of stellar photospheres}",
         year = 1976,
       adsurl = {https://ui.adsabs.harvard.edu/abs/1976oasp.book.....G}
}

@ARTICLE{howarth2011,
       author = {{Howarth}, Ian D.},
        title = "{New limb-darkening coefficients and synthetic photometry for model-atmosphere grids at Galactic, LMC and SMC abundances}",
      journal = {\mnras},
         year = 2011,
        month = may,
       volume = {413},
       number = {3},
        pages = {1515-1523},
          doi = {10.1111/j.1365-2966.2011.18122.x},
archivePrefix = {arXiv},
       eprint = {1011.2631},
 primaryClass = {astro-ph.SR},
       adsurl = {https://ui.adsabs.harvard.edu/abs/2011MNRAS.413.1515H}
}

@ARTICLE{reeve2016,
       author = {{Reeve}, D.~C. and {Howarth}, I.~D.},
        title = "{Limb-darkening coefficients from line-blanketed non-LTE hot-star model atmospheres}",
      journal = {\mnras},
         year = 2016,
        month = feb,
       volume = {456},
       number = {2},
        pages = {1294-1298},
          doi = {10.1093/mnras/stv2631},
archivePrefix = {arXiv},
       eprint = {1511.02029},
 primaryClass = {astro-ph.SR},
       adsurl = {https://ui.adsabs.harvard.edu/abs/2016MNRAS.456.1294R}
}

@ARTICLE{groh2013,
       author = {{Groh}, Jose H. and {Meynet}, Georges and {Georgy}, Cyril and {Ekstr{\"o}m}, Sylvia},
        title = "{Fundamental properties of core-collapse supernova and GRB progenitors: predicting the look of massive stars before death}",
      journal = {\aap},
         year = 2013,
        month = oct,
       volume = {558},
          eid = {A131},
        pages = {A131},
          doi = {10.1051/0004-6361/201321906},
archivePrefix = {arXiv},
       eprint = {1308.4681},
 primaryClass = {astro-ph.SR},
       adsurl = {https://ui.adsabs.harvard.edu/abs/2013A&A...558A.131G}
}

@ARTICLE{maeder2000,
       author = {{Maeder}, Andr{\'e} and {Meynet}, Georges},
        title = "{The Evolution of Rotating Stars}",
      journal = {\araa},
         year = 2000,
        month = jan,
       volume = {38},
        pages = {143-190},
          doi = {10.1146/annurev.astro.38.1.143},
archivePrefix = {arXiv},
       eprint = {astro-ph/0004204},
 primaryClass = {astro-ph},
       adsurl = {https://ui.adsabs.harvard.edu/abs/2000ARA&A..38..143M}
}

@ARTICLE{yoon2005,
       author = {{Yoon}, S. -C. and {Langer}, N.},
        title = "{Evolution of rapidly rotating metal-poor massive stars towards gamma-ray bursts}",
      journal = {\aap},
         year = 2005,
        month = nov,
       volume = {443},
       number = {2},
        pages = {643-648},
          doi = {10.1051/0004-6361:20054030},
archivePrefix = {arXiv},
       eprint = {astro-ph/0508242},
 primaryClass = {astro-ph},
       adsurl = {https://ui.adsabs.harvard.edu/abs/2005A&A...443..643Y}
}

@ARTICLE{jin2024,
       author = {{Jin}, Harim and {Langer}, Norbert and {Lennon}, Daniel J. and {Proffitt}, Charles R.},
        title = "{Boron depletion in Galactic early B-type stars reveals two different main sequence star populations}",
      journal = {\aap},
         year = 2024,
        month = oct,
       volume = {690},
          eid = {A135},
        pages = {A135},
          doi = {10.1051/0004-6361/202450896},
archivePrefix = {arXiv},
       eprint = {2405.18266},
 primaryClass = {astro-ph.SR},
       adsurl = {https://ui.adsabs.harvard.edu/abs/2024A&A...690A.135J}
}

@ARTICLE{bouret2013,
       author = {{Bouret}, J. -C. and {Lanz}, T. and {Martins}, F. and {Marcolino}, W.~L.~F. and {Hillier}, D.~J. and {Depagne}, E. and {Hubeny}, I.},
        title = "{Massive stars at low metallicity. Evolution and surface abundances of O dwarfs in the SMC}",
      journal = {\aap},
         year = 2013,
        month = jul,
       volume = {555},
          eid = {A1},
        pages = {A1},
          doi = {10.1051/0004-6361/201220798},
archivePrefix = {arXiv},
       eprint = {1304.6923},
 primaryClass = {astro-ph.SR},
       adsurl = {https://ui.adsabs.harvard.edu/abs/2013A&A...555A...1B}
}

@ARTICLE{hunter2008a,
       author = {{Hunter}, I. and {Brott}, I. and {Lennon}, D.~J. and {Langer}, N. and {Dufton}, P.~L. and {Trundle}, C. and {Smartt}, S.~J. and {de Koter}, A. and {Evans}, C.~J. and {Ryans}, R.~S.~I.},
        title = "{The VLT FLAMES Survey of Massive Stars: Rotation and Nitrogen Enrichment as the Key to Understanding Massive Star Evolution}",
      journal = {\apjl},
         year = 2008,
        month = mar,
       volume = {676},
       number = {1},
        pages = {L29},
          doi = {10.1086/587436},
archivePrefix = {arXiv},
       eprint = {0711.2267},
 primaryClass = {astro-ph},
       adsurl = {https://ui.adsabs.harvard.edu/abs/2008ApJ...676L..29H}
}

@ARTICLE{hunter2008b,
       author = {{Hunter}, I. and {Lennon}, D.~J. and {Dufton}, P.~L. and {Trundle}, C. and {Sim{\'o}n-D{\'\i}az}, S. and {Smartt}, S.~J. and {Ryans}, R.~S.~I. and {Evans}, C.~J.},
        title = "{The VLT-FLAMES survey of massive stars: atmospheric parameters and rotational velocity distributions for B-type stars in the Magellanic Clouds}",
      journal = {\aap},
         year = 2008,
        month = feb,
       volume = {479},
       number = {2},
        pages = {541-555},
          doi = {10.1051/0004-6361:20078511},
archivePrefix = {arXiv},
       eprint = {0711.2264},
 primaryClass = {astro-ph},
       adsurl = {https://ui.adsabs.harvard.edu/abs/2008A&A...479..541H}
}

@ARTICLE{hunter2009,
       author = {{Hunter}, I. and {Brott}, I. and {Langer}, N. and {Lennon}, D.~J. and {Dufton}, P.~L. and {Howarth}, I.~D. and {Ryans}, R.~S.~I. and {Trundle}, C. and {Evans}, C.~J. and {de Koter}, A. and {Smartt}, S.~J.},
        title = "{The VLT-FLAMES survey of massive stars: constraints on stellar evolution from the chemical compositions of rapidly rotating Galactic and Magellanic Cloud B-type stars}",
      journal = {\aap},
         year = 2009,
        month = mar,
       volume = {496},
       number = {3},
        pages = {841-853},
          doi = {10.1051/0004-6361/200809925},
archivePrefix = {arXiv},
       eprint = {0901.3853},
 primaryClass = {astro-ph.SR},
       adsurl = {https://ui.adsabs.harvard.edu/abs/2009A&A...496..841H}
}

@ARTICLE{wolff2006,
       author = {{Wolff}, S.~C. and {Strom}, S.~E. and {Dror}, D. and {Lanz}, L. and {Venn}, K.},
        title = "{Stellar Rotation: A Clue to the Origin of High-Mass Stars?}",
      journal = {\aj},
         year = 2006,
        month = aug,
       volume = {132},
       number = {2},
        pages = {749-755},
          doi = {10.1086/505534},
archivePrefix = {arXiv},
       eprint = {astro-ph/0604533},
 primaryClass = {astro-ph},
       adsurl = {https://ui.adsabs.harvard.edu/abs/2006AJ....132..749W}
}

@ARTICLE{oscar2013,
       author = {{Ram{\'\i}rez-Agudelo}, O.~H. and {Sim{\'o}n-D{\'\i}az}, S. and {Sana}, H. and {de Koter}, A. and {Sab{\'\i}n-Sanjul{\'\i}an}, C. and {de Mink}, S.~E. and {Dufton}, P.~L. and {Gr{\"a}fener}, G. and {Evans}, C.~J. and {Herrero}, A. and {Langer}, N. and {Lennon}, D.~J. and {Ma{\'\i}z Apell{\'a}niz}, J. and {Markova}, N. and {Najarro}, F. and {Puls}, J. and {Taylor}, W.~D. and {Vink}, J.~S.},
        title = "{The VLT-FLAMES Tarantula Survey. XII. Rotational velocities of the single O-type stars}",
      journal = {\aap},
         year = 2013,
        month = dec,
       volume = {560},
          eid = {A29},
        pages = {A29},
          doi = {10.1051/0004-6361/201321986},
archivePrefix = {arXiv},
       eprint = {1309.2929},
 primaryClass = {astro-ph.SR},
       adsurl = {https://ui.adsabs.harvard.edu/abs/2013A&A...560A..29R}
}

@ARTICLE{demink2013,
       author = {{de Mink}, S.~E. and {Langer}, N. and {Izzard}, R.~G. and {Sana}, H. and {de Koter}, A.},
        title = "{The Rotation Rates of Massive Stars: The Role of Binary Interaction through Tides, Mass Transfer, and Mergers}",
      journal = {\apj},
         year = 2013,
        month = feb,
       volume = {764},
       number = {2},
          eid = {166},
        pages = {166},
          doi = {10.1088/0004-637X/764/2/166},
archivePrefix = {arXiv},
       eprint = {1211.3742},
 primaryClass = {astro-ph.SR},
       adsurl = {https://ui.adsabs.harvard.edu/abs/2013ApJ...764..166D}
}

@ARTICLE{schneider2019,
       author = {{Schneider}, Fabian R.~N. and {Ohlmann}, Sebastian T. and {Podsiadlowski}, Philipp and {R{\"o}pke}, Friedrich K. and {Balbus}, Steven A. and {Pakmor}, R{\"u}diger and {Springel}, Volker},
        title = "{Stellar mergers as the origin of magnetic massive stars}",
      journal = {\nat},
         year = 2019,
        month = oct,
       volume = {574},
       number = {7777},
        pages = {211-214},
          doi = {10.1038/s41586-019-1621-5},
archivePrefix = {arXiv},
       eprint = {1910.14058},
 primaryClass = {astro-ph.SR},
       adsurl = {https://ui.adsabs.harvard.edu/abs/2019Natur.574..211S}
}

@ARTICLE{renzo2019b,
       author = {{Renzo}, M. and {Zapartas}, E. and {de Mink}, S.~E. and {G{\"o}tberg}, Y. and {Justham}, S. and {Farmer}, R.~J. and {Izzard}, R.~G. and {Toonen}, S. and {Sana}, H.},
        title = "{Massive runaway and walkaway stars. A study of the kinematical imprints of the physical processes governing the evolution and explosion of their binary progenitors}",
      journal = {\aap},
         year = 2019,
        month = apr,
       volume = {624},
          eid = {A66},
        pages = {A66},
          doi = {10.1051/0004-6361/201833297},
archivePrefix = {arXiv},
       eprint = {1804.09164},
 primaryClass = {astro-ph.SR},
       adsurl = {https://ui.adsabs.harvard.edu/abs/2019A&A...624A..66R}
}

@ARTICLE{xu2025,
	author = {{Xu}, X.-T. and {Sch{\"u}rmann}, C. and {Langer}, N. and {Wang}, C. and {Schootemeijer}, A. and {Shenar}, T. and {Ercolino}, A. and {Haberl}, F. and {Hastings}, B. and {Jin}, H. and {Kramer}, M. and {Lennon}, D. and {Marchant}, P. and {Sen}, K. and {Tauris}, T.~M. and {de Mink}, S.~E.},
	title = "{Populations of evolved massive binary stars in the Small Magellanic Cloud: I. Predictions from detailed evolution models}",
	journal = {\aap},
	year = 2025,
	month = dec,
	volume = {704},
	eid = {A218},
	pages = {A218},
	doi = {10.1051/0004-6361/202554786},
	archivePrefix = {arXiv},
	eprint = {2503.23876},
	primaryClass = {astro-ph.SR},
	adsurl = {https://ui.adsabs.harvard.edu/abs/2025A&A...704A.218X}
}

@ARTICLE{schuermann2025,
	author = {{Sch{\"u}rmann}, C. and {Xu}, X.-T. and {Langer}, N. and {Lennon}, D. and {Kruckow}, M.~U. and {Antoniadis}, J. and {Haberl}, F. and {Herrero}, A. and {Kramer}, M. and {Schootemeijer}, A. and {Shenar}, T. and {Tauris}, T.~M. and {Wang}, C.},
	title = "{Populations of evolved massive binary stars in the Small Magellanic Cloud: II. Predictions from rapid binary evolution}",
	journal = {\aap},
	year = 2025,
	month = dec,
	volume = {704},
	eid = {A219},
	pages = {A219},
	doi = {10.1051/0004-6361/202554874},
	archivePrefix = {arXiv},
	eprint = {2503.23878},
	primaryClass = {astro-ph.SR},
	adsurl = {https://ui.adsabs.harvard.edu/abs/2025A&A...704A.219S}
}

@INPROCEEDINGS{howarth2004,
       author = {{Howarth}, I.~D.},
        title = "{Rotation and Line Broadening in OBA Stars (Invited Review)}",
    booktitle = {Stellar Rotation},
         year = 2004,
       editor = {{Maeder}, Andre and {Eenens}, Philippe},
       series = {IAU Symposium},
       volume = {215},
        month = jun,
        pages = {33},
       adsurl = {https://ui.adsabs.harvard.edu/abs/2004IAUS..215...33H}
}

@ARTICLE{holgado2022,
       author = {{Holgado}, G. and {Sim{\'o}n-D{\'\i}az}, S. and {Herrero}, A. and {Barb{\'a}}, R.~H.},
        title = "{The IACOB project. VII. The rotational properties of Galactic massive O-type stars revisited}",
      journal = {\aap},
         year = 2022,
        month = sep,
       volume = {665},
          eid = {A150},
        pages = {A150},
          doi = {10.1051/0004-6361/202243851},
archivePrefix = {arXiv},
       eprint = {2207.12776},
 primaryClass = {astro-ph.SR},
       adsurl = {https://ui.adsabs.harvard.edu/abs/2022A&A...665A.150H}
}

@ARTICLE{braganca2012,
       author = {{Bragan{\c{c}}a}, G.~A. and {Daflon}, S. and {Cunha}, K. and {Bensby}, T. and {Oey}, M.~S. and {Walth}, G.},
        title = "{Projected Rotational Velocities and Stellar Characterization of 350 B Stars in the Nearby Galactic Disk}",
      journal = {\aj},
         year = 2012,
        month = nov,
       volume = {144},
       number = {5},
          eid = {130},
        pages = {130},
          doi = {10.1088/0004-6256/144/5/130},
archivePrefix = {arXiv},
       eprint = {1208.1674},
 primaryClass = {astro-ph.SR},
       adsurl = {https://ui.adsabs.harvard.edu/abs/2012AJ....144..130B}
}

@ARTICLE{huang2006,
       author = {{Huang}, W. and {Gies}, D.~R.},
        title = "{Stellar Rotation in Young Clusters. I. Evolution of Projected Rotational Velocity Distributions}",
      journal = {\apj},
         year = 2006,
        month = sep,
       volume = {648},
       number = {1},
        pages = {580-590},
          doi = {10.1086/505782},
archivePrefix = {arXiv},
       eprint = {astro-ph/0510450},
 primaryClass = {astro-ph},
       adsurl = {https://ui.adsabs.harvard.edu/abs/2006ApJ...648..580H}
}

@ARTICLE{berlanas2025,
       author = {{Berlanas}, S.~R. and {Mahy}, L. and {Herrero}, A. and {Ma{\'\i}z Apell{\'a}niz}, J. and {Blomme}, R. and {Comer{\'o}n}, F. and {Negueruela}, I. and {Molina Lera}, J.~A. and {Pantaleoni Gonz{\'a}lez}, M. and {Daflon}, S. and {Santos}, W. and {Kalari}, V.~M.},
        title = "{Gaia-ESO survey: Massive stars in the Carina Nebula: II. The spectroscopic analysis of the O-star population}",
      journal = {\aap},
         year = 2025,
        month = mar,
       volume = {695},
          eid = {A248},
        pages = {A248},
          doi = {10.1051/0004-6361/202453269},
archivePrefix = {arXiv},
       eprint = {2501.16508},
 primaryClass = {astro-ph.SR},
       adsurl = {https://ui.adsabs.harvard.edu/abs/2025A&A...695A.248B}
}

@ARTICLE{fraser2010,
       author = {{Fraser}, M. and {Dufton}, P.~L. and {Hunter}, I. and {Ryans}, R.~S.~I.},
        title = "{Atmospheric parameters and rotational velocities for a sample of Galactic B-type supergiants}",
      journal = {\mnras},
         year = 2010,
        month = may,
       volume = {404},
       number = {3},
        pages = {1306-1320},
          doi = {10.1111/j.1365-2966.2010.16392.x},
archivePrefix = {arXiv},
       eprint = {1001.3337},
 primaryClass = {astro-ph.SR},
       adsurl = {https://ui.adsabs.harvard.edu/abs/2010MNRAS.404.1306F}
}

@ARTICLE{deburgos2023,
       author = {{de Burgos}, A. and {Sim{\'o}n-D{\'\i}az}, S. and {Urbaneja}, M.~A. and {Negueruela}, I.},
        title = "{The IACOB project. IX. Building a modern empirical database of Galactic O9 - B9 supergiants: Sample selection, description, and completeness}",
      journal = {\aap},
         year = 2023,
        month = jun,
       volume = {674},
          eid = {A212},
        pages = {A212},
          doi = {10.1051/0004-6361/202346179},
archivePrefix = {arXiv},
       eprint = {2305.00305},
 primaryClass = {astro-ph.SR},
       adsurl = {https://ui.adsabs.harvard.edu/abs/2023A&A...674A.212D}
}

@ARTICLE{dufton2019,
       author = {{Dufton}, P.~L. and {Evans}, C.~J. and {Hunter}, I. and {Lennon}, D.~J. and {Schneider}, F.~R.~N.},
        title = "{A census of massive stars in NGC 346. Stellar parameters and rotational velocities}",
      journal = {\aap},
         year = 2019,
        month = jun,
       volume = {626},
          eid = {A50},
        pages = {A50},
          doi = {10.1051/0004-6361/201935415},
archivePrefix = {arXiv},
       eprint = {1905.03359},
 primaryClass = {astro-ph.SR},
       adsurl = {https://ui.adsabs.harvard.edu/abs/2019A&A...626A..50D}
}

@ARTICLE{mokiem2006,
       author = {{Mokiem}, M.~R. and {de Koter}, A. and {Evans}, C.~J. and {Puls}, J. and {Smartt}, S.~J. and {Crowther}, P.~A. and {Herrero}, A. and {Langer}, N. and {Lennon}, D.~J. and {Najarro}, F. and {Villamariz}, M.~R. and {Yoon}, S. -C.},
        title = "{The VLT-FLAMES survey of massive stars: mass loss and rotation of early-type stars in the SMC}",
      journal = {\aap},
         year = 2006,
        month = sep,
       volume = {456},
       number = {3},
        pages = {1131-1151},
          doi = {10.1051/0004-6361:20064995},
archivePrefix = {arXiv},
       eprint = {astro-ph/0606403},
 primaryClass = {astro-ph},
       adsurl = {https://ui.adsabs.harvard.edu/abs/2006A&A...456.1131M}
}

@ARTICLE{penny2009,
       author = {{Penny}, Laura R. and {Gies}, Douglas R.},
        title = "{A FUSE Survey of the Rotation Rates of Very Massive Stars in the Small and Large Magellanic Clouds}",
      journal = {\apj},
         year = 2009,
        month = jul,
       volume = {700},
       number = {1},
        pages = {844-858},
          doi = {10.1088/0004-637X/700/1/844},
archivePrefix = {arXiv},
       eprint = {0905.3681},
 primaryClass = {astro-ph.SR},
       adsurl = {https://ui.adsabs.harvard.edu/abs/2009ApJ...700..844P}
}

@ARTICLE{martayan2007,
       author = {{Martayan}, C. and {Fr{\'e}mat}, Y. and {Hubert}, A. -M. and {Floquet}, M. and {Zorec}, J. and {Neiner}, C.},
        title = "{Effects of metallicity, star-formation conditions, and evolution in B and Be stars. II. Small Magellanic Cloud, field of NGC{\,}330}",
      journal = {\aap},
         year = 2007,
        month = feb,
       volume = {462},
       number = {2},
        pages = {683-694},
          doi = {10.1051/0004-6361:20065076},
archivePrefix = {arXiv},
       eprint = {astro-ph/0609677},
 primaryClass = {astro-ph},
       adsurl = {https://ui.adsabs.harvard.edu/abs/2007A&A...462..683M}
}

@ARTICLE{ramachandran2019,
       author = {{Ramachandran}, V. and {Hamann}, W. -R. and {Oskinova}, L.~M. and {Gallagher}, J.~S. and {Hainich}, R. and {Shenar}, T. and {Sander}, A.~A.~C. and {Todt}, H. and {Fulmer}, L.},
        title = "{Testing massive star evolution, star formation history, and feedback at low metallicity. Spectroscopic analysis of OB stars in the SMC Wing}",
      journal = {\aap},
         year = 2019,
        month = may,
       volume = {625},
          eid = {A104},
        pages = {A104},
          doi = {10.1051/0004-6361/201935365},
archivePrefix = {arXiv},
       eprint = {1903.01762},
 primaryClass = {astro-ph.SR},
       adsurl = {https://ui.adsabs.harvard.edu/abs/2019A&A...625A.104R}
}

@ARTICLE{lamb2016,
       author = {{Lamb}, J.~B. and {Oey}, M.~S. and {Segura-Cox}, D.~M. and {Graus}, A.~S. and {Kiminki}, D.~C. and {Golden-Marx}, J.~B. and {Parker}, J. Wm.},
        title = "{The Runaways and Isolated O-Type Star Spectroscopic Survey of the SMC (RIOTS4)}",
      journal = {\apj},
         year = 2016,
        month = feb,
       volume = {817},
       number = {2},
          eid = {113},
        pages = {113},
          doi = {10.3847/0004-637X/817/2/113},
archivePrefix = {arXiv},
       eprint = {1512.01233},
 primaryClass = {astro-ph.GA},
       adsurl = {https://ui.adsabs.harvard.edu/abs/2016ApJ...817..113L}
}

@ARTICLE{shenar2024,
       author = {{Shenar}, T. and {Bodensteiner}, J. and {Sana}, H. and {Crowther}, P.~A. and {Lennon}, D.~J. and {Abdul-Masih}, M. and {Almeida}, L.~A. and {Backs}, F. and {Berlanas}, S.~R. and {Bernini-Peron}, M. and {Bestenlehner}, J.~M. and {Bowman}, D.~M. and {Bronner}, V.~A. and {Britavskiy}, N. and {de Koter}, A. and {de Mink}, S.~E. and {Deshmukh}, K. and {Evans}, C.~J. and {Fabry}, M. and {Gieles}, M. and {Gilkis}, A. and {Gonz{\'a}lez-Tor{\`a}}, G. and {Gr{\"a}fener}, G. and {G{\"o}tberg}, Y. and {Hawcroft}, C. and {H{\'e}nault-Brunet}, V. and {Herrero}, A. and {Holgado}, G. and {Janssens}, S. and {Johnston}, C. and {Josiek}, J. and {Justham}, S. and {Kalari}, V.~M. and {Katabi}, Z.~Z. and {Keszthelyi}, Z. and {Klencki}, J. and {Kub{\'a}t}, J. and {Kub{\'a}tov{\'a}}, B. and {Langer}, N. and {Lefever}, R.~R. and {Ludwig}, B. and {Mackey}, J. and {Mahy}, L. and {Ma{\'\i}z Apell{\'a}niz}, J. and {Mandel}, I. and {Maravelias}, G. and {Marchant}, P. and {Menon}, A. and {Najarro}, F. and {Oskinova}, L.~M. and {O'Grady}, A.~J.~G. and {Ovadia}, R. and {Patrick}, L.~R. and {Pauli}, D. and {Pawlak}, M. and {Ramachandran}, V. and {Renzo}, M. and {Rocha}, D.~F. and {Sander}, A.~A.~C. and {Sayada}, T. and {Schneider}, F.~R.~N. and {Schootemeijer}, A. and {Sch{\"o}sser}, E.~C. and {Sch{\"u}rmann}, C. and {Sen}, K. and {Shahaf}, S. and {Sim{\'o}n-D{\'\i}az}, S. and {Stoop}, M. and {Toonen}, S. and {Tramper}, F. and {van Loon}, J. Th. and {Valli}, R. and {van Son}, L.~A.~C. and {Vigna-G{\'o}mez}, A. and {Villase{\~n}or}, J.~I. and {Vink}, J.~S. and {Wang}, C. and {Willcox}, R.},
        title = "{Binarity at LOw Metallicity (BLOeM): A spectroscopic VLT monitoring survey of massive stars in the SMC}",
      journal = {\aap},
         year = 2024,
        month = oct,
       volume = {690},
          eid = {A289},
        pages = {A289},
          doi = {10.1051/0004-6361/202451586},
archivePrefix = {arXiv},
       eprint = {2407.14593},
 primaryClass = {astro-ph.SR},
       adsurl = {https://ui.adsabs.harvard.edu/abs/2024A&A...690A.289S}
}

@ARTICLE{dorigojones2020,
       author = {{Dorigo Jones}, J. and {Oey}, M.~S. and {Paggeot}, K. and {Castro}, N. and {Moe}, M.},
        title = "{Runaway OB Stars in the Small Magellanic Cloud: Dynamical versus Supernova Ejections}",
      journal = {\apj},
         year = 2020,
        month = nov,
       volume = {903},
       number = {1},
          eid = {43},
        pages = {43},
          doi = {10.3847/1538-4357/abbc6b},
archivePrefix = {arXiv},
       eprint = {2009.03571},
 primaryClass = {astro-ph.SR},
       adsurl = {https://ui.adsabs.harvard.edu/abs/2020ApJ...903...43D}
}

@ARTICLE{britavskiy2025,
       author = {{Britavskiy}, N. and {Mahy}, L. and {Lennon}, D.~J. and {Patrick}, L.~R. and {Sana}, H. and {Villase{\~n}or}, J.~I. and {Shenar}, T. and {Bodensteiner}, J. and {Bernini-Peron}, M. and {Berlanas}, S.~R. and {Bowman}, D.~M. and {Crowther}, P.~A. and {de Mink}, S.~E. and {Evans}, C.~J. and {G{\"o}tberg}, Y. and {Holgado}, G. and {Johnston}, C. and {Keszthelyi}, Z. and {Klencki}, J. and {Langer}, N. and {Mandel}, I. and {Menon}, A. and {Moe}, M. and {Oskinova}, L.~M. and {Pauli}, D. and {Pawlak}, M. and {Ramachandran}, V. and {Renzo}, M. and {Sander}, A.~A.~C. and {Schneider}, F.~R.~N. and {Schootemeijer}, A. and {Sen}, K. and {Sim{\'o}n-D{\'\i}az}, S. and {van Loon}, J. Th. and {Vink}, J.~S.},
        title = "{Binarity at LOw Metallicity (BLOeM): Multiplicity of early B-type supergiants in the Small Magellanic Cloud}",
      journal = {\aap},
         year = 2025,
        month = jun,
       volume = {698},
          eid = {A40},
        pages = {A40},
          doi = {10.1051/0004-6361/202452963},
archivePrefix = {arXiv},
       eprint = {2502.12239},
 primaryClass = {astro-ph.SR},
       adsurl = {https://ui.adsabs.harvard.edu/abs/2025A&A...698A..40B}
}

@ARTICLE{clark2013,
       author = {{Clark}, J.~S. and {Bartlett}, E.~S. and {Coe}, M.~J. and {Dorda}, R. and {Haberl}, F. and {Lamb}, J.~B. and {Negueruela}, I. and {Udalski}, A.},
        title = "{The supergiant B[e] star LHA 115-S 18 - binary and/or luminous blue variable?}",
      journal = {\aap},
         year = 2013,
        month = dec,
       volume = {560},
          eid = {A10},
        pages = {A10},
          doi = {10.1051/0004-6361/201321216},
archivePrefix = {arXiv},
       eprint = {1305.0459},
 primaryClass = {astro-ph.SR},
       adsurl = {https://ui.adsabs.harvard.edu/abs/2013A&A...560A..10C}
}

@INPROCEEDINGS{langer1998,
       author = {{Langer}, N. and {Heger}, A.},
        title = "{B[e] Supergiants: what is Their Evolutionary Status?}",
    booktitle = {B[e] stars},
         year = 1998,
       editor = {{Hubert}, Anne Marie and {Jaschek}, Carlos},
       series = {Astrophysics and Space Science Library},
       volume = {233},
        month = jan,
        pages = {235},
          doi = {10.1007/978-94-015-9014-3_33},
archivePrefix = {arXiv},
       eprint = {astro-ph/9711297},
 primaryClass = {astro-ph},
       adsurl = {https://ui.adsabs.harvard.edu/abs/1998ASSL..233..235L}
}

@INPROCEEDINGS{zickgraf2006,
       author = {{Zickgraf}, F. -J.},
        title = "{B[e] Supergiants in the Magellanic Clouds}",
    booktitle = {Stars with the B[e] Phenomenon},
         year = 2006,
       editor = {{Kraus}, Michaela and {Miroshnichenko}, Anatoly S.},
       series = {Astronomical Society of the Pacific Conference Series},
       volume = {355},
        month = dec,
        pages = {135},
       adsurl = {https://ui.adsabs.harvard.edu/abs/2006ASPC..355..135Z}
}

@INPROCEEDINGS{podsiadlowski2006,
       author = {{Podsiadlowski}, Ph. and {Morris}, T.~S. and {Ivanova}, N.},
        title = "{Massive Binary Mergers: A Unique Scenario for the sgB[e] Phenomenon?}",
    booktitle = {Stars with the B[e] Phenomenon},
         year = 2006,
       editor = {{Kraus}, Michaela and {Miroshnichenko}, Anatoly S.},
       series = {Astronomical Society of the Pacific Conference Series},
       volume = {355},
        month = dec,
        pages = {259},
       adsurl = {https://ui.adsabs.harvard.edu/abs/2006ASPC..355..259P}
}

@ARTICLE{zickgraf1996,
       author = {{Zickgraf}, F. -J. and {Kovacs}, J. and {Wolf}, B. and {Stahl}, O. and {Kaufer}, A. and {Appenzeller}, I.},
        title = "{R4 in the Small Magellanic Cloud: a spectroscopic binary with a B[e]/LBV-type component.}",
      journal = {\aap},
         year = 1996,
        month = may,
       volume = {309},
        pages = {505-514},
       adsurl = {https://ui.adsabs.harvard.edu/abs/1996A&A...309..505Z}
}

@ARTICLE{lennon2024,
       author = {{Lennon}, D.~J. and {Dufton}, P.~L. and {Villase{\~n}or}, J.~I. and {Langer}, N. and {Evans}, C.~J. and {Sana}, H. and {Taylor}, W.~D.},
        title = "{Rotational synchronisation of B-type binaries in 30 Doradus}",
      journal = {\aap},
         year = 2024,
        month = aug,
       volume = {688},
          eid = {A141},
        pages = {A141},
          doi = {10.1051/0004-6361/202450583},
archivePrefix = {arXiv},
       eprint = {2405.17608},
 primaryClass = {astro-ph.SR},
       adsurl = {https://ui.adsabs.harvard.edu/abs/2024A&A...688A.141L}
}

@ARTICLE{walborn2014,
       author = {{Walborn}, N.~R. and {Sana}, H. and {Sim{\'o}n-D{\'\i}az}, S. and {Ma{\'\i}z Apell{\'a}niz}, J. and {Taylor}, W.~D. and {Evans}, C.~J. and {Markova}, N. and {Lennon}, D.~J. and {de Koter}, A.},
        title = "{The VLT-FLAMES Tarantula Survey. XIV. The O-type stellar content of 30 Doradus}",
      journal = {\aap},
         year = 2014,
        month = apr,
       volume = {564},
          eid = {A40},
        pages = {A40},
          doi = {10.1051/0004-6361/201323082},
archivePrefix = {arXiv},
       eprint = {1402.6969},
 primaryClass = {astro-ph.SR},
       adsurl = {https://ui.adsabs.harvard.edu/abs/2014A&A...564A..40W}
}

@ARTICLE{sana2022,
       author = {{Sana}, H. and {Ram{\'\i}rez-Agudelo}, O.~H. and {H{\'e}nault-Brunet}, V. and {Mahy}, L. and {Almeida}, L.~A. and {de Koter}, A. and {Bestenlehner}, J.~M. and {Evans}, C.~J. and {Langer}, N. and {Schneider}, F.~R.~N. and {Crowther}, P.~A. and {de Mink}, S.~E. and {Herrero}, A. and {Lennon}, D.~J. and {Gieles}, M. and {Ma{\'\i}z Apell{\'a}niz}, J. and {Renzo}, M. and {Sabbi}, E. and {van Loon}, J. Th. and {Vink}, J.~S.},
        title = "{The VLT-FLAMES Tarantula Survey. Observational evidence for two distinct populations of massive runaway stars in 30 Doradus}",
      journal = {\aap},
         year = 2022,
        month = dec,
       volume = {668},
          eid = {L5},
        pages = {L5},
          doi = {10.1051/0004-6361/202244677},
archivePrefix = {arXiv},
       eprint = {2211.13476},
 primaryClass = {astro-ph.SR},
       adsurl = {https://ui.adsabs.harvard.edu/abs/2022A&A...668L...5S}
}

@ARTICLE{evans2011,
       author = {{Evans}, C.~J. and {Taylor}, W.~D. and {H{\'e}nault-Brunet}, V. and {Sana}, H. and {de Koter}, A. and {Sim{\'o}n-D{\'\i}az}, S. and {Carraro}, G. and {Bagnoli}, T. and {Bastian}, N. and {Bestenlehner}, J.~M. and {Bonanos}, A.~Z. and {Bressert}, E. and {Brott}, I. and {Campbell}, M.~A. and {Cantiello}, M. and {Clark}, J.~S. and {Costa}, E. and {Crowther}, P.~A. and {de Mink}, S.~E. and {Doran}, E. and {Dufton}, P.~L. and {Dunstall}, P.~R. and {Friedrich}, K. and {Garcia}, M. and {Gieles}, M. and {Gr{\"a}fener}, G. and {Herrero}, A. and {Howarth}, I.~D. and {Izzard}, R.~G. and {Langer}, N. and {Lennon}, D.~J. and {Ma{\'\i}z Apell{\'a}niz}, J. and {Markova}, N. and {Najarro}, F. and {Puls}, J. and {Ramirez}, O.~H. and {Sab{\'\i}n-Sanjuli{\'a}n}, C. and {Smartt}, S.~J. and {Stroud}, V.~E. and {van Loon}, J. Th. and {Vink}, J.~S. and {Walborn}, N.~R.},
        title = "{The VLT-FLAMES Tarantula Survey. I. Introduction and observational overview}",
      journal = {\aap},
         year = 2011,
        month = jun,
       volume = {530},
          eid = {A108},
        pages = {A108},
          doi = {10.1051/0004-6361/201116782},
archivePrefix = {arXiv},
       eprint = {1103.5386},
 primaryClass = {astro-ph.SR},
       adsurl = {https://ui.adsabs.harvard.edu/abs/2011A&A...530A.108E}
}

@ARTICLE{bestenlehner2025,
       author = {{Bestenlehner}, J.~M. and {Crowther}, Paul A. and {Bronner}, V.~A. and {Sim{\'o}n-D{\'\i}az}, S. and {Lennon}, D.~J. and {Bodensteiner}, J. and {Langer}, N. and {Marchant}, P. and {Sana}, H. and {Schneider}, F.~R.~N. and {Shenar}, T.},
        title = "{Binarity at LOw Metallicity (BLOeM): pipeline-determined physical properties of OB stars}",
      journal = {\mnras},
         year = 2025,
        month = jul,
       volume = {540},
       number = {4},
        pages = {3523-3548},
          doi = {10.1093/mnras/staf900},
archivePrefix = {arXiv},
       eprint = {2506.00117},
 primaryClass = {astro-ph.SR},
       adsurl = {https://ui.adsabs.harvard.edu/abs/2025MNRAS.540.3523B}
}

@ARTICLE{patrick2025,
       author = {{Patrick}, L.~R. and {Lennon}, D.~J. and {Najarro}, F. and {Shenar}, T. and {Bodensteiner}, J. and {Sana}, H. and {Crowther}, P.~A. and {Britavskiy}, N. and {Langer}, N. and {Schootemeijer}, A. and {Evans}, C.~J. and {Mahy}, L. and {G{\"o}tberg}, Y. and {de Mink}, S.~E. and {Schneider}, F.~R.~N. and {O'Grady}, A.~J.~G. and {Villase{\~n}or}, J.~I. and {Bernini-Peron}, M. and {Bowman}, D.~M. and {de Koter}, A. and {Deshmukh}, K. and {Gilkis}, A. and {Gonz{\'a}lez-Tor{\`a}}, G. and {Kalari}, V.~M. and {Keszthelyi}, Z. and {Mandel}, I. and {Menon}, A. and {Moe}, M. and {Oskinova}, L.~M. and {Pauli}, D. and {Renzo}, M. and {Sander}, A.~A.~C. and {Sen}, K. and {Stoop}, M. and {van Loon}, J. Th. and {Toonen}, S. and {Tramper}, F. and {Vink}, J.~S. and {Wang}, C.},
        title = "{Binarity at LOw Metallicity (BLOeM): The multiplicity properties and evolution of BAF-type supergiants}",
      journal = {\aap},
         year = 2025,
        month = jun,
       volume = {698},
          eid = {A39},
        pages = {A39},
          doi = {10.1051/0004-6361/202452949},
archivePrefix = {arXiv},
       eprint = {2502.02644},
 primaryClass = {astro-ph.SR},
       adsurl = {https://ui.adsabs.harvard.edu/abs/2025A&A...698A..39P}
}

@ARTICLE{evans2008,
       author = {{Evans}, Christopher J. and {Howarth}, Ian D.},
        title = "{Kinematics of massive stars in the Small Magellanic Cloud}",
      journal = {\mnras},
         year = 2008,
        month = may,
       volume = {386},
       number = {2},
        pages = {826-834},
          doi = {10.1111/j.1365-2966.2008.13012.x},
archivePrefix = {arXiv},
       eprint = {0801.3460},
 primaryClass = {astro-ph},
       adsurl = {https://ui.adsabs.harvard.edu/abs/2008MNRAS.386..826E}
}

@ARTICLE{frost2024,
       author = {{Frost}, A.~J. and {Sana}, H. and {Mahy}, L. and {Wade}, G. and {Barron}, J. and {Le Bouquin}, J. -B. and {M{\'e}rand}, A. and {Schneider}, F.~R.~N. and {Shenar}, T. and {Barb{\'a}}, R.~H. and {Bowman}, D.~M. and {Fabry}, M. and {Farhang}, A. and {Marchant}, P. and {Morrell}, N.~I. and {Smoker}, J.~V.},
        title = "{A magnetic massive star has experienced a stellar merger}",
      journal = {Science},
         year = 2024,
        month = apr,
       volume = {384},
       number = {6692},
        pages = {214-217},
          doi = {10.1126/science.adg7700},
archivePrefix = {arXiv},
       eprint = {2404.10167},
 primaryClass = {astro-ph.SR},
       adsurl = {https://ui.adsabs.harvard.edu/abs/2024Sci...384..214F}
}

@ARTICLE{ud-Doula2009,
       author = {{Ud-Doula}, Asif and {Owocki}, Stanley P. and {Townsend}, Richard H.~D.},
        title = "{Dynamical simulations of magnetically channelled line-driven stellar winds - III. Angular momentum loss and rotational spin-down}",
      journal = {\mnras},
         year = 2009,
        month = jan,
       volume = {392},
       number = {3},
        pages = {1022-1033},
          doi = {10.1111/j.1365-2966.2008.14134.x},
archivePrefix = {arXiv},
       eprint = {0810.4247},
 primaryClass = {astro-ph},
       adsurl = {https://ui.adsabs.harvard.edu/abs/2009MNRAS.392.1022U}
}

@ARTICLE{fossati2015,
       author = {{Fossati}, L. and {Castro}, N. and {Sch{\"o}ller}, M. and {Hubrig}, S. and {Langer}, N. and {Morel}, T. and {Briquet}, M. and {Herrero}, A. and {Przybilla}, N. and {Sana}, H. and {Schneider}, F.~R.~N. and {de Koter}, A. and {BOB Collaboration}},
        title = "{B fields in OB stars (BOB): Low-resolution FORS2 spectropolarimetry of the first sample of 50 massive stars}",
      journal = {\aap},
         year = 2015,
        month = oct,
       volume = {582},
          eid = {A45},
        pages = {A45},
          doi = {10.1051/0004-6361/201526725},
archivePrefix = {arXiv},
       eprint = {1508.00750},
 primaryClass = {astro-ph.SR},
       adsurl = {https://ui.adsabs.harvard.edu/abs/2015A&A...582A..45F}
}

@ARTICLE{petit2019,
       author = {{Petit}, V. and {Wade}, G.~A. and {Schneider}, F.~R.~N. and {Fossati}, L. and {Kamp}, K. and {Neiner}, C. and {David-Uraz}, A. and {Alecian}, E. and {MiMeS Collaboration}},
        title = "{The MiMeS survey of magnetism in massive stars: magnetic properties of the O-type star population}",
      journal = {\mnras},
         year = 2019,
        month = nov,
       volume = {489},
       number = {4},
        pages = {5669-5687},
          doi = {10.1093/mnras/stz2469},
archivePrefix = {arXiv},
       eprint = {1909.00877},
 primaryClass = {astro-ph.SR},
       adsurl = {https://ui.adsabs.harvard.edu/abs/2019MNRAS.489.5669P}
}

@ARTICLE{vink2010,
       author = {{Vink}, Jorick S. and {Brott}, I. and {Gr{\"a}fener}, G. and {Langer}, N. and {de Koter}, A. and {Lennon}, D.~J.},
        title = "{The nature of B supergiants: clues from a steep drop in rotation rates at 22 000 K. The possibility of Bi-stability braking}",
      journal = {\aap},
         year = 2010,
        month = mar,
       volume = {512},
          eid = {L7},
        pages = {L7},
          doi = {10.1051/0004-6361/201014205},
archivePrefix = {arXiv},
       eprint = {1003.1280},
 primaryClass = {astro-ph.SR},
       adsurl = {https://ui.adsabs.harvard.edu/abs/2010A&A...512L...7V}
}

@ARTICLE{trundle2004,
       author = {{Trundle}, C. and {Lennon}, D.~J. and {Puls}, J. and {Dufton}, P.~L.},
        title = "{Understanding B-type supergiants in the low metallicity environment of the SMC}",
      journal = {\aap},
         year = 2004,
        month = apr,
       volume = {417},
        pages = {217-234},
          doi = {10.1051/0004-6361:20034325},
archivePrefix = {arXiv},
       eprint = {astro-ph/0312233},
 primaryClass = {astro-ph},
       adsurl = {https://ui.adsabs.harvard.edu/abs/2004A&A...417..217T}
}

@ARTICLE{trundle2005,
       author = {{Trundle}, C. and {Lennon}, D.~J.},
        title = "{Understanding B-type supergiants in the low metallicity environment of the SMC II}",
      journal = {\aap},
         year = 2005,
        month = may,
       volume = {434},
       number = {2},
        pages = {677-689},
          doi = {10.1051/0004-6361:20042061},
archivePrefix = {arXiv},
       eprint = {astro-ph/0501228},
 primaryClass = {astro-ph},
       adsurl = {https://ui.adsabs.harvard.edu/abs/2005A&A...434..677T}
}

@ARTICLE{verhamme2024,
       author = {{Verhamme}, O. and {Sundqvist}, J. and {de Koter}, A. and {Sana}, H. and {Backs}, F. and {Brands}, S.~A. and {Najarro}, F. and {Puls}, J. and {Vink}, J.~S. and {Crowther}, P.~A. and {Kub{\'a}tov{\'a}}, B. and {Sander}, A.~A.~C. and {Bernini-Peron}, M. and {Kuiper}, R. and {Prinja}, R.~K. and {Schillemans}, P. and {Shenar}, T. and {van Loon}, J. Th. and {XShootu collaboration}},
        title = "{X-Shooting ULLYSES: Massive Stars at low metallicity: IX. Empirical constraints on mass-loss rates and clumping parameters for OB supergiants in the Large Magellanic Cloud}",
      journal = {\aap},
         year = 2024,
        month = dec,
       volume = {692},
          eid = {A91},
        pages = {A91},
          doi = {10.1051/0004-6361/202451169},
archivePrefix = {arXiv},
       eprint = {2410.14937},
 primaryClass = {astro-ph.SR},
       adsurl = {https://ui.adsabs.harvard.edu/abs/2024A&A...692A..91V}
}

@ARTICLE{deburgos2024,
       author = {{de Burgos}, A. and {Keszthelyi}, Z. and {Sim{\'o}n-D{\'\i}az}, S. and {Urbaneja}, M.~A.},
        title = "{The IACOB project. XI. No increase in mass-loss rates over the bistability region}",
      journal = {\aap},
         year = 2024,
        month = jul,
       volume = {687},
          eid = {L16},
        pages = {L16},
          doi = {10.1051/0004-6361/202450301},
archivePrefix = {arXiv},
       eprint = {2405.09868},
 primaryClass = {astro-ph.SR},
       adsurl = {https://ui.adsabs.harvard.edu/abs/2024A&A...687L..16D}
}

@ARTICLE{deburgos2025,
       author = {{de Burgos}, A. and {Sim{\'o}n-D{\'\i}az}, S. and {Urbaneja}, M.~A. and {Holgado}, G. and {Ekstr{\"o}m}, S. and {Ram{\'\i}rez-Tannus}, M.~C. and {Zari}, E.},
        title = "{The IACOB project: XIV. New clues on the location of the TAMS in the massive star domain}",
      journal = {\aap},
         year = 2025,
        month = mar,
       volume = {695},
          eid = {A87},
        pages = {A87},
          doi = {10.1051/0004-6361/202453242},
archivePrefix = {arXiv},
       eprint = {2501.17984},
 primaryClass = {astro-ph.SR},
       adsurl = {https://ui.adsabs.harvard.edu/abs/2025A&A...695A..87D}
}

@ARTICLE{vink2025,
       author = {{Vink}, Jorick S. and {Oudmaijer}, Rene D.},
        title = "{The Blue Supergiant Problem and the Main-Sequence Width}",
      journal = {Galaxies},
         year = 2025,
        month = mar,
       volume = {13},
       number = {2},
          eid = {19},
        pages = {19},
          doi = {10.3390/galaxies13020019},
archivePrefix = {arXiv},
       eprint = {2502.04461},
 primaryClass = {astro-ph.SR},
       adsurl = {https://ui.adsabs.harvard.edu/abs/2025Galax..13...19V}
}

@ARTICLE{crowther2006,
       author = {{Crowther}, P.~A. and {Lennon}, D.~J. and {Walborn}, N.~R.},
        title = "{Physical parameters and wind properties of galactic early B supergiants}",
      journal = {\aap},
         year = 2006,
        month = jan,
       volume = {446},
       number = {1},
        pages = {279-293},
          doi = {10.1051/0004-6361:20053685},
archivePrefix = {arXiv},
       eprint = {astro-ph/0509436},
 primaryClass = {astro-ph},
       adsurl = {https://ui.adsabs.harvard.edu/abs/2006A&A...446..279C}
}

@ARTICLE{villasenor2025,
       author = {{Villase{\~n}or}, J.~I. and {Sana}, H. and {Mahy}, L. and {Shenar}, T. and {Bodensteiner}, J. and {Britavskiy}, N. and {Lennon}, D.~J. and {Moe}, M. and {Patrick}, L.~R. and {Pawlak}, M. and {Bowman}, D.~M. and {Crowther}, P.~A. and {de Mink}, S.~E. and {Deshmukh}, K. and {Evans}, C.~J. and {Fabry}, M. and {Fouesneau}, M. and {Holgado}, G. and {Langer}, N. and {Ma{\'\i}z Apell{\'a}niz}, J. and {Mandel}, I. and {Oskinova}, L.~M. and {Pauli}, D. and {Ramachandran}, V. and {Renzo}, M. and {Rix}, H. -W. and {Rocha}, D.~F. and {Sander}, A.~A.~C. and {Schneider}, F.~R.~N. and {Sen}, K. and {Sim{\'o}n-D{\'\i}az}, S. and {van Loon}, J. Th. and {Toonen}, S. and {Vink}, J.~S.},
        title = "{Binarity at LOw Metallicity (BLOeM): Enhanced multiplicity of early B-type dwarfs and giants at Z = 0.2 Z$_{{\ensuremath{\odot}}}$}",
      journal = {\aap},
         year = 2025,
        month = jun,
       volume = {698},
          eid = {A41},
        pages = {A41},
          doi = {10.1051/0004-6361/202453166},
archivePrefix = {arXiv},
       eprint = {2503.21936},
 primaryClass = {astro-ph.SR},
       adsurl = {https://ui.adsabs.harvard.edu/abs/2025A&A...698A..41V}
}

@ARTICLE{schootemeijer2019,
       author = {{Schootemeijer}, A. and {Langer}, N. and {Grin}, N.~J. and {Wang}, C.},
        title = "{Constraining mixing in massive stars in the Small Magellanic Cloud}",
      journal = {\aap},
         year = 2019,
        month = may,
       volume = {625},
          eid = {A132},
        pages = {A132},
          doi = {10.1051/0004-6361/201935046},
archivePrefix = {arXiv},
       eprint = {1903.10423},
 primaryClass = {astro-ph.SR},
       adsurl = {https://ui.adsabs.harvard.edu/abs/2019A&A...625A.132S}
}

@ARTICLE{bodensteiner2025,
       author = {{Bodensteiner}, J. and {Shenar}, T. and {Sana}, H. and {Britavskiy}, N. and {Crowther}, P.~A. and {Langer}, N. and {Lennon}, D.~J. and {Mahy}, L. and {Patrick}, L.~R. and {Villase{\~n}or}, J.~I. and {Abdul-Masih}, M. and {Bowman}, D.~M. and {de Koter}, A. and {de Mink}, S.~E. and {Deshmukh}, K. and {Fabry}, M. and {Gilkis}, A. and {G{\"o}tberg}, Y. and {Holgado}, G. and {Izzard}, R.~G. and {Janssens}, S. and {Kalari}, V.~M. and {Keszthelyi}, Z. and {Kub{\'a}t}, J. and {Mandel}, I. and {Maravelias}, G. and {Oskinova}, L.~M. and {Pauli}, D. and {Ramachandran}, V. and {Rocha}, D.~F. and {Renzo}, M. and {Sander}, A.~A.~C. and {Schneider}, F.~R.~N. and {Schootemeijer}, A. and {Sen}, K. and {Stoop}, M. and {Toonen}, S. and {van Loon}, J. Th. and {Valli}, R. and {Vigna-G{\'o}mez}, A. and {Vink}, J.~S. and {Wang}, C. and {Xu}, X. -T.},
        title = "{Binarity at LOw Metallicity (BLOeM): Multiplicity properties of Oe and Be stars}",
      journal = {\aap},
         year = 2025,
        month = jun,
       volume = {698},
          eid = {A38},
        pages = {A38},
          doi = {10.1051/0004-6361/202452623},
archivePrefix = {arXiv},
       eprint = {2502.02641},
 primaryClass = {astro-ph.SR},
       adsurl = {https://ui.adsabs.harvard.edu/abs/2025A&A...698A..38B}
}

@INPROCEEDINGS{lennon2010,
       author = {{Lennon}, D.~J. and {Trundle}, C. and {Hunter}, I. and {Smartt}, S. and {Dufton}, P. and {Evans}, C. and {Langer}, N. and {Brott}, I.},
        title = "{B-Type Supergiants in the Magellanic Clouds}",
    booktitle = {Hot and Cool: Bridging Gaps in Massive Star Evolution},
         year = 2010,
       editor = {{Leitherer}, C. and {Bennett}, P.~D. and {Morris}, P.~W. and {Van Loon}, J. Th.},
       series = {Astronomical Society of the Pacific Conference Series},
       volume = {425},
        month = jun,
        pages = {23},
       adsurl = {https://ui.adsabs.harvard.edu/abs/2010ASPC..425...23L}
}

@ARTICLE{verdugo1999,
       author = {{Verdugo}, E. and {Talavera}, A. and {G{\'o}mez de Castro}, A.~I.},
        title = "{Understanding A-type supergiants. II. Atmospheric parameters and rotational velocities of Galactic A-type supergiants}",
      journal = {\aap},
         year = 1999,
        month = jun,
       volume = {346},
        pages = {819-830},
       adsurl = {https://ui.adsabs.harvard.edu/abs/1999A&A...346..819V}
}

@ARTICLE{royer2002,
       author = {{Royer}, F. and {Gerbaldi}, M. and {Faraggiana}, R. and {G{\'o}mez}, A.~E.},
        title = "{Rotational velocities of A-type stars. I. Measurement of v sin i in the southern hemisphere}",
      journal = {\aap},
         year = 2002,
        month = jan,
       volume = {381},
        pages = {105-121},
          doi = {10.1051/0004-6361:20011422},
archivePrefix = {arXiv},
       eprint = {astro-ph/0110490},
 primaryClass = {astro-ph},
       adsurl = {https://ui.adsabs.harvard.edu/abs/2002A&A...381..105R}
}

@ARTICLE{pawlak2016,
       author = {{Pawlak}, M. and {Soszy{\'n}ski}, I. and {Udalski}, A. and {Szyma{\'n}ski}, M.~K. and {Wyrzykowski}, {\L}. and {Ulaczyk}, K. and {Poleski}, R. and {Pietrukowicz}, P. and {Koz{\l}owski}, S. and {Skowron}, D.~M. and {Skowron}, J. and {Mr{\'o}z}, P. and {Hamanowicz}, A.},
        title = "{The OGLE Collection of Variable Stars. Eclipsing Binaries in the Magellanic System}",
      journal = {\actaa},
         year = 2016,
        month = dec,
       volume = {66},
       number = {4},
        pages = {421-432},
          doi = {10.48550/arXiv.1612.06394},
archivePrefix = {arXiv},
       eprint = {1612.06394},
 primaryClass = {astro-ph.SR},
       adsurl = {https://ui.adsabs.harvard.edu/abs/2016AcA....66..421P}
}

@ARTICLE{walborn2000,
       author = {{Walborn}, Nolan R. and {Lennon}, Daniel J. and {Heap}, Sara R. and {Lindler}, Don J. and {Smith}, Linda J. and {Evans}, Christopher J. and {Parker}, Joel Wm.},
        title = "{The Ultraviolet and Optical Spectra of Metal-deficient O Stars in the Small Magellanic Cloud}",
      journal = {\pasp},
         year = 2000,
        month = sep,
       volume = {112},
       number = {775},
        pages = {1243-1261},
          doi = {10.1086/316617},
       adsurl = {https://ui.adsabs.harvard.edu/abs/2000PASP..112.1243W}
}

@ARTICLE{evans2004,
       author = {{Evans}, Christopher J. and {Howarth}, Ian D. and {Irwin}, Michael J. and {Burnley}, Adam W. and {Harries}, Timothy J.},
        title = "{A 2dF survey of the Small Magellanic Cloud}",
      journal = {\mnras},
         year = 2004,
        month = sep,
       volume = {353},
       number = {2},
        pages = {601-623},
          doi = {10.1111/j.1365-2966.2004.08096.x},
archivePrefix = {arXiv},
       eprint = {astro-ph/0406409},
 primaryClass = {astro-ph},
       adsurl = {https://ui.adsabs.harvard.edu/abs/2004MNRAS.353..601E}
}

@ARTICLE{howarth2007,
       author = {{Howarth}, Ian D. and {Walborn}, Nolan R. and {Lennon}, Danny J. and {Puls}, Joachim and {Naz{\'e}}, Ya{\"e}l and {Annuk}, K. and {Antokhin}, I. and {Bohlender}, D. and {Bond}, H. and {Donati}, J. -F. and {Georgiev}, L. and {Gies}, D. and {Harmer}, D. and {Herrero}, A. and {Kolka}, I. and {McDavid}, D. and {Morel}, T. and {Negueruela}, I. and {Rauw}, G. and {Reig}, P.},
        title = "{Towards an understanding of the Of?p star HD 191612: optical spectroscopy}",
      journal = {\mnras},
         year = 2007,
        month = sep,
       volume = {381},
       number = {2},
        pages = {433-446},
          doi = {10.1111/j.1365-2966.2007.12178.x},
archivePrefix = {arXiv},
       eprint = {0707.0594},
 primaryClass = {astro-ph},
       adsurl = {https://ui.adsabs.harvard.edu/abs/2007MNRAS.381..433H}
}

@ARTICLE{dufton2020,
       author = {{Dufton}, P.~L. and {Evans}, C.~J. and {Lennon}, D.~J. and {Hunter}, I.},
        title = "{The NGC 346 massive star census. Nitrogen abundances for apparently single, narrow lined, hydrogen core burning B-type stars}",
      journal = {\aap},
         year = 2020,
        month = feb,
       volume = {634},
          eid = {A6},
        pages = {A6},
          doi = {10.1051/0004-6361/201936921},
archivePrefix = {arXiv},
       eprint = {1912.07539},
 primaryClass = {astro-ph.SR},
       adsurl = {https://ui.adsabs.harvard.edu/abs/2020A&A...634A...6D}
}

@ARTICLE{dufton2005,
       author = {{Dufton}, P.~L. and {Ryans}, R.~S.~I. and {Trundle}, C. and {Lennon}, D.~J. and {Hubeny}, I. and {Lanz}, T. and {Allende Prieto}, C.},
        title = "{B-type supergiants in the SMC: Chemical compositions and comparison of static and unified models}",
      journal = {\aap},
         year = 2005,
        month = may,
       volume = {434},
       number = {3},
        pages = {1125-1137},
          doi = {10.1051/0004-6361:20042530},
archivePrefix = {arXiv},
       eprint = {astro-ph/0412367},
 primaryClass = {astro-ph},
       adsurl = {https://ui.adsabs.harvard.edu/abs/2005A&A...434.1125D}
}

@ARTICLE{lanz2007,
   author = {{Lanz}, T. and {Hubeny}, I.},
    title = "{A Grid of NLTE Line-blanketed Model Atmospheres of Early B-Type Stars}",
  journal = {\apjs},
   eprint = {arXiv:astro-ph/0611891},
     year = 2007,
    month = mar,
   volume = 169,
    pages = {83-104},
      doi = {10.1086/511270},
   adsurl = {http://adsabs.harvard.edu/abs/2007ApJS..169...83L}
}

@ARTICLE{hubeny1988,
	Adsurl = {http://adsabs.harvard.edu/abs/1988CoPhC..52..103H},
	Author = {{Hubeny}, I.},
	Doi = {10.1016/0010-4655(88)90177-4},
	Journal = {Computer Physics Communications},
	Month = dec,
	Pages = {103-132},
	Title = {{A computer program for calculating non-LTE model stellar atmospheres}},
	Volume = 52,
	Year = 1988
}

@ARTICLE{sana2025,
       author = {{Sana}, H. and {Shenar}, T. and {Bodensteiner}, J. and {Britavskiy}, N. and {Langer}, N. and {Lennon}, D.~J. and {Mahy}, L. and {Mandel}, I. and {de Mink}, S.~E. and {Patrick}, L.~R. and {Villase{\~n}or}, J.~I. and {Dirickx}, M. and {Abdul-Masih}, M. and {Almeida}, L.~A. and {Backs}, F. and {Berlanas}, S.~R. and {Bernini-Peron}, M. and {Bowman}, D.~M. and {Bronner}, V.~A. and {Crowther}, P.~A. and {Deshmukh}, K. and {Evans}, C.~J. and {Fabry}, M. and {Gieles}, M. and {Gilkis}, A. and {Gonz{\'a}lez-Tor{\`a}}, G. and {Gr{\"a}fener}, G. and {G{\"o}tberg}, Y. and {Hawcroft}, C. and {H{\'e}nault-Brunet}, V. and {Herrero}, A. and {Holgado}, G. and {Izzard}, R.~G. and {de Koter}, A. and {Janssens}, S. and {Johnston}, C. and {Josiek}, J. and {Justham}, S. and {Kalari}, V.~M. and {Klencki}, J. and {Kub{\'a}t}, J. and {Kub{\'a}tov{\'a}}, B. and {Lefever}, R.~R. and {van Loon}, J. Th. and {Ludwig}, B. and {Mackey}, J. and {Ma{\'\i}z Apell{\'a}niz}, J. and {Maravelias}, G. and {Marchant}, P. and {Mazeh}, T. and {Menon}, A. and {Moe}, M. and {Najarro}, F. and {Oskinova}, L.~M. and {Ovadia}, R. and {Pauli}, D. and {Pawlak}, M. and {Ramachandran}, V. and {Renzo}, M. and {Rocha}, D.~F. and {Sander}, A.~A.~C. and {Schneider}, F.~R.~N. and {Schootemeijer}, A. and {Sch{\"o}sser}, E.~C. and {Sch{\"u}rmann}, C. and {Sen}, K. and {Shahaf}, S. and {Sim{\'o}n-D{\'\i}az}, S. and {van Son}, L.~A.~C. and {Stoop}, M. and {Toonen}, S. and {Tramper}, F. and {Valli}, R. and {Vigna-G{\'o}mez}, A. and {Vink}, J.~S. and {Wang}, C. and {Willcox}, R.},
        title = "{A high fraction of close massive binary stars at low metallicity}",
      journal = {Nature Astronomy},
         year = 2025,
        month = sep,
       volume = {9},
        pages = {1337-1346},
          doi = {10.1038/s41550-025-02610-x},
archivePrefix = {arXiv},
       eprint = {2509.12488},
 primaryClass = {astro-ph.SR},
       adsurl = {https://ui.adsabs.harvard.edu/abs/2025NatAs...9.1337S}
}

@ARTICLE{lennon2003,
       author = {{Lennon}, D.~J. and {Dufton}, P.~L. and {Crowley}, C.},
        title = "{More nitrogen rich B-type stars in the SMC cluster, NGC 330}",
      journal = {\aap},
         year = 2003,
        month = feb,
       volume = {398},
        pages = {455-466},
          doi = {10.1051/0004-6361:20021194},
archivePrefix = {arXiv},
       eprint = {astro-ph/0207031},
 primaryClass = {astro-ph},
       adsurl = {https://ui.adsabs.harvard.edu/abs/2003A&A...398..455L}
}

@ARTICLE{gies1992,
       author = {{Gies}, Douglas R. and {Lambert}, David L.},
        title = "{Carbon, Nitrogen, and Oxygen Abundances in Early B-Type Stars}",
      journal = {\apj},
         year = 1992,
        month = mar,
       volume = {387},
        pages = {673},
          doi = {10.1086/171116},
       adsurl = {https://ui.adsabs.harvard.edu/abs/1992ApJ...387..673G}
}

@ARTICLE{morel2008,
       author = {{Morel}, T. and {Hubrig}, S. and {Briquet}, M.},
        title = "{Nitrogen enrichment, boron depletion and magnetic fields in slowly-rotating B-type dwarfs}",
      journal = {\aap},
         year = 2008,
        month = apr,
       volume = {481},
       number = {2},
        pages = {453-463},
          doi = {10.1051/0004-6361:20078999},
archivePrefix = {arXiv},
       eprint = {0801.4491},
 primaryClass = {astro-ph},
       adsurl = {https://ui.adsabs.harvard.edu/abs/2008A&A...481..453M}
}

@ARTICLE{wang2022,
       author = {{Wang}, Chen and {Langer}, Norbert and {Schootemeijer}, Abel and {Milone}, Antonino and {Hastings}, Ben and {Xu}, Xiao-Tian and {Bodensteiner}, Julia and {Sana}, Hugues and {Castro}, Norberto and {Lennon}, D.~J. and {Marchant}, Pablo and {de Koter}, A. and {de Mink}, Selma E.},
        title = "{Stellar mergers as the origin of the blue main-sequence band in young star clusters}",
      journal = {Nature Astronomy},
         year = 2022,
        month = feb,
       volume = {6},
        pages = {480-487},
          doi = {10.1038/s41550-021-01597-5},
archivePrefix = {arXiv},
       eprint = {2202.05552},
 primaryClass = {astro-ph.SR},
       adsurl = {https://ui.adsabs.harvard.edu/abs/2022NatAs...6..480W}
}

@ARTICLE{sana2024,
       author = {{Sana}, H. and {Tramper}, F. and {Abdul-Masih}, M. and {Blomme}, R. and {Dsilva}, K. and {Maravelias}, G. and {Martins}, L. and {Mehner}, A. and {Wofford}, A. and {Banyard}, G. and {Barbosa}, C.~L. and {Bestenlehner}, J. and {Hawcroft}, C. and {John Hillier}, D. and {Todt}, H. and {Larkin}, C.~J.~K. and {Mahy}, L. and {Najarro}, F. and {Ramachandran}, V. and {Ram{\'\i}rez-Tannus}, M.~C. and {Rubio-D{\'\i}ez}, M.~M. and {Sander}, A.~A.~C. and {Shenar}, T. and {Vink}, J.~S. and {Backs}, F. and {Brands}, S.~A. and {Crowther}, P. and {Decin}, L. and {de Koter}, A. and {Hamann}, W.-R. and {Kehrig}, C. and {Kuiper}, R. and {Oskinova}, L. and {Pauli}, D. and {Sundqvist}, J. and {Verhamme}, O. and {Xshoot-U Collaboration}},
        title = "{X-Shooting ULLYSES: Massive stars at low metallicity. II. DR1: Advanced optical data products for the Magellanic Clouds}",
      journal = {\aap},
         year = 2024,
        month = aug,
       volume = {688},
          eid = {A104},
        pages = {A104},
          doi = {10.1051/0004-6361/202347479},
archivePrefix = {arXiv},
       eprint = {2402.16987},
 primaryClass = {astro-ph.SR},
       adsurl = {https://ui.adsabs.harvard.edu/abs/2024A&A...688A.104S}
}

@ARTICLE{bernini2024,
       author = {{Bernini-Peron}, M. and {Sander}, A.~A.~C. and {Ramachandran}, V. and {Oskinova}, L.~M. and {Vink}, J.~S. and {Verhamme}, O. and {Najarro}, F. and {Josiek}, J. and {Brands}, S.~A. and {Crowther}, P.~A. and {G{\'o}mez-Gonz{\'a}lez}, V.~M.~A. and {Gormaz-Matamala}, A.~C. and {Hawcroft}, C. and {Kuiper}, R. and {Mahy}, L. and {Marcolino}, W.~L.~F. and {Martins}, L.~P. and {Mehner}, A. and {Parsons}, T.~N. and {Pauli}, D. and {Shenar}, T. and {Schootemeijer}, A. and {Todt}, H. and {van Loon}, J. Th. and {XShootU Collaboration}},
        title = "{X-Shooting ULLYSES: Massive stars at low metallicity: VII. Stellar and wind properties of B supergiants in the Small Magellanic Cloud}",
      journal = {\aap},
         year = 2024,
        month = dec,
       volume = {692},
          eid = {A89},
        pages = {A89},
          doi = {10.1051/0004-6361/202450475},
archivePrefix = {arXiv},
       eprint = {2407.14216},
 primaryClass = {astro-ph.SR},
       adsurl = {https://ui.adsabs.harvard.edu/abs/2024A&A...692A..89B}
}

@ARTICLE{mahy2020,
       author = {{Mahy}, L. and {Sana}, H. and {Abdul-Masih}, M. and {Almeida}, L.~A. and {Langer}, N. and {Shenar}, T. and {de Koter}, A. and {de Mink}, S.~E. and {de Wit}, S. and {Grin}, N.~J. and {Evans}, C.~J. and {Moffat}, A.~F.~J. and {Schneider}, F.~R.~N. and {Barb{\'a}}, R. and {Clark}, J.~S. and {Crowther}, P. and {Gr{\"a}fener}, G. and {Lennon}, D.~J. and {Tramper}, F. and {Vink}, J.~S.},
        title = "{The Tarantula Massive Binary Monitoring. III. Atmosphere analysis of double-lined spectroscopic systems}",
      journal = {\aap},
         year = 2020,
        month = feb,
       volume = {634},
          eid = {A118},
        pages = {A118},
          doi = {10.1051/0004-6361/201936151},
archivePrefix = {arXiv},
       eprint = {1912.08107},
 primaryClass = {astro-ph.SR},
       adsurl = {https://ui.adsabs.harvard.edu/abs/2020A&A...634A.118M}
}

@ARTICLE{bowman2020,
       author = {{Bowman}, Dominic M.},
        title = "{Asteroseismology of high-mass stars: new insights of stellar interiors with space telescopes}",
      journal = {Frontiers in Astronomy and Space Sciences},
         year = 2020,
        month = oct,
       volume = {7},
          eid = {70},
        pages = {70},
          doi = {10.3389/fspas.2020.578584},
archivePrefix = {arXiv},
       eprint = {2008.11162},
 primaryClass = {astro-ph.SR},
       adsurl = {https://ui.adsabs.harvard.edu/abs/2020FrASS...7...70B}
}

@article{gaia,
	adsurl = {http://adsabs.harvard.edu/abs/2016A%26A...595A...1G},
	archiveprefix = {arXiv},
	author = {{Gaia Collaboration} and {Prusti}, T. and {de Bruijne}, J.~H.~J. and {Brown}, A.~G.~A. and {Vallenari}, A. and {Babusiaux}, C. and {Bailer-Jones}, C.~A.~L. and {Bastian}, U. and {Biermann}, M. and {Evans}, D.~W. and et al.},
	doi = {10.1051/0004-6361/201629272},
	eid = {A1},
	eprint = {1609.04153},
	journal = {\aap},
	month = nov,
	pages = {A1},
	primaryclass = {astro-ph.IM},
	title = {{The Gaia mission}},
	volume = 595,
	year = 2016
}

@ARTICLE{hirschi2025,
       author = {{Hirschi}, R. and {Goodman}, K. and {Meynet}, G. and {Maeder}, A. and {Ekstr{\"o}m}, S. and {Eggenberger}, P. and {Georgy}, C. and {Sibony}, Y. and {Yusof}, N. and {Martinet}, S. and {Varma}, Vishnu and {Nomoto}, K.},
        title = "{The fate of rotating massive stars across cosmic times}",
      journal = {\mnras},
         year = 2025,
        month = nov,
       volume = {543},
       number = {3},
        pages = {2796-2815},
          doi = {10.1093/mnras/staf1470},
archivePrefix = {arXiv},
       eprint = {2508.21233},
 primaryClass = {astro-ph.SR},
       adsurl = {https://ui.adsabs.harvard.edu/abs/2025MNRAS.543.2796H}
}

@ARTICLE{schneider2020,
       author = {{Schneider}, F.~R.~N. and {Ohlmann}, S.~T. and {Podsiadlowski}, Ph and {R{\"o}pke}, F.~K. and {Balbus}, S.~A. and {Pakmor}, R.},
        title = "{Long-term evolution of a magnetic massive merger product}",
      journal = {\mnras},
         year = 2020,
        month = jul,
       volume = {495},
       number = {3},
        pages = {2796-2812},
          doi = {10.1093/mnras/staa1326},
archivePrefix = {arXiv},
       eprint = {2005.05335},
 primaryClass = {astro-ph.SR},
       adsurl = {https://ui.adsabs.harvard.edu/abs/2020MNRAS.495.2796S}
}

@ARTICLE{podsiadlowski1990,
       author = {{Podsiadlowski}, P. and {Joss}, P.~C. and {Rappaport}, S.},
        title = "{A merger model for SN 1987A.}",
      journal = {\aap},
         year = 1990,
        month = jan,
       volume = {227},
        pages = {L9-L12},
       adsurl = {https://ui.adsabs.harvard.edu/abs/1990A&A...227L...9P}
}

@ARTICLE{bastian2020,
       author = {{Bastian}, Nate and {Kamann}, Sebastian and {Amard}, Louis and {Charbonnel}, Corinne and {Haemmerl{\'e}}, Lionel and {Matt}, Sean P.},
        title = "{On the origin of the bimodal rotational velocity distribution in stellar clusters: rotation on the pre-main sequence}",
      journal = {\mnras},
         year = 2020,
        month = jun,
       volume = {495},
       number = {2},
        pages = {1978-1983},
          doi = {10.1093/mnras/staa1332},
archivePrefix = {arXiv},
       eprint = {2005.01779},
 primaryClass = {astro-ph.SR},
       adsurl = {https://ui.adsabs.harvard.edu/abs/2020MNRAS.495.1978B}
}

@ARTICLE{blex2024,
       author = {{Blex}, Susanne and {Haas}, Martin and {Chini}, Rolf},
        title = "{The rotation rate of single- and double-lined southern O stars: Determining what increases the rotation rate in binaries}",
      journal = {\aap},
         year = 2024,
        month = dec,
       volume = {692},
          eid = {A192},
        pages = {A192},
          doi = {10.1051/0004-6361/202450176},
       adsurl = {https://ui.adsabs.harvard.edu/abs/2024A&A...692A.192B}
}

@ARTICLE{shenar2020,
       author = {{Shenar}, T. and {Bodensteiner}, J. and {Abdul-Masih}, M. and {Fabry}, M. and {Mahy}, L. and {Marchant}, P. and {Banyard}, G. and {Bowman}, D.~M. and {Dsilva}, K. and {Hawcroft}, C. and {Reggiani}, M. and {Sana}, H.},
        title = "{The ``hidden'' companion in LB-1 unveiled by spectral disentangling}",
      journal = {\aap},
         year = 2020,
        month = jul,
       volume = {639},
          eid = {L6},
        pages = {L6},
          doi = {10.1051/0004-6361/202038275},
archivePrefix = {arXiv},
       eprint = {2004.12882},
 primaryClass = {astro-ph.SR},
       adsurl = {https://ui.adsabs.harvard.edu/abs/2020A&A...639L...6S}
}

@ARTICLE{bodensteiner2020,
       author = {{Bodensteiner}, J. and {Shenar}, T. and {Mahy}, L. and {Fabry}, M. and {Marchant}, P. and {Abdul-Masih}, M. and {Banyard}, G. and {Bowman}, D.~M. and {Dsilva}, K. and {Frost}, A.~J. and {Hawcroft}, C. and {Reggiani}, M. and {Sana}, H.},
        title = "{Is HR 6819 a triple system containing a black hole?. An alternative explanation}",
      journal = {\aap},
         year = 2020,
        month = sep,
       volume = {641},
          eid = {A43},
        pages = {A43},
          doi = {10.1051/0004-6361/202038682},
archivePrefix = {arXiv},
       eprint = {2006.10770},
 primaryClass = {astro-ph.SR},
       adsurl = {https://ui.adsabs.harvard.edu/abs/2020A&A...641A..43B}
}

@ARTICLE{klencki2022,
       author = {{Klencki}, Jakub and {Istrate}, Alina and {Nelemans}, Gijs and {Pols}, Onno},
        title = "{Partial-envelope stripping and nuclear-timescale mass transfer from evolved supergiants at low metallicity}",
      journal = {\aap},
         year = 2022,
        month = jun,
       volume = {662},
          eid = {A56},
        pages = {A56},
          doi = {10.1051/0004-6361/202142701},
archivePrefix = {arXiv},
       eprint = {2111.10271},
 primaryClass = {astro-ph.SR},
       adsurl = {https://ui.adsabs.harvard.edu/abs/2022A&A...662A..56K}
}

@ARTICLE{bodensteiner2023,
       author = {{Bodensteiner}, J. and {Sana}, H. and {Dufton}, P.~L. and {Wang}, C. and {Langer}, N. and {Banyard}, G. and {Mahy}, L. and {de Koter}, A. and {de Mink}, S.~E. and {Evans}, C.~J. and {G{\"o}tberg}, Y. and {H{\'e}nault-Brunet}, V. and {Patrick}, L.~R. and {Schneider}, F.~R.~N.},
        title = "{The young massive SMC cluster NGC 330 seen by MUSE. III. Stellar parameters and rotational velocities}",
      journal = {\aap},
         year = 2023,
        month = dec,
       volume = {680},
          eid = {A32},
        pages = {A32},
          doi = {10.1051/0004-6361/202345950},
archivePrefix = {arXiv},
       eprint = {2308.14799},
 primaryClass = {astro-ph.SR},
       adsurl = {https://ui.adsabs.harvard.edu/abs/2023A&A...680A..32B}
}

@ARTICLE{maeder1999,
       author = {{Maeder}, Andr{\'e} and {Grebel}, Eva K. and {Mermilliod}, Jean-Claude},
        title = "{Differences in the fractions of Be stars in galaxies}",
      journal = {\aap},
         year = 1999,
        month = jun,
       volume = {346},
        pages = {459-464},
          doi = {10.48550/arXiv.astro-ph/9904008},
archivePrefix = {arXiv},
       eprint = {astro-ph/9904008},
 primaryClass = {astro-ph},
       adsurl = {https://ui.adsabs.harvard.edu/abs/1999A&A...346..459M}
}

@ARTICLE{schootemeijer2022,
       author = {{Schootemeijer}, A. and {Lennon}, D.~J. and {Garcia}, M. and {Langer}, N. and {Hastings}, B. and {Sch{\"u}rmann}, C.},
        title = "{A census of OBe stars in nearby metal-poor dwarf galaxies reveals a high fraction of extreme rotators}",
      journal = {\aap},
         year = 2022,
        month = nov,
       volume = {667},
          eid = {A100},
        pages = {A100},
          doi = {10.1051/0004-6361/202244730},
archivePrefix = {arXiv},
       eprint = {2209.04943},
 primaryClass = {astro-ph.GA},
       adsurl = {https://ui.adsabs.harvard.edu/abs/2022A&A...667A.100S}
}

@ARTICLE{cornett1997,
       author = {{Cornett}, Robert H. and {Greason}, Michael R. and {Hill}, Jesse K. and {Parker}, Joel Wm. and {Waller}, William H. and {Bohlin}, R.~C. and {Cheng}, K.-P. and {Neff}, S.~G. and {O'Connell}, R.~W. and {Roberts}, M.~S. and {Smith}, A.~M. and {Stecher}, T.~P.},
        title = "{UIT: Ultraviolet Observations of the Small Magellanic Cloud}",
      journal = {\aj},
         year = 1997,
        month = mar,
       volume = {113},
        pages = {1011-1021},
          doi = {10.1086/118317},
       adsurl = {https://ui.adsabs.harvard.edu/abs/1997AJ....113.1011C}
}

@ARTICLE{schneider2018,
       author = {{Schneider}, F.~R.~N. and {Sana}, H. and {Evans}, C.~J. and {Bestenlehner}, J.~M. and {Castro}, N. and {Fossati}, L. and {Gr{\"a}fener}, G. and {Langer}, N. and {Ram{\'\i}rez-Agudelo}, O.~H. and {Sab{\'\i}n-Sanjuli{\'a}n}, C. and {Sim{\'o}n-D{\'\i}az}, S. and {Tramper}, F. and {Crowther}, P.~A. and {de Koter}, A. and {de Mink}, S.~E. and {Dufton}, P.~L. and {Garcia}, M. and {Gieles}, M. and {H{\'e}nault-Brunet}, V. and {Herrero}, A. and {Izzard}, R.~G. and {Kalari}, V. and {Lennon}, D.~J. and {Ma{\'\i}z Apell{\'a}niz}, J. and {Markova}, N. and {Najarro}, F. and {Podsiadlowski}, Ph. and {Puls}, J. and {Taylor}, W.~D. and {van Loon}, J. Th. and {Vink}, J.~S. and {Norman}, C.},
        title = "{An excess of massive stars in the local 30 Doradus starburst}",
      journal = {Science},
         year = 2018,
        month = jan,
       volume = {359},
       number = {6371},
        pages = {69-71},
          doi = {10.1126/science.aan0106},
archivePrefix = {arXiv},
       eprint = {1801.03107},
 primaryClass = {astro-ph.SR},
       adsurl = {https://ui.adsabs.harvard.edu/abs/2018Sci...359...69S}
}

@ARTICLE{dufton2018,
       author = {{Dufton}, P.~L. and {Thompson}, A. and {Crowther}, P.~A. and {Evans}, C.~J. and {Schneider}, F.~R.~N. and {de Koter}, A. and {de Mink}, S.~E. and {Garland}, R. and {Langer}, N. and {Lennon}, D.~J. and {McEvoy}, C.~M. and {Ram{\'\i}rez-Agudelo}, O.~H. and {Sana}, H. and {S{\'\i}mon D{\'\i}az}, S. and {Taylor}, W.~D. and {Vink}, J.~S.},
        title = "{The VLT-FLAMES Tarantula Survey. XXVIII. Nitrogen abundances for apparently single dwarf and giant B-type stars with small projected rotational velocities}",
      journal = {\aap},
         year = 2018,
        month = jul,
       volume = {615},
          eid = {A101},
        pages = {A101},
          doi = {10.1051/0004-6361/201732440},
archivePrefix = {arXiv},
       eprint = {1804.02025},
 primaryClass = {astro-ph.SR},
       adsurl = {https://ui.adsabs.harvard.edu/abs/2018A&A...615A.101D}
}

@ARTICLE{garland2017,
       author = {{Garland}, R. and {Dufton}, P.~L. and {Evans}, C.~J. and {Crowther}, P.~A. and {Howarth}, I.~D. and {de Koter}, A. and {de Mink}, S.~E. and {Grin}, N.~J. and {Langer}, N. and {Lennon}, D.~J. and {McEvoy}, C.~M. and {Sana}, H. and {Schneider}, F.~R.~N. and {S{\'\i}mon D{\'\i}az}, S. and {Taylor}, W.~D. and {Thompson}, A. and {Vink}, J.~S.},
        title = "{The VLT-FLAMES Tarantula Survey. XXVII. Physical parameters of B-type main-sequence binary systems in the Tarantula nebula}",
      journal = {\aap},
         year = 2017,
        month = jul,
       volume = {603},
          eid = {A91},
        pages = {A91},
          doi = {10.1051/0004-6361/201629982},
archivePrefix = {arXiv},
       eprint = {1704.07131},
 primaryClass = {astro-ph.SR},
       adsurl = {https://ui.adsabs.harvard.edu/abs/2017A&A...603A..91G}
}

@ARTICLE{mcevoy2015,
       author = {{McEvoy}, C.~M. and {Dufton}, P.~L. and {Evans}, C.~J. and {Kalari}, V.~M. and {Markova}, N. and {Sim{\'o}n-D{\'\i}az}, S. and {Vink}, J.~S. and {Walborn}, N.~R. and {Crowther}, P.~A. and {de Koter}, A. and {de Mink}, S.~E. and {Dunstall}, P.~R. and {H{\'e}nault-Brunet}, V. and {Herrero}, A. and {Langer}, N. and {Lennon}, D.~J. and {Ma{\'\i}z Apell{\'a}niz}, J. and {Najarro}, F. and {Puls}, J. and {Sana}, H. and {Schneider}, F.~R.~N. and {Taylor}, W.~D.},
        title = "{The VLT-FLAMES Tarantula Survey. XIX. B-type supergiants: Atmospheric parameters and nitrogen abundances to investigate the role of binarity and the width of the main sequence}",
      journal = {\aap},
         year = 2015,
        month = mar,
       volume = {575},
          eid = {A70},
        pages = {A70},
          doi = {10.1051/0004-6361/201425202},
archivePrefix = {arXiv},
       eprint = {1412.2705},
 primaryClass = {astro-ph.SR},
       adsurl = {https://ui.adsabs.harvard.edu/abs/2015A&A...575A..70M}
}

\begin{appendix}

\section{Fourier transform uncertainties}
\label{app:errors}
Some typical Fourier transforms are shown in Fig.\,\ref{fig:plotfts}, illustrating the reasonable agreement between results for different lines. 
For these sources the FWHM values are within 10\% of the FT values.
As noted in Sect. \ref{results}, the FT and FWHM values of \vsini\ may be used to gauge reliability.
For example, three stars (5-001, 5-051 and 6-113) have FT uncertainties greater than 100\,\kms. These are the result of poorly defined line profiles due to intrinsic emission (6-113 is an extreme Oe star), poor s/n (5-051), and line profile distortion (5-001 is flagged as a possible SB2 by \citealt{lamb2016}).
An additional $\sim$20 sources uncertainties above $\sim$50\,kms and inspection of these spectra reveals a number of similar causes: low s/n, nebular emission, OBe emission, a possible triple system, etc.
For most of these cases the FWHM value appears to be more robust, however we have retained the FT values in our analysis as the small number of these outliers has little impact on the overall picture.
Nevertheless, for a specific source, of sources, we emphasize the need to examine both FT and FWHM values, and their uncertainties.

\begin{figure*}
    \centering
    \includegraphics[width=0.45\linewidth]{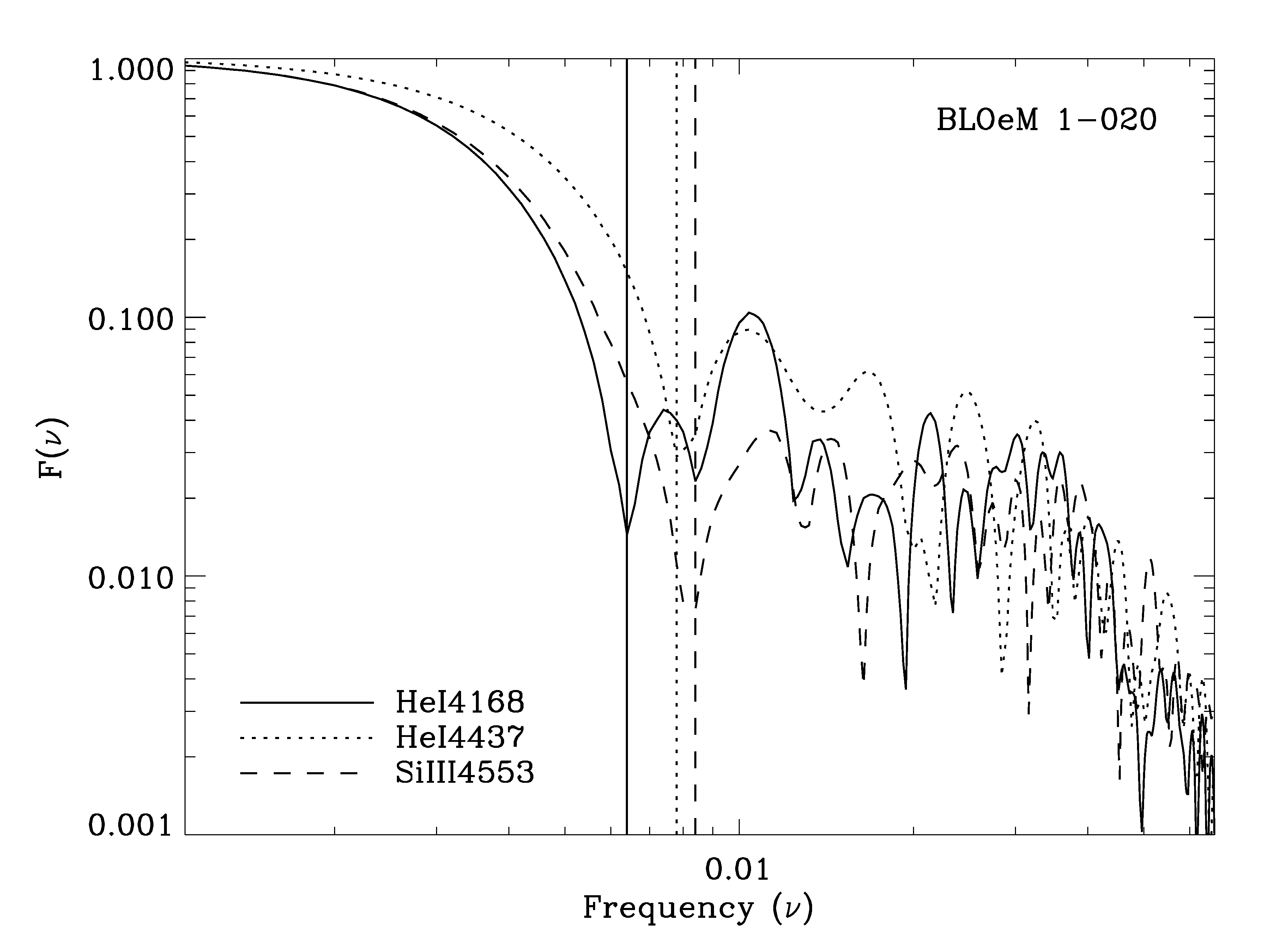}
    \includegraphics[width=0.45\linewidth]{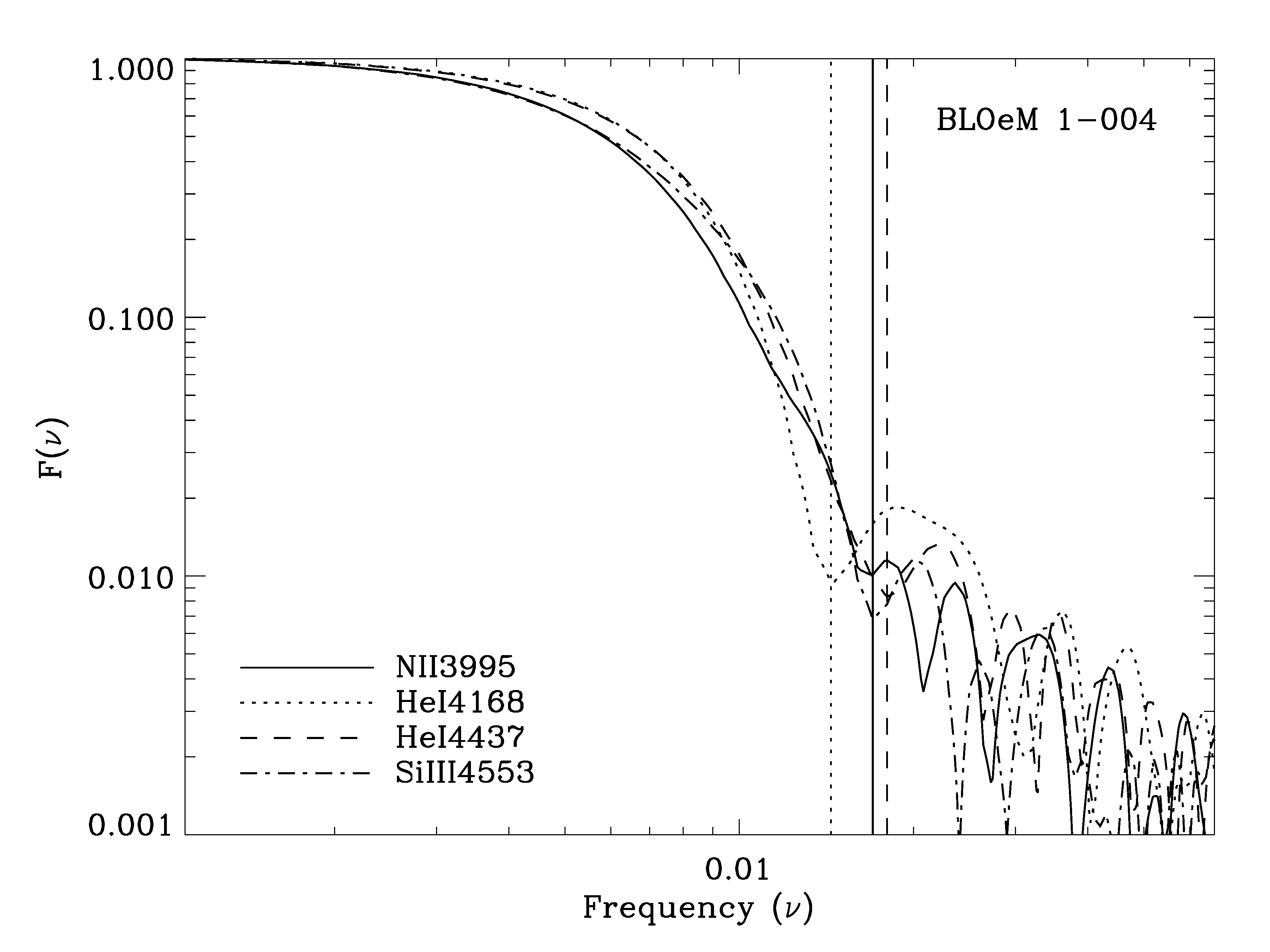}
    \includegraphics[width=0.45\linewidth]{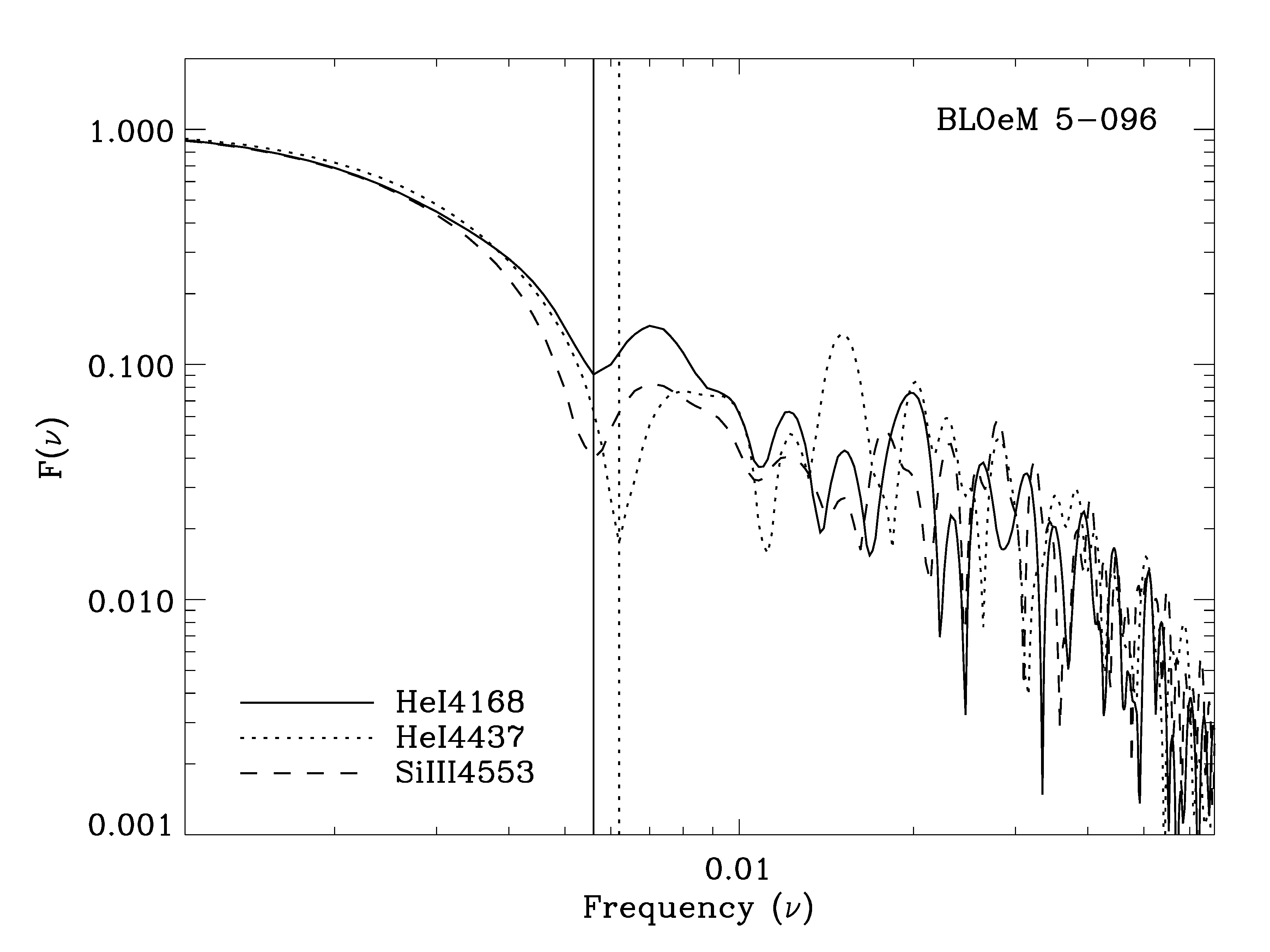}
    \includegraphics[width=0.45\linewidth]{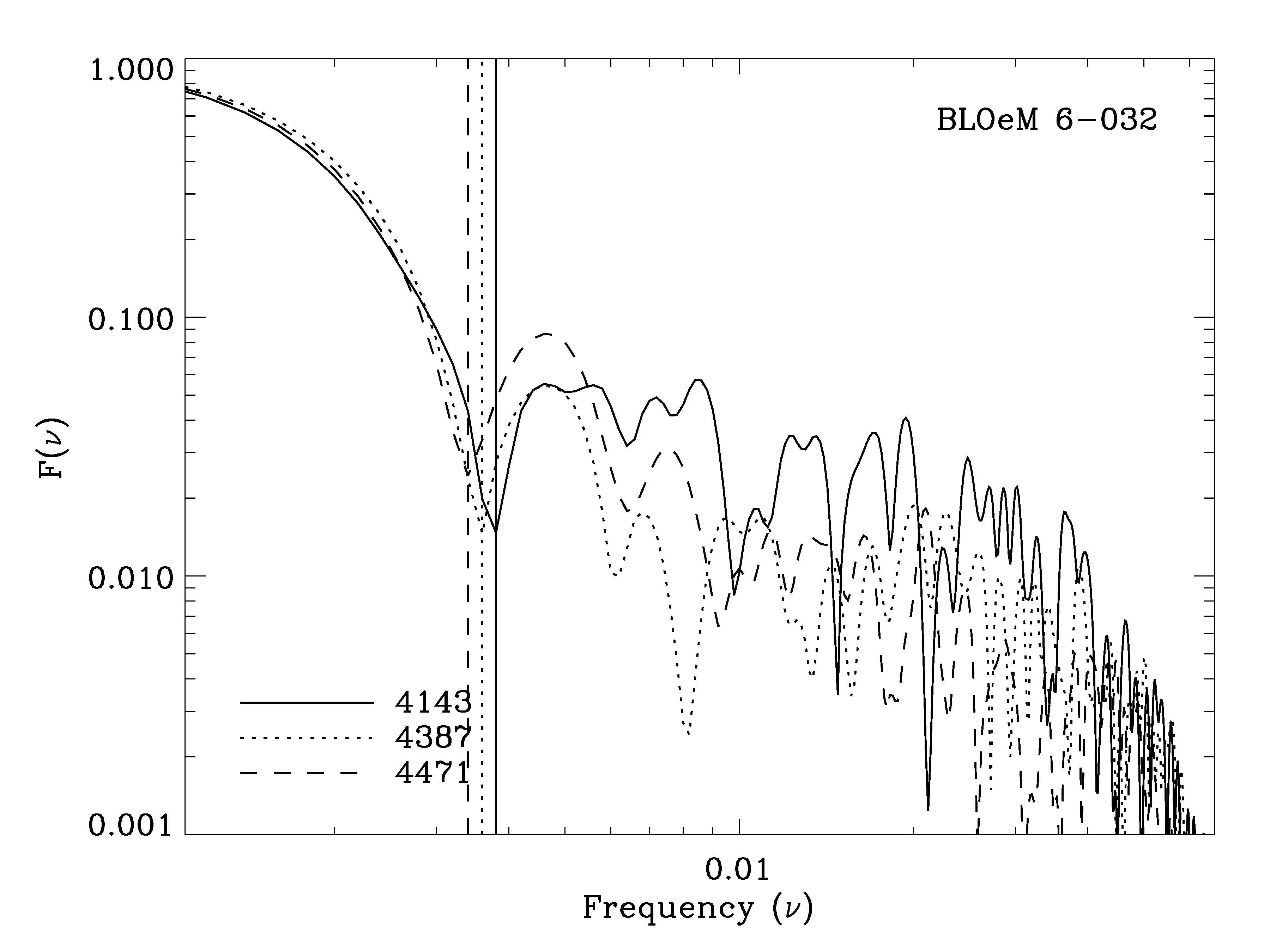}
    \caption{Example Fourier transforms, with `zeros' (or minima) indicated by vertical lines, the BLOeM identifier and line identifications are inset. Spectral types (and mean \vsini\ in parenthesis), in clockwise order from top left are B0\,III (88\,\kms), B1\,Ib (40\,\kms), O9\,V:(n) (187\,\kms), and B1.5\,II=III (117\,\kms). }
    \label{fig:plotfts}
\end{figure*}

\section{Peculiar/individual objects}
\label{peculiar}

\subsection{The SgB[e] stars}
\label{sgbe}

The SgB[e] stars BLOeM 2-116, 3-012 and 4-055 are more commonly referred to in the literature by their aliases, LHA\,115-S\,18 (or AzV\,154), LHA\,115-S\,6 (or RMC\,4, AzV\,16, Sk\,11) and LHA\,115-S\,29 (or RMC\,15, Sk\,79) respectively. 
We will therefore refer to these systems as S\,18, S\,6 and S\,29 in this short discussion.
Their spectra are dominated by allowed and forbidden emission lines, although S\,6 and S\,29 also exhibit a B-type absorption line spectrum that contains sharp lines of ions such \ion{He}{i}, \ion{N}{ii}, \ion{O}{ii}, \ion{Mg}{ii}, \ion{Si}{ii} and \ion{Si}{iv}.
All sources exhibit radial velocity variations for the emission and absorption components, with the absorption lines S\,6 and S\,29 yielding \vsinift\ values of 62 and 50\,\kms respectively, although these values may be upper limits due to the inability to detect a secondary minimum in the FT.
Note that S\,6 was also found to be a binary by \citet{zickgraf1996}, who  obtained a period of $\sim$21\,yr, and also detected an A-type secondary, although the BLOeM data show no sign of this secondary at any epoch.
As noted, S\,18 exhibits no reliable absorption lines in our wavelength range, we nevertheless measured \vsini\ from several weak, isolated, emission lines, obtaining values in the range 30--50\,\kms, though these values may not reflect the rotation rate of the central star.

The low \vsini\ for S\,6 and S\,29 are perhaps a little surprising as SgB[e] stars are commonly supposed to be either interacting binaries, or the products of a stellar merger \citep{langer1998,pasquali2000,podsiadlowski2006} 
The stellar merger scenario would appear to be inconsistent with the apparent binary nature of  of these three systems (S\,18 is thought to be a binary due to its X-ray luminosity driven by a colliding wind; \citealt{clark2013}).
There are only a few published measurements of \vsini\ for SgB[e] stars, \citet{zickgraf2006} gives approximate values of 150, 50 and 65\,\kms for RMC\,50, RMC\,66 and Hen\,S93\,65 respectively. 
The present low \vsini\ results also appear to suggest that on average the primary stars of these systems have rotational velocities well below critical values, perhaps suggesting that these are the donor stars in the systems.

\subsection{The O\,f?p stars}

Two BLOeM sources in the OBe group have previously been classified as rare Of?p stars rather than classical Be stars; 2-104 (alias 2dFS\,936) and 4-039 (AzV\, 220) that were both classified as O6.5\,f?p by \citet{evans2004} and \citet{walborn2000} respectively.
Both stars exhibit Balmer emission, being classified as  O5.5\,f?pe and O6.5\,f?pe in \citet{shenar2024}, reflecting the presence of Balmer emission in their spectra (as in previous references quoted above).
These stars are known to be magnetic, very slow rotators that exhibit significant variability correlated with their long rotation period \citep[see for example][]{howarth2007}. Therefore the change in spectral type for 2-104 from O6.5 to O5.5 is not unusual, but reflects the presence of infilling of the \ion{He}{i} 4471\,\AA\ line by emission in the BLOeM data, which also prohibits the use of this line for \vsini\ measurement.
Its \vsini\ is 90\,\kms, based on three measurements of 68, 83 and 150\,\kms for the lines \ion{He}{ii} 4026\,\AA and \ion{He}{i} 4143, 4387\,\AA. 
BLOeM 4-039 on the other hand has a \vsini\ of 45\,\kms determined from 7 absorption lines, compared with 62\,\kms from a GOF determination by \citet{walborn2000} from high resolution (R=25\,000) data.
Both BLOeM measurements, and that of \citet{walborn2000}, should be considered as upper limits and while consistent with these stars' expected very long rotation periods do not provide meaningful constraints on their values.

\subsection{Fast rotators}

As \citet{walborn2014} noted some correlation between high \vsini\ and potential runaway status we briefly consider a small subset of the highest \vsini\ systems in BLOeM.
Four targets have \vsini\ measurements in excess of 400\,\kms; BLOeM 3-002 (O6\,V:nn), 5-051 (B0.2\,III), 7-072 (O8\,Vnn) and 8-021 (O7\,V-IIInnn\,pe).
BLOeM 3-002 (Simbad name [M2002] SMC 5880) resides in the SMC bar, is not especially isolated, has a rather normal radial velocity (168\,\kms).
While formally classed as `single' the star is faint, $G$=15.44, and some individual epochs have a s/n as low as 20.
A temporal analysis of its line profiles might be informative as some O-type SB2 systems in the past have been classified as On stars \citep{mahy2020}.
BLOeM 5-051 has been flagged as RVvar by \citet{villasenor2025}, but it is also faint, $G$=16.31, and the s/n of each epoch is approximately 20, which may account for the reported variability.
BLOeM 7-072 (AzV 251) has an anomalously low radial velocity for the SMC, we derive $\sim$80\,\kms \citep[as do][within the X-shooting ULLYSES project]{sana2024}, while \citet{dorigojones2020} report this star as a potential runaway with a peculiar transverse velocity of 62\,\kms\ from $Gaia$ proper motion data.
BLOeM 8-021 (AzV 113) is an Oe star exhibiting typical OBe double peaked emission in its Balmer lines. Its \ion{He}{ii} 4200,4542\,\AA\ lines are in absorption however and yield a mean radial velocity of $\sim$130\,\kms, not a significant outlier given the SMC's radial velocity dispersion \citep{evans2008}.
A more detailed search for runaway stars will be presented in a forthcoming paper.

\section{Master table}

Much of the information used in discussing the various trends revealed here is presented in other papers of the BLOeM series \citep{shenar2024,villasenor2025,britavskiy2025,bodensteiner2025,patrick2025,bestenlehner2025}. Therefore we collect relevant information into a master table, Table C.1, as a convenience for interested readers. This table, available at the CDS contains the following information. 
\newline
Column 1: BLOeM identifier \newline
Column 2: Spectral type \newline
Column 3: Binary status \newline
Column 4: \vsini\ (\kms) \newline
Column 5: Uncertainty in \vsini\ (upper) \newline
Column 6: Uncertainty in \vsini\ (lower) \newline
Column 7: \teff \newline
Column 8: $\log L/L_{\odot}$ \newline
Column 9: Evolutionary mass ($M_\odot$) \newline
Column 10: Stellar radius ($R_\odot$) \newline
Column 11: \vcrit\ (\kms) \newline
Column 12: Field number in which the source lies \newline
Column 13: pflag is the parameters flag (see below) \newline
Column 14: mflag is the multiplicity flag (see below) \newline
Column 15: gflag is the group flag (see below) \newline \newline
The pflag denotes the source used for the stellar parameters, in columns 7--11, as follows: \newline
-1 -- No parameters are available \newline
0 -- \citet{shenar2024} \newline
1 -- \citet{bestenlehner2025} \newline
2 -- \citet{patrick2025} \newline \newline
The mflag denotes the multiplicity adopted for sources as described at the beginning of Sect. \ref{results}: \newline
-1 -- not set \newline
0 -- single \newline
1 -- not used \newline
2 -- SB1 \newline
3 -- SB2 \newline \newline
The gflag denotes BLOeM groups as originally defined in \citet{shenar2024}: \newline
0 -- O-type stars \newline
1 -- B-type dwarfs and giants \newline
2 -- B-type supergiants and bright giants \newline
3 -- OBe stars \newline
4 -- BAF supergiants \newline

\section{Nitrogen in low \vsini\ systems}
\label{app:nii}
\begin{figure}
        \includegraphics[width=1.0\linewidth]{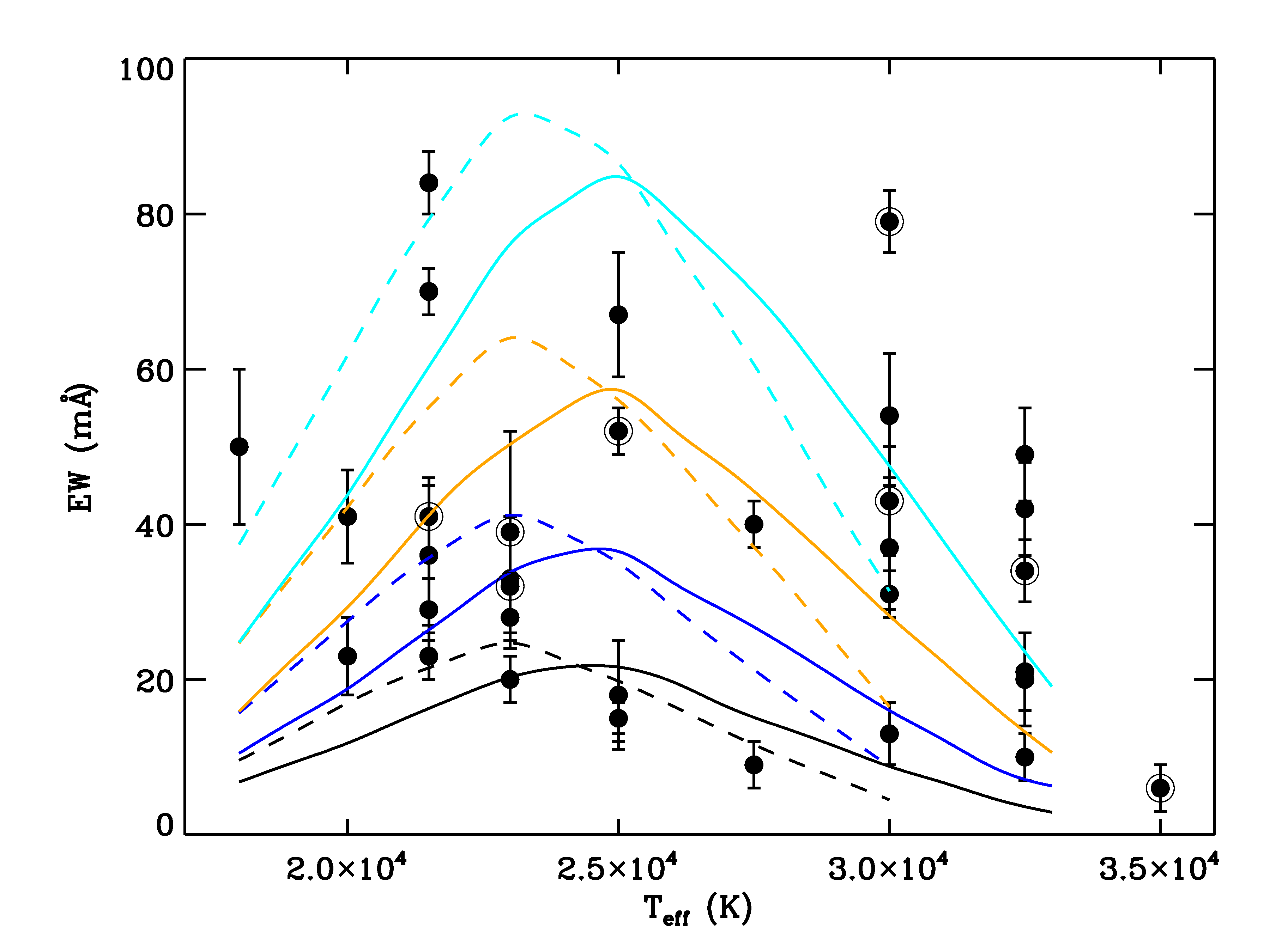}     
        \caption{Estimates of the equivalent widths for the \ion{N}{ii}\ line at 3995\AA\ for sources with \vsini$\leq$40\,\kms\ plotted against effective temperature. Formal uncertainties in \teff\ are half the grid separation, which can be inferred from inspection of the figure. Also shown are theoretical predictions for two gravities; \logg=4.0 dex (solid lines) and 3.5 dex (dotted lines). Black curves are for the SMC baseline nitrogen abundance of 6.5\,dex, with other curves showing enhancements of 0.3\,dex (blue), 0.6\,dex (orange) and 0.9\,dex (cyan). SB1 systems are circled.}
        \label{fig:plot_nii}
\end{figure} 

To facilitate comparison with previous results we have re-estimated the atmospheric parameters using the same model-atmosphere grid that was calculated with the {\sc tlusty} and {\sc synspec} codes \citep{hubeny1988,lanz2007} as described in \citet{dufton2005}.
These models cover an effective temperature range of 10\,000\,K $\leq$\teff$\leq$35\,000\,K in steps of typically 1\,500\,K.  
Logarithmic gravities (in cm s$^{-2}$) range from 4.5 dex to the Eddington limit in steps of 0.25 dex, and microturbulence velocities are from 0-30 \kms\ in steps of 5\,\kms.
In an analogous approach to that adopted by \cite{patrick2025} for the BAF supergiants, approximate stellar parameters were determined by cross-correlating each spectrum with the SMC metallicity subset of models after convolution with instrumental resolution and appropriate rotational broadening for each star.
In addition to enabling a direct comparison with published SMC results, this approach has the additional advantage of closing the `gap' found in \cite{bestenlehner2025}, with few B-type stars having \teff\ around 25\,000\,K. 
As discussed in that paper, the \teff\ estimates in this region depend on the adopted weighting of different spectral features; for example, effective temperature estimates differing by more than 6\,000\,K were found for star BLOeM 1-105, depending on the adopted weighting.
The present revised effective temperature estimates are now more uniformly spread with seven targets having estimates between 24\,000 and 30\,000\,K. Gravity estimates were normally either 3.5 or 4.0\,dex with higher values being found for some of the hotter stars. Microturbulence values of 0 (30 targets), 5 (4 targets) and 10\,\kms (9 targets) were deduced.

Of the 43 stars in this sample, reliable equivalent width measurements of the \ion{N}{ii} 3995\AA\ line are possible in 34 cases; we used the same techniques as in \citet{dufton2018,dufton2020} where further details can be found. These equivalent widths are plotted in Fig.\,\ref{fig:plot_nii} as a function of our \teff\ estimates.
\citet{dufton2020} has reviewed nitrogen abundances found in stars and \ion{H}{ii}\ regions in the SMC and adopted a baseline abundance of 6.5\,dex. We have used this baseline in Fig.\ \ref{fig:plot_nii} and show the loci of theoretical equivalent widths deduced from our grid of models for a microturbulence of 5\kms and for \logg\ of 3.5 and 4.0\,dex. Also plotted are loci for nitrogen enhancements of up to 0.9 dex.

Inspection of Fig.\,\ref{fig:plot_nii} implies that most targets have significant nitrogen enhancements. In Table \ref{t_nii}, we summarize the number of stars with different levels of enhancement. Also listed are results for the SMC region of NGC346 \citep{dufton2020} and for 30 Doradus in the LMC \citep{dufton2018}. The percentages of nitrogen enhanced stars are very similar for the two SMC studies but are larger than those for the LMC survey; this is not surprising given the different baseline nitrogen abundances. 
Also noteworthy is that roughly one fifth of this subsample are SB1 systems, and a preliminary check of their radial velocities implies that their periods are long, and of order tens of days.
While a full orbital and atmospheric analysis of the sample is beyond the scope of present paper, inspection of Fig.\,\ref{fig:plot_nii} implies that $\sim$30\% (17/43) of the slowly rotating single and binary B-type stars are enhanced by at least 0.6 dex. 

\begin{table}
        \caption{Number and percentages of nitrogen enriched stars with \vsini$\leq$40\,\kms, where $N$\ is the number of B-type stars with luminosity classes III-V in each survey.}\label{t_nii}
        \begin{center}
                \begin{tabular}{lcrrrcrrrrrrr}
                        \hline
                        Survey & $N$ & \multicolumn{6}{c}{Nitrogen enhancements}\\
                        & & \multicolumn{2}{c}{$>$0.3 dex} & \multicolumn{2}{c}{$>$0.6 dex} & \multicolumn{2}{c}{$>$0.9 dex}\\ \hline
                        BLOeM & 319 & 20 & 6.3\% & 14 & 4.4\% & 7 & 2.2\% \\
                        NGC346 & 211 & 14 & 6.6\% & 7 & 3.3\% & 4 & 1.9\% \\
                        VFTS & 255 & 7 & 2.7\% & 5 & 2.0\% & - & - \\
                        \hline
                \end{tabular}
        \end{center}
\end{table}

\end{appendix}

\end{document}